%% file: main.tex
\documentclass[twocolumn,tighten]{aastex63}
\usepackage{natbib}
\usepackage{epsfig}
\usepackage{amsmath}
\usepackage{xcolor}
\usepackage{hyperref}
\usepackage[encapsulated]{CJK}
\usepackage{ucs}
\usepackage[utf8x]{inputenc}
\usepackage{url}

\newcommand{\tabdiv}{~---~}

\newcommand{\hi}{\mbox{\rm \ion{H}{1}}}

\newcommand{\cotwo}{\mbox{\rm CO($2\text{--}1$)}}

\newcommand{\OSU}{\affil{Department of Astronomy, The Ohio State University, 140 West 18th Avenue, Columbus, Ohio 43210, USA}}

\newcommand{\Alberta}{\affil{Department of Physics, University of Alberta, Edmonton, AB T6G 2E1, Canada}}

\newcommand{\ANU}{\affil{Research School of Astronomy and Astrophysics, Australian National University, Canberra, ACT 2611, Australia}}

\newcommand{\IPAC}{\affil{Caltech-IPAC, 1200 E. California Blvd. Pasadena, CA 91125, USA}}

\newcommand{\Carnegie}{\affil{Observatories of the Carnegie Institution for Science, 813 Santa Barbara Street, Pasadena, CA 91101, USA}}

\newcommand{\CCAPP}{\affil{Center for Cosmology and Astroparticle Physics, 191 West Woodruff Avenue, Columbus, OH 43210, USA}}

\newcommand{\CNRS}{\affil{CNRS, IRAP, 9 Av. du Colonel Roche, BP 44346, F-31028 Toulouse cedex 4, France}}

\newcommand{\ESO}{\affil{European Southern Observatory, Karl-Schwarzschild Stra{\ss}e 2, D-85748 Garching bei M\"{u}nchen, Germany}}

\newcommand{\Heidelberg}{\affil{Astronomisches Rechen-Institut, Zentrum f\"{u}r Astronomie der Universit\"{a}t Heidelberg, M\"{o}nchhofstra\ss e 12-14, D-69120 Heidelberg, Germany}}

\newcommand{\ICRAR}{\affil{International Centre for Radio Astronomy Research, University of Western Australia, 35 Stirling Highway, Crawley, WA 6009, Australia}}

\newcommand{\IRAM}{\affil{Institut de Radioastronomie Millim\'{e}trique (IRAM), 300 Rue de la Piscine, F-38406 Saint Martin d'H\`{e}res, France}}

\newcommand{\ITA}{\affil{Universit\"{a}t Heidelberg, Zentrum f\"{u}r Astronomie, Institut f\"{u}r Theoretische Astrophysik, Albert-Ueberle-Str 2, D-69120 Heidelberg, Germany}}

\newcommand{\IWR}{\affil{Universit\"{a}t Heidelberg, Interdisziplin\"{a}res Zentrum f\"{u}r Wissenschaftliches Rechnen, Im Neuenheimer Feld 205, D-69120 Heidelberg, Germany}}

\newcommand{\JHU}{\affil{Department of Physics and Astronomy, The Johns Hopkins University, Baltimore, MD 21218, USA}}

\newcommand{\Leiden}{\affil{Leiden Observatory, Leiden University, P.O. Box 9513, 2300 RA Leiden, The Netherlands}}

\newcommand{\Maryland}{\affil{Department of Astronomy, University of Maryland, College Park, MD 20742, USA}}

\newcommand{\MPE}{\affil{Max-Planck-Institut f\"{u}r extraterrestrische Physik, Giessenbachstra{\ss}e 1, D-85748 Garching, Germany}}

\newcommand{\MPIA}{\affil{Max-Planck-Institut f\"{u}r Astronomie, K\"{o}nigstuhl 17, D-69117, Heidelberg, Germany}}

\newcommand{\NRAO}{\affil{National Radio Astronomy Observatory, 520 Edgemont Road, Charlottesville, VA 22903-2475, USA}}

\newcommand{\OAN}{\affil{Observatorio Astron\'{o}mico Nacional (IGN), C/Alfonso XII, 3, E-28014 Madrid, Spain}}

\newcommand{\ObsParis}{\affil{Sorbonne Universit\'{e}, Observatoire de Paris, Universit\'{e} PSL, CNRS, LERMA, F-75014, Paris, France}}

\newcommand{\UToledo}{\affil{University of Toledo, 2801 W. Bancroft St., Mail Stop 111, Toledo, OH, 43606}}

\newcommand{\Toulouse}{\affil{Universit\'{e} de Toulouse, UPS-OMP, IRAP, F-31028 Toulouse cedex 4, France}}

\newcommand{\UBonn}{\affil{Argelander-Institut f\"ur Astronomie, Universit\"at Bonn, Auf dem H\"ugel 71, 53121 Bonn, Germany}}

\newcommand{\UChile}{\affil{Departamento de Astronom\'{i}a, Universidad de Chile, Camino del Observatorio 1515, Las Condes, Santiago, Chile}}

\newcommand{\UCSD}{\affil{Center for Astrophysics and Space Sciences, Department of Physics,  University of California,\\ San Diego, 9500 Gilman Drive, La Jolla, CA 92093, USA}}

\newcommand{\UGent}{\affil{Sterrenkundig Observatorium, Universiteit Gent, Krijgslaan 281 S9, B-9000 Gent, Belgium}}

\newcommand{\ULyon}{\affil{Univ Lyon, Univ Lyon 1, ENS de Lyon, CNRS, Centre de Recherche Astrophysique de Lyon UMR5574,\\ F-69230 Saint-Genis-Laval, France}}

\newcommand{\UMass}{\affil{University of Massachusetts—Amherst, 710 N. Pleasant Street, Amherst, MA 01003, USA}}

\newcommand{\UWyoming}{\affil{Department of Physics and Astronomy, University of Wyoming, Laramie, WY 82071, USA}}

\newcommand{\LAM}{\affil{
Aix Marseille Univ, CNRS, CNES, LAM (Laboratoire d’Astrophysique de Marseille), Marseille,
France}}

\newcommand{\UHawaii}{\affil{Institute for Astronomy, University of Hawaii, 2680 Woodlawn Drive, Honolulu, HI 96822, USA}}

\newcommand{\UCM}{\affil{Departamento de F\'{\i}sica de la Tierra y Astrof\'{\i}sica, Universidad Complutense de Madrid, E-28040, Spain}}

\newcommand{\IPARC}{\affil{Instituto de F\'{\i}sica de Part\'{\i}culas y del Cosmos IPARCOS, Facultad de Ciencias F\'{\i}sicas, Universidad Complutense de Madrid, E-28040, Spain}}

\newcommand{\STScI}{\affil{Space Telescope Science Institute, 3700 San Martin Drive, Baltimore, MD 21218, USA}}

\newcommand{\McMaster}{\affil{Department of Physics and Astronomy, McMaster University, Hamilton, ON L8S 4M1, Canada}}

\newcommand{\INAF}{\affil{INAF -- Osservatorio Astrofisico di Arcetri, Largo E. Fermi 5, I-50157, Firenze, Italy}}

\newcommand{\Sydney}{\affil{Sydney Institute for Astronomy, School of Physics A28, The University of Sydney, NSW 2006, Australia}}

\newcommand{\UA}{\affil{Centro de Astronomía (CITEVA), Universidad de Antofagasta, Avenida Angamos 601, Antofagasta, Chile}}

\graphicspath{{./Figures/}}

\shorttitle{PHANGS--ALMA Data Processing and Pipeline}
\shortauthors{Leroy et al.}

\begin{document}

\title{PHANGS--ALMA Data Processing and Pipeline}

% Order is TBD - please add your name ORCID and affiliation below.

\correspondingauthor{Adam K. Leroy}
\email{leroy.42@osu.edu}

\author[0000-0002-2545-1700]{Adam~K.~Leroy}
\footnote{This paper represents a collective effort by the PHANGS--ALMA data reduction team. Please see a description of contributions of individual team members in Appendix \ref{sec:contrib}.}
\OSU

\author[0000-0002-9181-1161]{Annie~Hughes}
\CNRS
\Toulouse

\author[0000-0001-9773-7479]{Daizhong~Liu}
\MPIA

\author[0000-0003-3061-6546]{J\'er\^ome~Pety}
\IRAM
\ObsParis

\author[0000-0002-5204-2259]{Erik~Rosolowsky}
\Alberta

\author[0000-0002-2501-9328]{Toshiki~Saito}
\MPIA

\author[0000-0002-3933-7677]{Eva~Schinnerer}
\MPIA

\author{Andreas~Schruba}
\MPE

\author[0000-0003-1242-505X]{Antonio~Usero}
\OAN

%%%%%%%%%%%%%%%%%%%%%%%%%%%%%%%%%%%%%%%%%%%%%%%%%%

\author[0000-0001-5310-467X]{Christopher M. Faesi}
\UMass

\author[0000-0001-6405-0785]{Cinthya~N.~Herrera}
\IRAM

%%%%%%%%%%%%%%%%%%%%%%%%%%%%%%%%%%%%%%%%%%%%%%%%%%

\author[0000-0002-5635-5180]{M\'{e}lanie~Chevance}
\Heidelberg

\author[0000-0002-6488-471X]{Alexander P.~S.~Hygate}
\Leiden

\author[0000-0002-3227-4917]{Amanda A. Kepley}
\NRAO

\author[0000-0001-9605-780X]{Eric W. Koch}
\Alberta

\author[0000-0002-0472-1011]{Miguel~Querejeta}
\OAN

\author{Kazimierz Sliwa}
\MPIA

\author{David Will}
\OSU

\author[0000-0001-5817-0991]{Christine D. Wilson}
\McMaster

%%%%%%%%%%%%%%%%%%%%%%%%%%%%%%%%%%%%%%%%%%%%%%%%%%

\author[0000-0002-5259-2314]{Gagandeep S. Anand}
\UHawaii

\author[0000-0003-0410-4504]{Ashley Barnes}
\UBonn

\author[0000-0002-2545-5752]{Francesco Belfiore}
\INAF

\author[0000-0003-0583-7363]{Ivana Be\v{s}li\'c}
\UBonn

\author[0000-0003-0166-9745]{Frank Bigiel}
\UBonn

\author[0000-0003-4218-3944]{Guillermo A. Blanc}
\Carnegie
\UChile

\author[0000-0002-5480-5686]{Alberto D. Bolatto}
\Maryland

\author[0000-0003-0946-6176]{Médéric Boquien}
\UA

\author[0000-0001-5301-1326]{Yixian Cao}
\LAM

\author[0000-0003-0085-4623]{Rupali Chandar}
\UToledo

\author[0000-0002-5235-5589]{J\'er\'emy Chastenet}
\UCSD

\author[0000-0003-2551-7148]{I-Da Chiang}
\UCSD

\author[0000-0002-8549-4083]{Enrico Congiu}
\UChile
\Carnegie

\author[0000-0002-5782-9093]{Daniel A. Dale}
\UWyoming

\author[0000-0003-1943-723X]{Sinan Deger}
\IPAC

\author[0000-0002-8760-6157]{Jakob S. den Brok}
\UBonn

\author[0000-0002-1185-2810]{Cosima Eibensteiner}
\UBonn

\author[0000-0002-6155-7166]{Eric Emsellem}
\ESO
\ULyon

\author[0000-0002-0697-0177]{Axel Garc\'{i}a-Rodr\'{i}guez}
\OAN

\author[0000-0001-6708-1317]{Simon~C.~O.~Glover}
\ITA

\author[0000-0002-3247-5321]{Kathryn Grasha}
\ANU

\author[0000-0002-9768-0246]{Brent Groves}
\ANU
\ICRAR

\author[0000-0001-9656-7682]{Jonathan D. Henshaw}
\MPIA

\author[0000-0002-9165-8080]{Mar\'ia J. Jim\'enez Donaire}
\OAN

\author[0000-0002-0432-6847]{Jenny~J.~Kim}
\Heidelberg

\author[0000-0002-0560-3172]{Ralf S.\ Klessen}
\ITA
\IWR

\author[0000-0001-6551-3091]{Kathryn~Kreckel}
\Heidelberg

\author[0000-0002-8804-0212]{J.~M.~Diederik Kruijssen}
\Heidelberg

\author[0000-0003-3917-6460]{Kirsten L. Larson}
\IPAC

\author[0000-0002-2278-9407]{Janice C. Lee}
\IPAC

\author[0000-0002-5993-6685]{Ness Mayker}
\OSU 
\CCAPP

\author[0000-0003-2290-7060]{Rebecca McElroy}
\Sydney

\author[0000-0002-6118-4048]{Sharon E. Meidt}
\UGent

\author[0000-0001-7413-7534]{Angus Mok}
\UToledo

\author[0000-0002-1370-6964]{Hsi-An Pan}
\MPIA

\author[0000-0003-1111-3951]{Johannes Puschnig}
\UBonn

\author[0000-0001-7876-1713]{Alessandro Razza}
\UChile

\author[0000-0003-0651-0098]{Patricia S\'anchez-Bl'azquez}
\UCM
\IPARC

\author[0000-0002-4378-8534]{Karin M. Sandstrom}
\UCSD

\author[0000-0002-6363-9851]{Francesco Santoro}
\MPIA

\author[0000-0002-5783-145X]{Amy Sardone}
\OSU 
\CCAPP

\author{Fabian Scheuermann}
\Heidelberg

\author[0000-0003-0378-4667]{Jiayi~Sun}
\OSU

\author[0000-0002-8528-7340]{David A. Thilker}
\JHU

\author[0000-0003-2261-5746]{Jordan A. Turner}
\UWyoming

\author{Leonardo Ubeda}
\STScI

\author[0000-0003-4161-2639]{Dyas Utomo}
\OSU
\NRAO
\CCAPP

\author[0000-0002-7365-5791]{Elizabeth~J.~Watkins}
\Heidelberg

\author[0000-0002-0012-2142]{Thomas G. Williams}
\MPIA

\begin{abstract}
We describe the processing of the PHANGS--ALMA survey and present the PHANGS--ALMA pipeline, a public software package that processes calibrated interferometric and total power data into science-ready data products. PHANGS--ALMA is a large, high-resolution survey of \cotwo\ emission from nearby galaxies. The observations combine ALMA's main 12-m array, the 7-m array, and total power observations and use mosaics of dozens to hundreds of individual pointings. We describe the processing of the $u{-}v$ data, imaging and deconvolution, linear mosaicking, combining interferometer and total power data, noise estimation, masking, data product creation, and quality assurance. Our pipeline has a general design and can also be applied to VLA and ALMA observations of other spectral lines and continuum emission. We highlight our recipe for deconvolution of complex spectral line observations, which combines multiscale clean, single scale clean, and automatic mask generation in a way that appears robust and effective. We also emphasize our two-track approach to masking and data product creation. We construct one set of ``broadly masked'' data products, which have high completeness but significant contamination by noise, and another set of ``strictly masked'' data products, which have high confidence but exclude faint, low signal-to-noise emission. Our quality assurance tests, supported by simulations, demonstrate that 12-m+7-m deconvolved data recover a total flux that is significantly closer to the total power flux than the 7-m deconvolved data alone. In the appendices, we measure the stability of the ALMA total power calibration in PHANGS--ALMA and test the performance of popular short-spacing correction algorithms. 
\end{abstract}


\input{introduction.tex}

\input{staging.tex}

\input{imaging.tex}

\input{totalpower.tex}

\input{postprocess.tex}

\input{products.tex}

\input{quality.tex}

\input{summary.tex}

\input{acknowledgments.tex}

\clearpage

\begin{appendix}

\input{contributions.tex}

\input{tpinternalcal.tex}

\input{arrays.tex}

\input{ssctest.tex}

\end{appendix}

%\bibliography{akl}

\input{biblio_pipelinepaper.bbl}
\end{document}

%% file: introduction.tex
\section{Introduction}
\label{sec:intro}

Modern radio interferometric data sets often include hundreds or thousands of distinct observations. They combine data from different arrays, including both total power and interferometric measurements, and both the visibility and image data have large volumes. Calibrating, imaging, and deconvolving these data to produce correct images of the sky can be challenging. Even after these steps, further processing is required to translate these images (or data cubes) into data products ready for scientific analysis.

Current observatories, especially the Atacama Large Millimeter/submillimeter Array (ALMA), have made amazing strides towards automated, high quality calibration of interferometric data.
For ALMA, this stems from hard work by the observatory and the success of the ALMA interferometric and total power pipelines (L.~Davis et al.\ in preparation), which in turn build on the \texttt{CASA} software project \citep[Common Astronomy Software Applications][]{MCMULLIN07}. Thanks to these efforts, ALMA delivers well-calibrated visibility ($u{-}v$) data to its users.

Turning these calibrated $u{-}v$ data into science-ready data products represents a complex task. In this paper, we focus on ALMA observations of CO line emission from nearby galaxies. This emission has complex spatial and velocity structure. It often spans across many individual telescope pointings, and requires both high angular resolution and short-spacing data to recover a full picture of the emission. Moreover, most scientific analysis does not make use of the full position-position-velocity data cube produced by imaging. Translating the visibility data into a science-ready form also involves producing a suite of higher level data products with well-understood properties and uncertainties.

This paper describes the post-calibration processing pipeline constructed to carry out these steps for PHANGS--ALMA, a CO survey of nearby galaxies. As part of this project, we encountered all the issues mentioned: large data volume, the need to reconstruct complex emission from observations using multiple arrays and telescopes, and the need to create high level data products for use in scientific analysis. We address them by adopting or developing appropriate algorithms and implementing them in a modular \texttt{python} and \texttt{CASA} pipeline. The result, described in this paper, is a suite of reproducible, automated methods for processing calibrated $u{-}v$ observations of galaxies into science-ready data products.

\subsection{PHANGS--ALMA}

PHANGS--ALMA is an ALMA survey of \cotwo\ emission from $90$ nearby galaxies. The sample selection, observations, and scientific motivation are described in A.~K.\ Leroy, E.~Schinnerer et al.\ (in preparation). Briefly, this is a large, multi-cycle program focused on mapping \cotwo\ emission at ${\sim}100$~pc resolution across the areas of active star formation in a large, cleanly selected sample of nearby galaxies. The core of the survey is an ALMA Cycle\,5 Large Program (P.I.\ E.~Schinnerer), which is supplemented by a series of smaller programs across five ALMA observing cycles.

Tables~\ref{tab:phangsalma} and \ref{tab:phangsimaging} summarize the properties of the PHANGS--ALMA data set. The survey combines observations with ALMA's main \mbox{12-m} array and both parts of the Morita Atacama Compact Array (ACA): the \mbox{7-m} array and the total power antennas. The \mbox{12-m} array observations used relatively compact configurations, corresponding to angular resolutions of $\sim 1''{-}1.5''$ at the frequency of \cotwo. ALMA's main array and \mbox{7-m} array observe independently, so that separate $u{-}v$ data exist for the main \mbox{12-m} array and the array of smaller \mbox{7-m} antennas. The \mbox{7-m} array consists of fewer antennas, twelve in total. As a result, the total integration time needed to achieve suitable surface brightness sensitivity using the \mbox{7-m} array is $3{-}7$ times longer than that of the main array (see ALMA Technical Handbook\footnote{\mbox{For Cycle~5, when the large program was executed:} \href{https://almascience.nrao.edu/documents-and-tools/cycle5/alma-technical-handbook}{https://almascience.nrao.edu/documents-and-tools/\linebreak[0]{}cycle5/\linebreak[0]{}alma-technical-handbook}}).

We covered each target using large, multi-field mosaics with sizes that frequently approach the observatory-imposed maximum of $150$ pointings. When one $150$-field mosaic could not cover the galaxy, we observed multiple, adjacent mosaics to cover the galaxy. The correlator setup covered $^{12}$\cotwo\ at high spectral resolution and one or more other lines at coarser spectral resolution. We devoted the remainder of the correlator resources to observe the continuum.

In more basic terms, the data for each PHANGS--ALMA target consist of single dish spectroscopic mapping and interferometric visibilities, or ``$u{-}v$ data,'' for dozens or hundreds of individual pointed fields. The \mbox{7-m} and \mbox{12-m} arrays map almost the same area on the sky, but do not share the same pointing centers. The total power data consist of individual spectra obtained using on-the-fly mapping techniques that cover the same spatial region mapped by the interferometer.

Based on the inspection described in Section~\ref{sec:staging}, we verified that, as expected, ALMA delivers reliable, well-calibrated $u{-}v$ data. These data products reflect the excellent performance of the ALMA interferometric calibration pipeline, the stability of the instrument, and the still-minimal impact of radio frequency interference (RFI) on millimeter-wave observations.

\subsection{From \texorpdfstring{$u{-}v$}{u-v} data to science-ready data products}

While calibration is handled by the observatory, the observatory does not deliver images that combine multiple arrays, interferometric and total power data, or multiple mosaics. Nor does the observatory currently provide derived data products beyond data cubes and images. This leaves the user with the task of translating the visibility and total power data into science-ready data products. 

This procedure begins with imaging and deconvolution. The $u{-}v$ data sample the Fourier transform of the sky emission at each frequency. They need to be gridded and Fourier transformed, or ``imaged,'' at each frequency to produce data cubes. Interferometers sample the $u{-}v$ plane incompletely. Producing accurate images of the sky requires reconstructing the true intensity distribution from these incomplete visibility data. This process is referred to as deconvolution or often simply as ``CLEANing'' in reference to the most commonly used algorithm \citep{HOGBOM74}. Modern methods include both versions of the classic CLEAN \citep{HOGBOM74}, which reconstructs the emission as a collection of point sources, and the more recent ``multiscale CLEAN'' \citep{CORNWELL08}, which uses a combination of Gaussian components with a range of scales to reconstruct the image. In parallel, the total power data need to have any frequency-dependent baseline structure removed and the data then combined from a spatially-sampled grid of individual spectra into data cubes \citep[e.g., see][]{MANGUM07}.

After deconvolution, the interferometric data need to be combined with the single dish data in order to correct for the interferometer's lack of sensitivity to extended emission. Approaches to this step vary, and include joint imaging of the interferometric and total power data \cite[e.g.,][]{KODA19}, image plane combination \citep{STANIMIROVIC99}, or Fourier-based processing \citep[``feathering'';][]{COTTON17}. For galaxies observed, imaged, and deconvolved in separate parts, the individual parts must also be stitched together after imaging. We use linear mosaicking to combine individual parts and yield a complete image of each galaxy.

The steps described above yield science-ready data cubes. The subsequent analysis often relies on higher level data products, for example maps of integrated line intensity, mean velocity, or line width, as well as more complex quantities. The first step towards creating such high-level products is usually signal identification. For line emission from well-resolved galaxies, the fraction of a data cube filled by real emission is often small, reflecting the wide bandwidth of the instrument compared to typical linewidths for the interstellar medium (ISM). Identifying the parts of the data cube likely to contain emission is critical to accurately measuring the moments of the emission distribution, particularly the higher moments like line width (``moment 2'').

The most common approach to signal identification is to ``mask'' the data cubes. In this procedure each voxel, i.e., each three dimensional volumetric pixel, is labeled ``True'' or ``False'' according to whether it is likely to contain line emission (``True'') or only noise (``False''). Choices made during the masking process can prioritize either high completeness, meaning inclusion of all emission, or a low false-positive rate, meaning that ``True'' pixels are very likely to contain real emission.

After identifying the part of a cube likely to contain signal, the mask is applied to the line data cube. The voxels containing signal are then ``collapsed'' to form maps that describe the line emission in ways directly relevant to scientific analysis. The resulting maps are usually referred to as ``moment'' maps, though this term frequently includes more than just the intensity-weighted velocity moments of the data. Commonly computed quantities include line-integrated intensity, measurements of the line width and spectral profile shape, and measurements of the characteristic velocity.

\subsection{The PHANGS--ALMA pipeline}

The ALMA imaging interferometric pipeline implements deconvolution and imaging of visibility data for individual arrays \citep[e.g.,][]{KEPLEY20}, but does not yet image combined data from different arrays or combine total power and interferometric data. These steps are all necessary to produce science-ready data products to achieve our science goals.  This paper describes the steps taken to post-process the PHANGS--ALMA data and details the motivations for our choices. We also describe the PHANGS--ALMA post-processing pipeline software, which combines \texttt{CASA} with \texttt{python} extended by additional packages (\S \ref{sec:implementation}).

Although the PHANGS pipeline was developed for the PHANGS--ALMA survey to produce \cotwo, \mbox{C$^{18}$O($2\text{--}1$)}, and continuum images, the software represents a general post-processing pipeline. We have used it to process \mbox{CO($3\text{--}2$)}, \mbox{CO($4\text{--}3$)}, $^{13}$\cotwo, \mbox{HCN($1\text{--}0$)}, \mbox{HCO$^+$($1\text{--}0$)}, \mbox{CS($2\text{--}1$)}, \mbox{[C\textsc{i}]($^3$P$_1$--$^3$P$_0$)}, and dust continuum data from ALMA as well as \hi\ \mbox{21-cm} data from the VLA. Altogether, we have processed of order $1{,}000$ interferometric observations using this software. The closely related total power calibration and imaging pipeline presented in \citet{HERRERA20} and summarized here has also processed of order $1{,}000$ total power observations.

Section~\ref{sec:overview} summarizes the workflow, notes the software used to implement the PHANGS pipeline, and defines key terms. Sections \ref{sec:staging} and~\ref{sec:imaging} describe the $u{-}v$ data processing, imaging, and deconvolution. Section~\ref{sec:totalpower} reports our total power processing procedures. Sections \ref{sec:postprocess} and~\ref{sec:products} explain our approaches to cube post-processing and product creation. Section~\ref{sec:quality} provides an overview of our quality assurance procedures, including end-to-end tests of our pipeline using simulated data. The Appendices list the contributions of members of the PHANGS--ALMA data reduction team and report on tests related to the combination of total power and interferometer data, the stability of the flux calibration in the total power data, and the relative performance of 7{-}m and combined 12{-}m+7{-}m array imaging.

\section{Workflow, Definitions, and Implementation}
\label{sec:overview}

\begin{figure*}[ht!]
\gridline{\fig{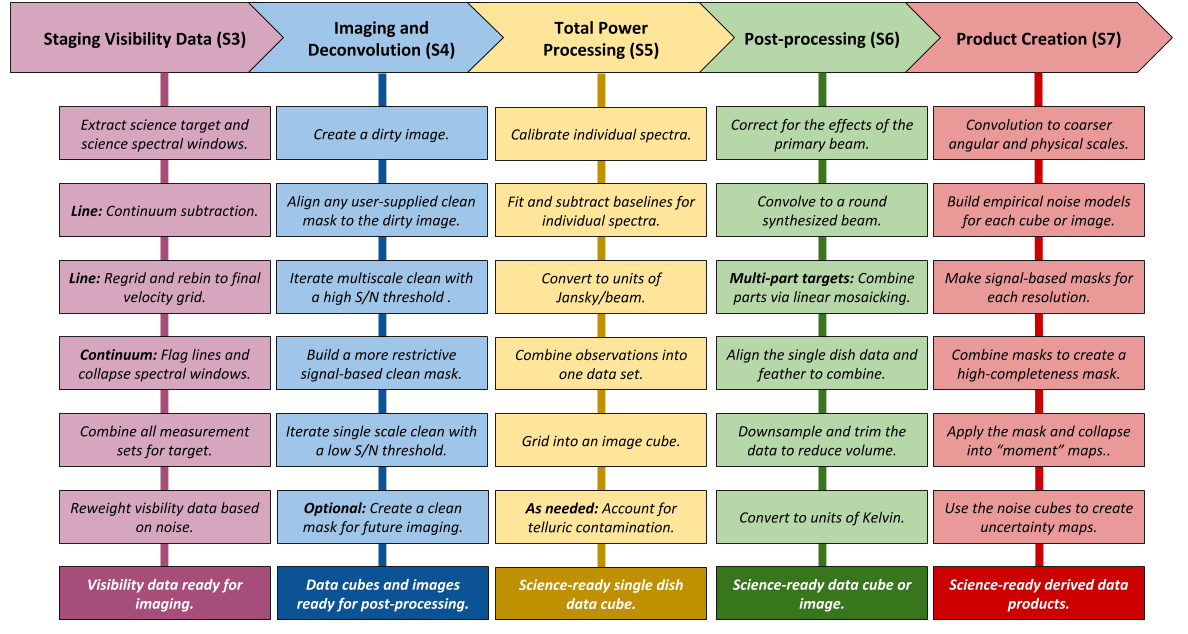}{1.0\textwidth}{}
}
\vspace{-24pt} % reduce excessive gaps between figure rows
\caption{
\textbf{Overall pipeline workflow.} A schematic view of the pipeline steps. We begin by staging the calibrated visibility data in a form appropriate for imaging (Section~\ref{sec:staging}). Then we image and deconvolve the data (Section~\ref{sec:imaging}). In parallel, we reduce and image the total power data (Section~\ref{sec:totalpower}). Next we post-process the imaged products into science-ready data cubes and images (Section~\ref{sec:postprocess}). Finally we process the images into more advanced science-ready products (Section~\ref{sec:products}).
\label{fig:flowchart}}
\end{figure*}

\begin{deluxetable}{lc}
\tabletypesize{\small}
\tablecaption{PHANGS--ALMA Data Summary \label{tab:phangsalma}}
\tablewidth{0pt}
\tablehead{
\colhead{Description} & 
\colhead{Value} 
}
\startdata
\hline
Targets & \\
... galaxies & $90$ \\
... targets (i.e., individual mosaics)  & $136$ \\
... galaxies observed using multiple mosaics & $26$ \\
\hline
Input measurement sets ... & \\
... ACA 7-m array & $479$ \\
... typical 7-m $u{-}v$ range & $8{-}43$m  ($6{-}33$k$\lambda$)\\
... typical 7-m beam & $7.2'' \times 4.4''$ \\
... main 12-m array & $184$ \\
... typical 12-m $u{-}v$ range & $13{-}380$m ($10{-}292$k$\lambda$)\\
... typical 12-m+7-m beam & $1.26'' \times 1.04''$ \\
... total power observations & 744 \\
... typical total power beam & $28.4''$\\
\hline
Standard spectral products ... & \\
... CO(2\text{--}1) native channel & $\sim 0.32$~km~s$^{-1}$ \\
... CO(2\text{--}1) working channel\tablenotemark{a} & $\sim 2.54$~km~s$^{-1}$ \\
... C$^{18}$O(2\text{--}1) native channel & $\sim 2.7$~km~s$^{-1}$ \\
... C$^{18}$O(2\text{--}1) working channel\tablenotemark{a} & $\sim 6.0$~km~s$^{-1}$ \\
... total bandwidth & $6.7$~GHz\\
\enddata
\tablenotetext{a}{Target velocity resolution. The pipeline aims to get as close as possible to this number without going over.}
\tablecomments{These numbers refer to all of the data processed by our team for the first public data release of PHANGS--ALMA, internal ``version 4.'' However, not all data are released to the public with this first release; some are part of projects still under proprietary period or recent archival data sets. A summary of observations and more details about the survey itself are presented elsewhere. The resolutions, $u{-}v$ range, and other details represent typical values. They vary slightly from target to target.}
\end{deluxetable}

\subsection{Workflow}

We begin with calibrated $u{-}v$ data of the sort produced by the ALMA (or VLA) interferometric calibration pipelines. This is stored in the \texttt{CASA} data format of a ``measurement set'' in which the visibilities have a calibrated phase and amplitude scale. Starting with these data, the pipeline carries out the following steps, which we summarize in Figure~\ref{fig:flowchart}:

\begin{enumerate}
\item \textbf{Stage the $u{-}v$ data.} The pipeline begins by processing the calibrated $u{-}v$ data into a form appropriate for imaging. It extracts the $u{-}v$ data associated with the science target and relevant spectral windows from the original measurement sets. Then it subtracts the continuum signal from the $u{-}v$ data. It then regrids and rebins all continuum-subtracted, line $u{-}v$ data onto a common velocity grid to be used in imaging. It also extracts the line-free regions of the spectrum from the original measurement set to make a continuum-only $u{-}v$ data set. This is described in Section~\ref{sec:staging}.

\item \textbf{Image and deconvolve the data.} This involves repeated calls to \texttt{CASA}'s \texttt{tclean} task interleaved with the creation of masks that guide the deconvolution and checks for convergence. We use a mixture of multi-scale and single-scale CLEAN calls during this process. This is described in Section~\ref{sec:imaging}.

In parallel, we reduce total power data via the calibration and imaging pipeline presented by \citet{HERRERA20}. We summarize these steps in Section~\ref{sec:totalpower}. There we also describe the issue of telluric ozone contamination. This issue specifically affects the PHANGS--ALMA \cotwo\ data.

\item \textbf{Post-process the imaged data.} The pipeline applies primary beam corrections, convolves the data to have a round synthesized beam, combines the interferometric and total power data, mosaicks together multi-part fields, converts the data to have units of Kelvin, and trims and down-samples the cubes to save disk space. Finally the images are exported into science-ready FITS cubes. These steps are described in Section~\ref{sec:postprocess}.

\item \textbf{Derive additional high-level data products.} The pipeline creates versions of these cubes at several angular and physical resolutions. For each cube and resolution, it creates a noise model that accounts for spectral and spatial variations. 

The pipeline uses this noise model to create masks that identify the location of likely signal. We create two sets of masks. The ``broad'' masks have high completeness, meaning that they include most of the emission in the cube. The ``strict'' masks have low false-positive rates, meaning that they include only regions where emission is detected at high confidence.

Using these masks, the pipeline produces maps of velocity, integrated intensity, and a suite of other ``moments'' of the intensity distribution, along with associated uncertainties. This is described in Section~\ref{sec:products}.

\end{enumerate}

\subsection{Definitions}

Mostly, this paper uses general radio astronomy terminology and jargon associated with the standard ALMA data reduction package, \texttt{CASA}. We also define a few pipeline-specific terms here:

\begin{enumerate}
\item The pipeline considers \textit{``targets''} to be regions of the sky that will be imaged or processed together. For PHANGS--ALMA, targets are either whole galaxies or parts of galaxies, and each target is a mosaic with tens to more than one hundred individual fields. Within the pipeline infrastructure, each target has an associated mean velocity, velocity width, and phase center.

Some targets only correspond to part of a galaxy. For example, PHANGS--ALMA observed the nearby galaxy NGC~2903 using three separate ${\sim} 150$-field mosaics. As described above this was required due to ALMA's 150 field limit on any individual observation. We imaged the three observations separately as targets named ``NGC2903\_1,'' ``NGC2903\_2,'' and ``NGC2903\_3.'' These galaxy parts account for the difference between the number of targets and smaller number of galaxies in Table \ref{tab:phangsalma}.

\item The pipeline makes images for a variety of spectral \textit{``products.''} These are either line products or continuum products. Line products are defined by a spectral line, which sets the rest frequency to be used, and a velocity channel width. For example, \cotwo\ at $\approx 2.54$~km~s$^{-1}$ channel width defines the main PHANGS--ALMA line product. \mbox{C$^{18}$O($2\text{--}1$)} at $6.0$~km~s$^{-1}$ channel width defines another. Continuum products represent the integrated continuum intensity after excluding all user-defined spectral lines of interest in the window.

\item Each input data set is tagged with an \textit{``array combination.''} This does not need to refer to a rigorous antenna or array setup (e.g., ALMA's \mbox{C43-1} configuration). The purpose of the array combination tag is to group data that will be imaged together. For example, PHANGS--ALMA processes data for all main array compact configurations as a single array combination, which we call ``\mbox{12-m}.'' We also process the ACA \mbox{7-m} data together as part of an array combination called ``\mbox{7-m}.'' Finally, we process the ACA and main array data together in an array combination called ``\mbox{12-m}+\mbox{7-m}.''

We also define ``feathered array combinations.'' These combine an interferometric array combination and total power data (``tp''). For PHANGS--ALMA, these are ``\mbox{12-m}+\mbox{7-m}+tp'' and ``\mbox{7-m}+tp.''

\end{enumerate}

\subsection{Implementation}
\label{sec:implementation}

As of the PHANGS pipeline ``version 2.0'' described in this paper, the pipeline consists of a series of linked programs designed to run in \texttt{CASA} \citep{MCMULLIN07} and a python environment equipped with \texttt{numpy} \citep{NUMPY2006}, \texttt{scipy} \citep{SCIPY2020}, \texttt{astropy} \citep{ASTROPY1,ASTROPY2} and several affiliated packages, most notably \texttt{reproject}\footnote{\href{https://reproject.readthedocs.io/}{https://reproject.readthedocs.io/}}, \texttt{spectral-cube}\footnote{\href{https://spectral-cube.readthedocs.io/}{https://spectral-cube.readthedocs.io/}}, and \texttt{radio-beam}. Currently the total power reduction scripts and quality assurance scripts still exist as separate packages. 

Both the total power pipeline and our ``version 2.0'' processing pipeline are publicly available on GitHub\footnote{\href{https://github.com/akleroy/phangs_imaging_scripts}{https://github.com/akleroy/phangs\_imaging\_scripts} and \href{https://github.com/PhangsTeam/TP_ALMA_data_reduction}{https://github.com/PhangsTeam/TP\_ALMA\_data\_reduction}}. Our intention is that development continue on this public version as long as the software remains useful, with ``version 2.0'' benchmarked as a release. Many of the quality assurance procedures are written in IDL and \texttt{python} and are specific to PHANGS--ALMA, so these are not part of the general publicly available pipeline.

We use several different versions of \texttt{CASA} for processing. We note which versions we use for each application in the relevant section. We did not impose a strict version requirement on the \texttt{astropy} packages but mostly used version $4.0$ of \texttt{astropy}, version $0.4$ and after for \texttt{spectral-cube}, and version $0.7$ and later for \texttt{reproject}. We draw the frequencies of spectral lines from \texttt{splatalogue} \citep{SPLATALOGUE}.

During prototyping and quality assurance we also made extensive use of IDL, including the astronomy user's library \citep{IDLASTRO}, \texttt{cprops} \citep{ROSOLOWSKY06}, and an updated version called \texttt{cpropstoo} \citep{LEROY15A}. For the total power data we made heavy use of the GILDAS package\footnote{\url{http://www.iram.fr/IRAMFR/GILDAS}. For more information about the GILDAS software, see \citet{PETY2005}.}, especially CLASS and ASTRO, to prototype, investigate the telluric contamination, and deal with challenging processing cases.

In practice, the PHANGS pipeline is built around a set of modules that are wrapped and called by a series of ``handler'' classes. The modules contain routines that can run on any input file. They implement tasks like linear mosaicking, spectral line extraction, mask creation, etc. These tasks do not require the rest of the pipeline infrastructure to run, and could be used in other applications.  The handlers are aware of the larger project. They interface with user-provided data files, manage directories and files, and loop over targets, spectral line setups, and array configurations. The handlers construct a series of calls to the task-oriented modules to implement the steps described in this paper.

The user establishes the input parameters for a project through a series of input text files, which are read and used by the handlers. In these files, the user lists the input calibrated measurement sets and associates each with a target name and array configuration. They also define the targets, specifying a phase center and velocity range for imaging, associating targets that should be linearly mosaicked, and inputing distances to each target. The user inputs also specify the spectral grid, target line, and array combinations for imaging and post-processing. Finally the user defines which data products to create, including choosing the angular and spatial scales to be analyzed. In principle, many of these choices could be automated, but we found that leaving them as input parameters worked well for a survey the size of PHANGS--ALMA. In practice, the PHANGS--ALMA choices serve as widely applicable defaults, and most of the customization to define new projects involves simply defining targets, listing input data, and choosing the relevant observed lines.

The pipeline is then executed through a master \texttt{python} script, either through the shell or a command line call. Staging, imaging, and post-processing are run inside of \texttt{CASA}. Derivation of data products is run outside of \texttt{CASA} in a pure python environment. For many applications, the pipeline is trivial to parallelize by simply starting multiple runs targeting different galaxies. 

More details and examples can be found with the software itself. The rest of this paper focuses on the procedures used to process the data rather than on the details of the software.

%% file: staging.tex
\section{Staging of Visibility Data} 
\label{sec:staging}

For each PHANGS--ALMA observation, we apply the observatory-provided calibration and flagging to produce a calibrated measurement set. Then, for each combination of target, spectral product, and interferometric array combination, we construct a ``staged'' visibility data set that will be used in imaging (Section~\ref{sec:imaging}). This staged data set combines all relevant visibility data, including data from different ALMA projects, into a single file on a common velocity grid.

The PHANGS--ALMA pipeline assumes calibrated input $u{-}v$ data. To verify that the input $u{-}v$ were correctly calibrated, we carried out a by-hand inspection of the calibrated Large Program data. We describe this briefly before discussing the other data processing steps.

\subsection{Starting point}

We begin by applying the calibration and flagging produced by the ALMA observatory interferometric pipeline (L.~Davis et al.\ in preparation) to the data. This step uses the same version of \texttt{CASA} as the original ALMA observatory pipeline run in order to avoid any potential issues arising from changes in calibration tables with \texttt{CASA} version. The ALMA observatory pipeline version changed over the course of the project. Data from the PHANGS--ALMA pilot projects (from Cycles 2 and~3) were mostly calibrated using the Cycle~3 pipeline available with \texttt{CASA} version $4.5.3$. Most data from the PHANGS--ALMA large program were calibrated using the Cycle~5 version of the pipeline available with \texttt{CASA} version $5.1.1$. Most of the extension projects were calibrated using the Cycle~6 version of the pipeline delivered with \texttt{CASA} $5.4.0$.

For PHANGS--ALMA, the ALMA interferometric calibration pipeline performance and observatory quality assurance was excellent. We did not find additional flagging to be necessary, which largely reflects that the data have already been quality assured by the observatory before delivery. To verify this, at several stages during the project we carried out the inspection described in the next section. These checks aimed to determine whether the pipeline either missed significant flagging or appeared to flag real signal. We did not find any problems serious enough to appreciably affect the final images, so we proceeded using the observatory-provided calibration. 

This paper focuses on ALMA observations, but the pipeline also works for other types of data. When we use the pipeline for data with less stringent quality assurance or less stable calibration, the process tends to be iterative. For example, we first image the data. Then this initial imaging often reveals defects or issues indicating bad data or imperfect calibration. We then improve the flagging, re-calibrate, and re-image the data. These flagging and re-calibration steps occur outside the PHANGS--ALMA pipeline. After improving the visibility data, the PHANGS--ALMA pipeline is re-run to stage and image the data again. This workflow is common for, e.g., VLA \mbox{21-cm} data in which radio frequency interference (RFI) can play a large role.

\medskip

\noindent \textbf{(No) Self calibration:} We did not apply self-calibration to the PHANGS--ALMA data, and we have not yet implemented self-calibration in the PHANGS--ALMA pipeline. The PHANGS--ALMA \cotwo\ images do not appear dynamic range limited, and our mosaic observing strategy does not lend itself to self-calibration. Most fields in most of our sources do not contain bright enough emission to allow for self-calibration. When bright sources are present, they tend to be confined to a small part of the mosaic, and so are visited only infrequently as part of the mosaic observations.

\subsection{Manual quality assessment of PHANGS--ALMA \texorpdfstring{$u{-}v$}{u-v} data}
\label{sec:uvquality}

\begin{figure*}[ht!]
\gridline{
\fig{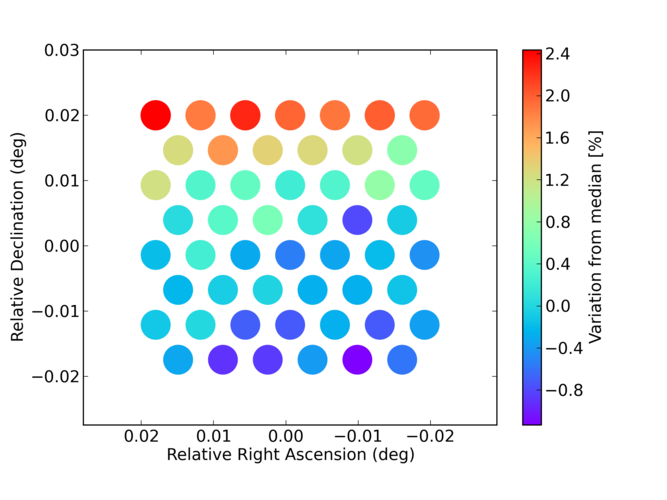}{0.5\textwidth}{}
\fig{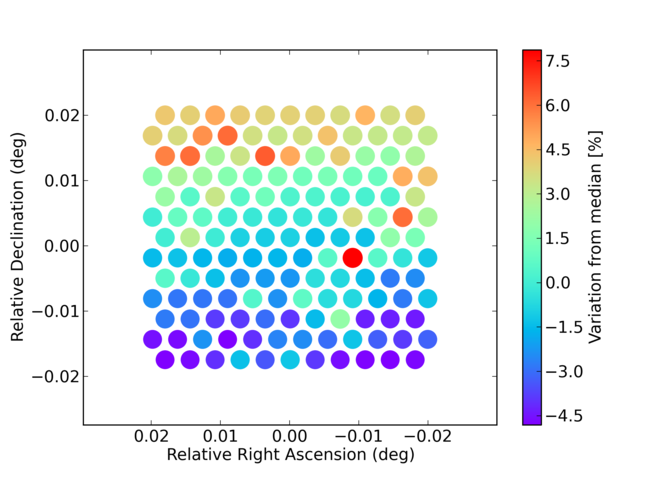}{0.5\textwidth}{}
}
\vspace{-30pt} % reduce excessive gaps between figure rows
\caption{
\textbf{Example of the variation of rms noise per pointing in the $u{-}v$ data.} The figure illustrates one of the checks that we carried out during manual quality assurance of the PHANGS--ALMA $u{-}v$ data (Section~\ref{sec:uvquality}). For each mosaic pointing of NGC~4303 we plot the rms noise in the $u{-}v$ data as a function of the phase center of that pointing. The \textit{left} panel shows results for the \mbox{7-m} observations and the \textit{right} panel for the \mbox{12-m} observations. The color scale indicates the variation of rms noise in that pointing from the median value across the map. The maximum deviations are ${\sim}3\%$ for the \mbox{7-m} data and ${\sim}12\%$ for the \mbox{12-m}~data. In general, noise correlates with elevation in our data. The mild gradient in declination in this example reflects the fact that for this particular galaxy, by chance the change in observation elevation correlates with  declination.
\label{fig:rms_per_pointing}}
\end{figure*}

As part of the PHANGS--ALMA data reduction process, we inspected the calibrated $u{-}v$ data from our pilot programs and the Large Program. This inspection focused on the calibrated data, i.e., the direct output from applying the observatory-provided calibration. We inspected:

\begin{enumerate}
\item \textbf{Observation set-up.} We checked the calibrated measurement sets and delivered weblog to confirm that our observational setup was correct. We verified that the observations contained the correct number of fields, total integration time, number of antennas, pointing position on the sky, $u{-}v$ coverage, and antenna positions. 

\item \textbf{Observing conditions.} We verified that the weather conditions and related parameters in the weblog were roughly constant across the observations and matched expectations. We checked the precipitable water vapour (PWV), air pressure, humidity, temperature, and wind speed and direction. 

We also inspected the antenna-based $T_{\rm sys}$ measurement versus frequency and compared these to the PWV of the observation. For PHANGS--ALMA \cotwo\ observations, the typical $T_{\rm sys}$ is ${\sim}70$~K with the highest $T_{\rm sys}$ of ${\sim}100$~K around the weak atmospheric absorption at $231.3$~GHz.

\item \textbf{Calibrator inspection.} For the pilot program and the first part of the Large Program, we examined the calibrated visibilities for the bandpass and phase calibrators. In this inspection, we aimed to identify outliers and assess the need for additional flagging in the calibrated measurement sets. We plotted time-averaged amplitude and phase as a function of frequency, frequency-averaged amplitude and phase as a function of time, and time- and frequency-averaged amplitude and phase as a function of $u{-}v$ distance. When we found deviations from the expected behavior in the plots, we manually investigated the $u{-}v$ data to find the cause of the aberrations. This investigation generated a candidate set of additional flagging commands.

Overall, we found that the observatory-provided calibrations yielded calibrated $u{-}v$ data with few visible pathologies. As described below, our tests suggested that adding additional flagging had negligible impact on the final images. Reflecting this, after the first part of the Large Program, we shifted our manual quality assurance efforts from the $u{-}v$ data to the imaged data (Section~\ref{sec:quality}). We did not manually inspect the calibrator data for the last part of the Large Program and follow up programs.

\item \textbf{Inspection of synthesized beam.} As an additional check on the $u{-}v$ coverage of the data after flagging, we created a map of the synthesized dirty beam at the observed \cotwo\ frequency using the \texttt{CASA} tool \texttt{imager}. We compared the size and axis ratio of the synthesized beam to expectations based on the $u{-}v$ coverage before flagging in order to further verify that the flagging did not have any pathological impact on the data.

\item \textbf{Quality across mosaic.} Finally, we examined the spatial structure of the noise across each mosaic. In particular, we calculated the rms $u{-}v$ amplitude noise at each individual mosaic pointing, considering all frequencies in the main \cotwo\ science window at the spectral resolution of $\sim 2.6$~km~s$^{-1}$. We used this test to check for missing data, e.g., due to flagging or other problems with individual fields. Figure~\ref{fig:rms_per_pointing} illustrates this check for the \mbox{12-m} and \mbox{7-m} data for one galaxy. In this figure, the field-to-field variations in $u{-}v$ amplitude noise are $0.3$\% on average and ${\sim}6$\% at most for the \mbox{7-m} data, and $1.7$\% on average and ${\sim}9$\% at most for the \mbox{12-m} data. These results are typical of the targets that we inspected.

\end{enumerate}

The tests described above would often suggest a modest amount of possible additional manual flagging. To assess the science impact of a final round of human flagging on the delivered data, we manually flagged the data for two cases. Then we compared the resulting images to those made using no additional flagging. 

We chose one galaxy with bright CO emission and one galaxy with faint CO emission for this experiment. Then we manually inspected the $u{-}v$ data as described above, identified an aggressive set of flagging commands, and applied the flags to the data. Finally, we imaged the data with and without this additional flagging. In these tests, the difference between the original and the manually-flagged data cubes is less than 5\% in total flux and 5\% in rms noise for both the bright and faint targets.

In addition to inspecting the quality of individual $u{-}v$ data, we searched for consistency in the overall calibration scale across the full data set. We examined images of the interferometric calibrators and the flux scale solved for by the pipeline. When we plot the derived flux of any specific secondary calibrator as a function of time, we find good overall consistency among the 12{-}m array, the 7{-}m array, and the ALMA calibrator database.  For the total power data, we find relatively stable gains, expressed as the observatory-provided Jansky-per-Kelvin (Jy/K or Jy-per-K) values, across the whole data set at any given time. We did observe that, for observations taking place on similar dates, there is a $\sim7\%$ difference in the observatory-provided Jy/K between data delivered before and after the last quarter of 2018 (see Appendix \ref{sec:tp-cal}). As this is an observatory-derived calibration, we accepted the provided values for ``version~4'' of the PHANGS--ALMA delivery. However, we note that based on consultation with the observatory future releases are likely to see the overall flux of some galaxies decline by $2{-}5\%$ (see discussion in Appendix~\ref{sec:tp-cal}). Only five galaxies have observations delivered both before and after the date of this transition. In Appendix~\ref{sec:tp-cal} we show that for all other galaxies, the total power data show excellent internal consistency with rms variation of about $\pm 3\%$.

The inspection steps described above repeatedly verified that the calibration and flagging delivered by ALMA are science ready, in good agreement with the observatory goals and our previous experience with the telescope. Given the minimal impact of additional flagging, we decided to adopt the observatory-delivered calibration for the PHANGS--ALMA processing. 

For the rest of the project, we trusted our detailed quality assurance on the imaged data (Section~\ref{sec:quality}) to reveal any remaining issues with the data.

\subsection{Staging and continuum subtraction}

The PHANGS--ALMA pipeline begins by extracting the calibrated data for each galaxy part and spectral line using the \texttt{CASA} task \texttt{split}. We select only the science target, as specified by either the ``scan intents'' recorded by ALMA, or we manually select a user-provided field or set of fields.

If the continuum subtraction requires multiple spectral windows, we select all spectral windows. Otherwise, for line products we select only the spectral windows overlapping the line of interest given the mean recessional velocity and velocity width of the source.

After extracting the science target and window of interest, we subtract the continuum using the \texttt{CASA} task \texttt{uvcontsub}. The pipeline is aware of a set of bright lines and the user-provided systemic velocity and velocity width for each target. The pipeline calculates the spectral footprint of each line in the $u{-}v$ data and excludes channels that contain line emission when determining the continuum level. For PHANGS--ALMA we excluded the regions around CO lines from continuum determination. These are the only bright spectral lines in our setup.

For PHANGS--ALMA, we used a single spectral window (spw) for continuum subtraction, which we carried out for all targets. We fit a polynomial of order zero (i.e., a constant) and fit the continuum in each individual integration. For \cotwo\ the observations used a spectral window with width $\approx 1{,}200$~km~s$^{-1}$, and for other lines like \mbox{C$^{18}$O($2\text{--}1$)} that were covered, the velocity coverage was even larger. This is much larger than the velocity width of any of our targets, leaving wide bandwidth for continuum subtraction. Given the low fractional bandwidth of a $1{,}200$~km~s$^{-1}$ spectral window and the low signal-to-noise of the continuum near the \cotwo\ line, we found that a zeroth order polynomial did a good job removing the continuum.

For cases with brighter continuum, the pipeline can fit polynomials of higher order, with the order set by the user. In this case, the user can specify the fit to span multiple spectral windows. This is useful, e.g., for ALMA data at Band~7 or above and for VLA data at \mbox{L-band}, where the continuum is strong and the slope is steep. Using multiple spectral windows is also useful when the spectral line of interest covers the entire window, leaving no free bandwidth to fit the continuum. In this case, the pipeline will extract all relevant spectral windows as part of the \texttt{split} call above and run \texttt{uvcontsub} on all of them, combining spectral windows for the fit.

\medskip

\noindent \textbf{Time binning:} Optionally at this stage, we also apply some time binning to the data. This is specified by the user when defining each interferometric configuration (e.g., ``\mbox{7-m}'', ``\mbox{12-m}'', ``\mbox{12-m} extended''). This allows the time binning to be defined in a way that avoids time smearing but compresses the data as much as possible. We did not use this option for PHANGS--ALMA, but this is a common step used in VLA \mbox{21-cm} data processing or processing of ALMA ACA data, especially at Band~3.

\subsection{Spectral regridding and rebinning}
\label{sec:spectral_regridding}

\begin{figure}[ht!]
\gridline{\fig{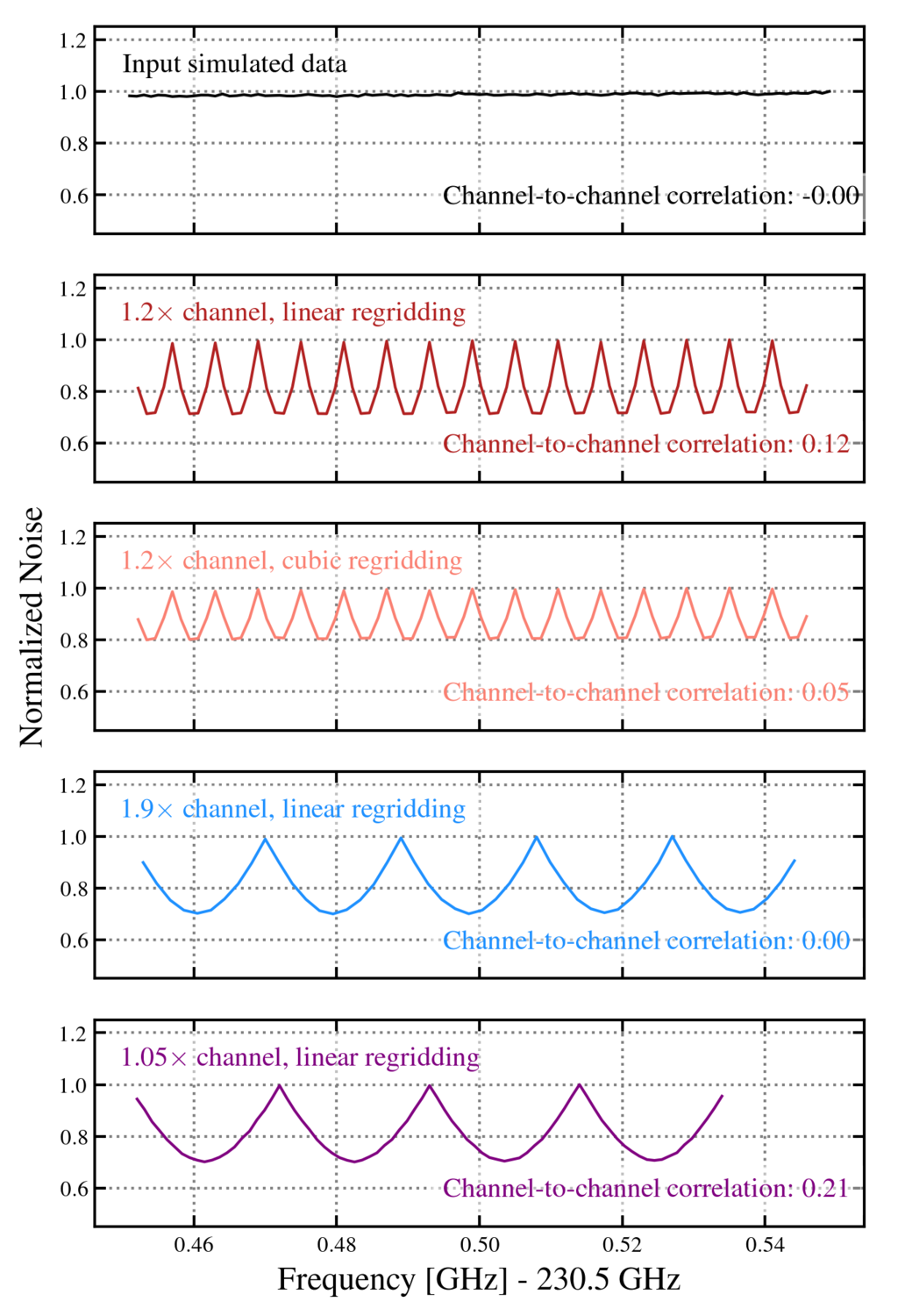}{0.45\textwidth}{}
}
\vspace{-24pt} % reduce excessive gaps between figure rows
\caption{
\textbf{Noise and regridding in \texttt{CASA}.} Illustration of regridding effects on the noise spectrum of $u{-}v$ data. We begin with a pure noise data set created by \texttt{CASA}'s \texttt{simalma}. This data set produces the noise spectrum seen in the top panel. Subsequent panels show the noise spectrum in the visibility data after regridding using \texttt{CASA}'s \texttt{mstransform} task and using \texttt{visstat} to measure the noise spectrum. The regridding introduces patterns into the noise due to the uneven amount of independent data contributing to each output channel. The frequency of the variations is set by the aliasing between the new and original channel width. For example, increasing the channel size by a factor of $1.05$ creates a pattern with $20$ channel periodicity. To see this, compare the second ($1.2\times$ increase in channel width), fourth ($1.9\times$), and bottom ($1.05\times$) panels. In the bottom right of each panel we quote the channel-to-channel correlation of the data. This is the linear correlation coefficient between noise values in adjacent channels of individual spectra and would ideally remain zero under interpolation. The correlation is a function of the statistical independence of the data contributing to the new channel and the interpolation method used. Because cubic interpolation draws on many nearby channels the correlation induced between immediate neighbors is weaker.
\label{fig:uvnoise}}
\end{figure}

After continuum subtraction, for each spectral line of interest we create a line-specific measurement set that combines all data on the common velocity grid to be used for imaging. This operation begins with the continuum-subtracted $u{-}v$ data.

The output spectral grid adopts the ``radio'' Doppler shift convention, in which $\left| \delta v \right| / c = \left| \delta \nu \right| / \nu_0$, and we work mostly in the kinematic Local Standard of Rest (LSRK) frame. The user provides the central velocity and width of the final frequency grid as an input. For PHANGS--ALMA, these were initially estimated from large extragalactic databases like NED and LEDA \citep{PATUREL03,MAKAROV14}. We then refined them after inspecting a first round of imaging. On average, the careful systemic velocity estimates using PHANGS--ALMA CO data in \citet{LANG20} differ from the radio velocity estimates in LEDA by $\sim \pm 5$~km~s$^{-1}$ and from the optical velocity estimates by $\sim \pm 10$~km~s$^{-1}$. This is small compared to the overall velocity widths used for PHANGS--ALMA cubes. This width for most cubes is $500{-}1{,}000$~km~s$^{-1}$, with larger values for more massive, heavily inclined galaxies and smaller values for face-on and low-mass galaxies.

To place the data on the final frequency grid, we first call the \texttt{CASA} task \texttt{mstransform} to place all observations onto a velocity grid with a common starting channel and channel width in the LSRK frame. This step converts from the topocentric frame, and so adjusts for changes in the Earth's motion compared to the LSRK frame. This operation reduces the data to only a moderate velocity range of interest around the line of interest. 

After this, we call the \texttt{CASA} task \texttt{mstransform} again to rebin the data to the final channel width of ${\sim}2.54$~km~s$^{-1}$ for PHANGS--ALMA. This rebinning averages together an integer number of channels, typically $5{-}6$ for PHANGS--ALMA \cotwo\ data, and uses no interpolation. The rebinning factor is picked to ensure that the final channel width is as close as possible to the desired spectral resolution for that configuration without exceeding the specified value. 

Next we combine all regridded and rebinned spectral windows for each target and spectral product into a single measurement set using \texttt{CASA}'s \texttt{concat} task.

We adopt this regrid-then-rebin approach in order to work around current limitations in \texttt{CASA}'s spectral regridding capabilities, which we describe below. For the PHANGS--ALMA \cotwo\ data this procedure yields a final channel width, and so a final spectral resolution, near $\Delta v \approx 2.54$~km~s$^{-1}$ for \cotwo\ and $^{13}$\cotwo, with minor variations from target to target. It yields a resolution near $\Delta v \approx 6.0$~km~s$^{-1}$ for \mbox{C$^{18}$O($2\text{--}1$)}, see Table~\ref{tab:phangsalma}.

After combining the data, the user has the option of re-weighting the visibilities by the measured noise using \texttt{CASA}'s \texttt{statwt}. This reweighting ensures self-consistent weights in each final line data set but risks introducing pathologies if real line or continuum emission contaminates the weight calculation. For PHANGS--ALMA this step occurs after continuum subtraction, so the main danger is contamination by broad line emission. We do apply this procedure to PHANGS--ALMA. We used a $50$~km~s$^{-1}$-wide window at each edge of the final spectral window for re-weighting with \texttt{statwt}. This process excludes channels associated with the line itself from the weight calculation. The new weights reflect noise measured from channels far from the systemic velocity of the galaxy. 

\medskip

\noindent \textbf{Noise and spectral regridding in \texttt{CASA}:} Our rebinning and regridding strategy introduces some frequency dependence into the noise in the final data products and also leads to some channel-to-channel correlation. While this is unfortunate, our strategy appears to reflect the best current option given the spectral regridding capabilities of \texttt{CASA}. We expect that this situation will improve in future versions of \texttt{CASA}. Since it leaves an imprint on our data and likely affects a significant amount of already published ALMA data, we explain the effect here.

The noise pattern arises from the interpolation carried out by \texttt{CASA}'s \texttt{mstransform} task. \texttt{mstransform} can only regrid to larger channel widths than those in the input data. In the case where the output channel width is not an integer multiple of the input channel width, this regridding leads to a varying number of independent data points contributing to different output channels.

We illustrate this effect in Figure~\ref{fig:uvnoise}. We begin with a pure noise, $100$ channel visibility data set created using \texttt{CASA}'s \texttt{simalma}. The nominal frequency and channel width are ${\sim}230$~GHz and $1$~MHz, but do not matter for this exercise. In the top panel, we plot the noise spectrum in the original visibility data, which is nearly flat. In the rest of the panels, we show the noise spectrum after regridding the data to new channel widths using \texttt{mstransform}. 

Figure~\ref{fig:uvnoise} shows periodic noise variations in the regridded data. The periodicity is set by the fractional difference between the output channel width and an integer multiple of the input channel width. For example, consider regridding to a new grid with a channel width $1.2$ times the original channel width. During regridding, sometimes a single input channel dominates the data in an output channel. In these cases other channels do contribute but might only receive, e.g., 20\% of the weight in the interpolation. Other times two input channels are equally weighted and averaged together to form the new output channel. This latter case effectively averages together twice as many independent data points and will thus have $\sqrt{2}$ times lower noise. 

When the output channel is only slightly different from an integer rebinning the position (in frequency) of output channels relative to the position of input channels ``slides'' across the output data set. As a result, the amount of independent data contributing to an output channel varies smoothly across the output data set. The periodicity of the variation is set by the fractional difference between channel size and integer rebinning. For example, when gridding to channels a factor of $1.05$ or $1.95$ larger than the original channel, the output grid steps are offset by $0.05$ initial channels at each channel, and periodicity over $20$ channels is expected.

In more extreme cases, the interpolation creates rapid variations and a ``sawtooth'' pattern in the output noise spectrum. For example, consider gridding from a $1$~km~s$^{-1}$ channel to a $1.2$~km~s$^{-1}$ channel. Every ${\sim}5$ output channels, the balance of independent input data will shift from $5$-to-$1$ to $1$-to-$1$ and then back.

In addition to noise variation, the interpolation also affects the correlation between the intensity in successive channels. Because of the variable amount of input data per output channel, the interpolation both introduces channel-to-channel correlation and leads to variations in this channel-to-channel correlation. In optical terms, this processing broadens the line spread function of the data and leads to some dependence of the line spread function on frequency. Figure~\ref{fig:uvnoise} notes the magnitude of the induced channel-to-channel correlation for each case.

These issues reflect current limitations of \texttt{CASA}, and we expect that the situation will improve in the future. The issue could be addressed by using the \texttt{fftshift} option in \texttt{mstransform} but that option was not functioning as intended in the versions of \texttt{CASA} that we used. Alternatively, the effect could be mitigated by allowing \texttt{mstransform} to oversample the line spread function (i.e., to move to smaller channel width). In this case, heavily oversampled data could be convolved with an appropriate kernel to produce an even amount of independent data per final, coarser output channel. This functionality is also currently not available.

\medskip

\noindent \textbf{Regridding in the pipeline:} To minimize the effects of the interpolation scheme, the pipeline picks an output channel size that leads to only slow noise variations, i.e., a much more ``stretched out'' version of the last panel in Figure~\ref{fig:uvnoise}. During the initial regridding step we increase the common channel size by a small factor, $\epsilon \approx 3 \times 10^{-4}$, compared to the largest channel in any input data set. This will lead to slow noise variations on scales of ${\gtrsim}1{,}000$ channels. After this regridding, we rebin the data.

The magnitude of this effect is damped out by the rebinning that follows our regridding. At this stage, many independent channels are averaged together to form each final output channel, e.g., for PHANGS--ALMA \cotwo\ we rebin by a factor of $5{-}6$. As a result of this rebinning the fractional difference in the amount of independent data in a final output channel varies only modestly across our data. 

Still, this effect is enough to induce gradual noise variations with magnitude of ${\sim}10\%$ and corresponding variations in the channel-to-channel covariance. These algorithm-induced variations combine with real receiver temperature variations and atmospheric effects to yield the final frequency dependence of the noise in our data cubes (see Section~\ref{sec:products}).

\subsection{Continuum extraction}

We also extract a line-free continuum measurement set. We begin by making a continuum measurement set that includes all spectral windows in each input measurement set. Then we cycle through a list of bright extragalactic emission lines. We use the user-supplied systemic velocity and width to calculate the frequency footprint of each bright line. Whenever a line would overlap the data, we flag all channels associated with the line. 

The user can choose which bright lines to consider for flagging. For most PHANGS--ALMA data, we flag only the \cotwo\ and \mbox{C$^{18}$O($2\text{--}1$)} lines. These represent the only bright lines in our bandpass. This flagging amounts to a flagged bandwidth of ${\sim}0.75$~GHz out of the total $6.75$~GHz bandwidth observed. For observations in later cycles that cover $^{13}$\cotwo, we also flag that line.

After this flagging, we combine all of these line-free measurement sets using the \texttt{CASA} task \texttt{concat}. At this stage, as for spectral line imaging, the user has the option to re-weight the combined data set according to the measured rms noise in the visibility data using the \texttt{CASA} \texttt{statwt}. This option runs the risk of down-weighting regions with bright emission. For PHANGS--ALMA, the signal-to-noise in the continuum is extremely low, and the scatter in amplitude for individual $u{-}v$ data will be determined mainly by the noise in the data. Therefore, we did apply this option for PHANGS--ALMA.

Finally, we use the \texttt{CASA} task \texttt{split} to collapse each spectral window in this continuum-only data set to have only a single channel. This step dramatically reduces the overall volume of the measurement set. For PHANGS--ALMA even after this averaging, the fractional bandwidth of the widest continuum channels is modest, $\lesssim1\%$, and bandwidth smearing is not a large concern given the low signal-to-noise of the continuum. 

\subsection{Staged \texorpdfstring{$u{-}v$}{u-v} data}

After the staging steps, we have a single, combined visibility measurement set for each combination of target, spectral product, and array combination. These measurement sets are usually significantly reduced in data volume from the input products. For example, for NGC~4303 the calibrated \mbox{12-m} and \mbox{7-m} data total ${\sim}48$~GB, while the staged visibility data set totals $2.4$~GB\footnote{Presently, the scaling down for \mbox{7-m}-only data is less dramatic because \texttt{CASA} measurement sets include a large pointing table that cannot be removed. This table represents the majority of the data for \mbox{7-m} observations, but not for \mbox{12-m} observations.}. They are on the desired spectral grid with appropriate weighting for imaging and deconvolution.

%% file: imaging.tex
\begin{figure*}[ht!]
\vspace{-0.2in}
\gridline{\fig{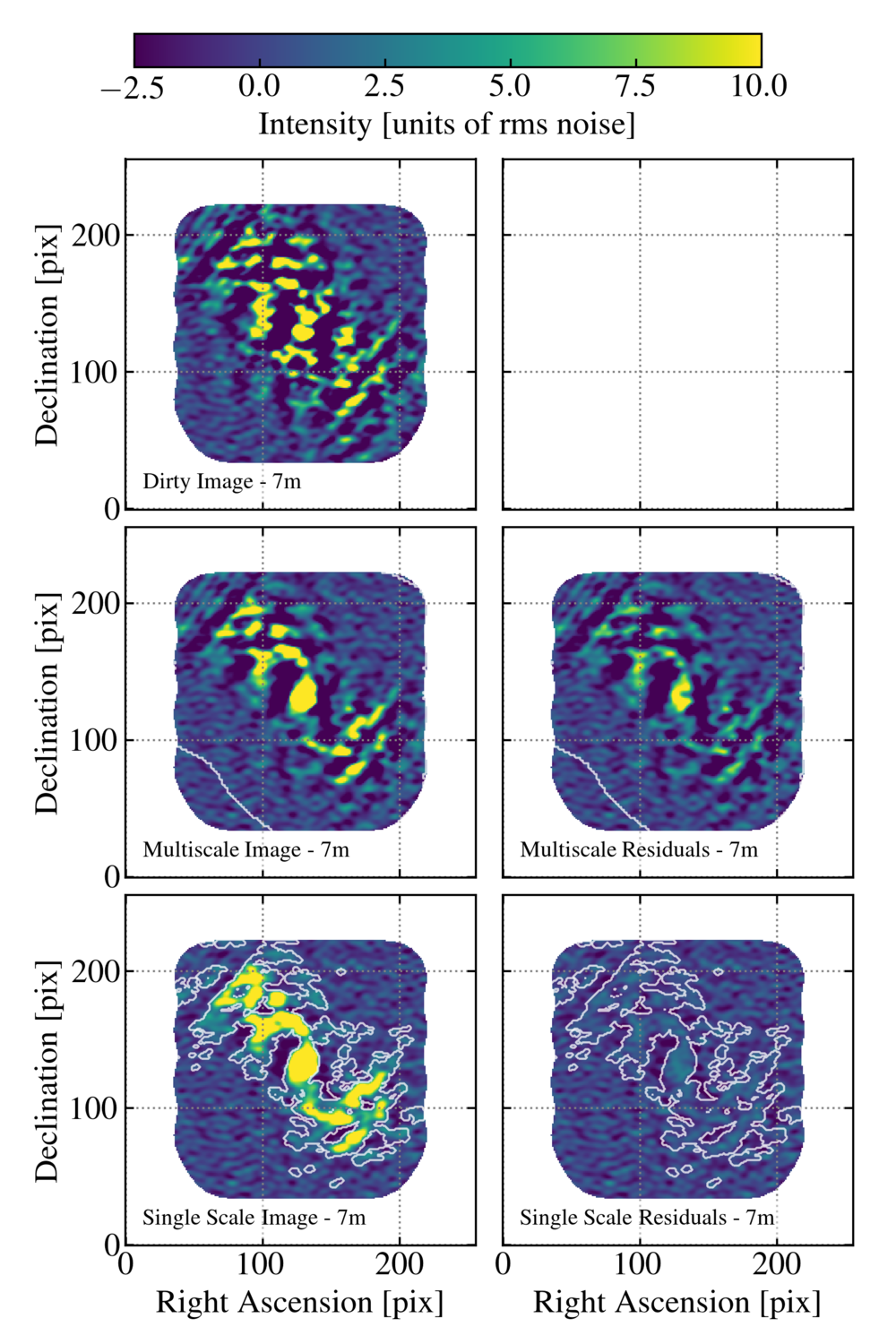}{0.75\textwidth}{}}
\vspace{-0.25in}
\caption{
\textbf{Example of deconvolution for PHANGS-ALMA \mbox{7-m} array data.} Integrated emission from a 20-channel thick slab in the \cotwo\ data cube for one PHANGS-ALMA target. The left column shows the image and the right column the residuals. From top to bottom, we show the dirty image, the image after our first stage of cleaning, which uses multi-scale clean, and the image after both rounds of cleaning, which follows the multi-scale clean with a single-scale clean. Contours show the clean mask at each stage. All images share the same stretch, which saturates at $10$ times the rms noise level in order to focus the image on the low-level, extended structure rather than the bright, easily cleaned sources. Along with Figure~\ref{fig:deconv_12m+7m}, the figure illustrates our approach to imaging and deconvolution. The white space indicates masking used in imaging. The second row shows the broad, inclusive nature of the user-supplied clean mask used in multi-scale clean. The third row shows the much more restrictive mask automatically generated and used in single-scale clean. The middle right panel shows the filamentary, clumpy nature of the residuals after multi-scale clean. The bottom right panel shows the clean residual image after single-scale clean.
\label{fig:deconv_7m}}
\end{figure*}

\begin{figure*}[ht!]
\vspace{-0.2in}
\gridline{\fig{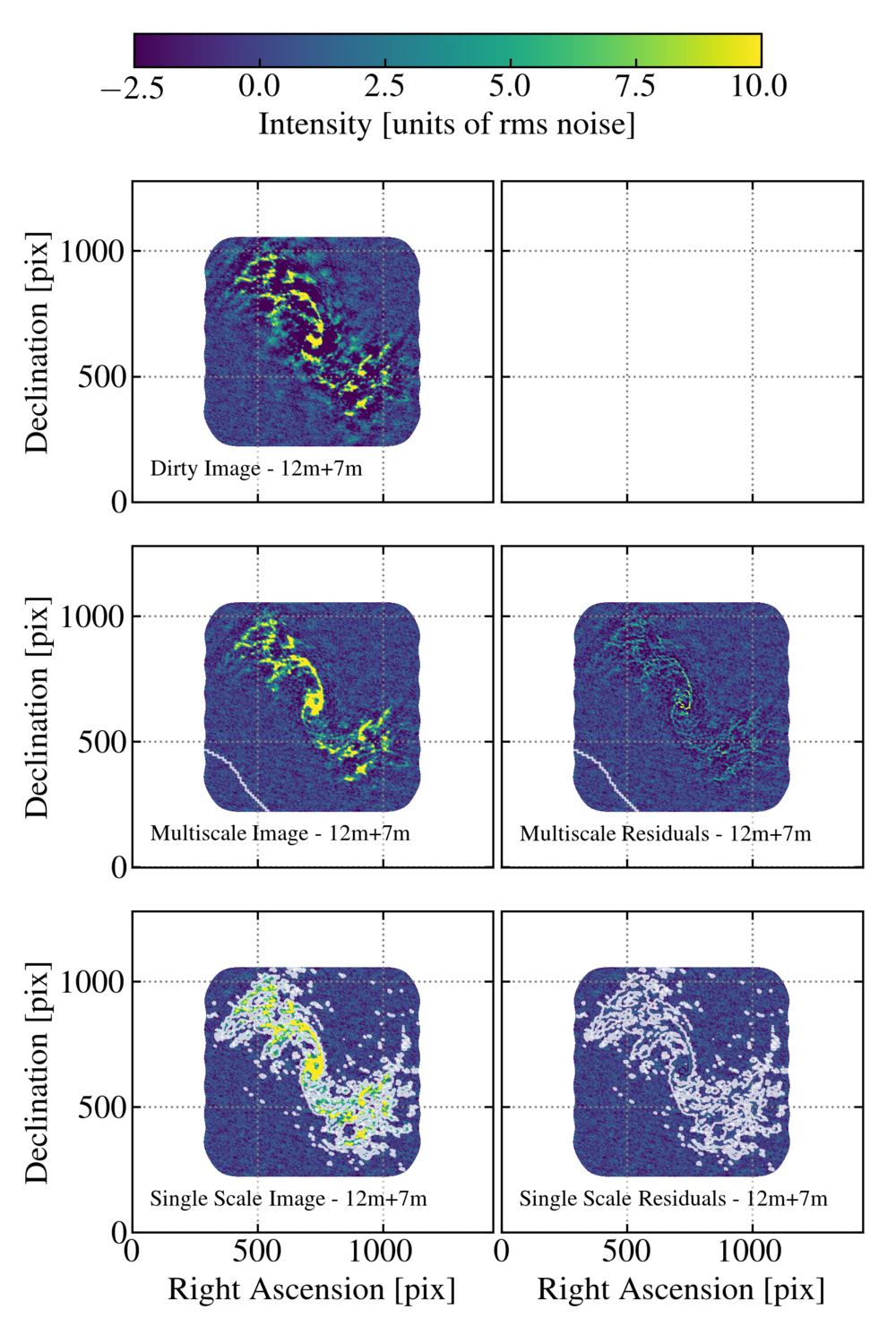}{0.75\textwidth}{}}
\vspace{-0.35in}
\caption{
\textbf{Example of deconvolution for PHANGS-ALMA combined \mbox{12-m} and \mbox{7-m} array data.} As Figure~\ref{fig:deconv_7m} but now showing results for imaging the \mbox{12-m}\,+\,\mbox{7-m} array data together.
\label{fig:deconv_12m+7m}}
\end{figure*}

\begin{deluxetable*}{lcc}
\tabletypesize{\small}
\tablecaption{PHANGS-ALMA CO(2--1) Imaging \label{tab:phangsimaging}}
\tablewidth{0pt}
\tablehead{
\colhead{Description} & 
\colhead{ACA 7-m Value} & 
\colhead{12-m\,+\,7-m Value} 
\\
\colhead{} & 
\multicolumn{2}{c}{(minimum~---~\textbf{16$^{\rm th}$ percentile~---~\underline{median}~---~84$^{\rm th}$ percentile}~---~maximum)} \\
}
\startdata
\hline
\multicolumn{3}{l}{\textbf{Beam}} \\
\hline
Major axis [$''$] & 
6.2\tabdiv\textbf{6.8\tabdiv\underline{7.2}\tabdiv7.9}\tabdiv9.7 & 
0.58\tabdiv\textbf{1.0\tabdiv\underline{1.2}\tabdiv1.6}\tabdiv1.9 \\
Position angle [$^\circ$] & 
69\tabdiv\textbf{82\tabdiv\underline{88}\tabdiv98}\tabdiv124 & 
5\tabdiv\textbf{59\tabdiv\underline{95}\tabdiv116}\tabdiv179 \\
Elongation [major/minor axis] & 
1.1\tabdiv\textbf{1.4\tabdiv\underline{1.7}\tabdiv2.0}\tabdiv2.3 & 
1.0\tabdiv\textbf{1.1\tabdiv\underline{1.2}\tabdiv1.4}\tabdiv1.9 \\
Pixels across beam minor axis & 
3.5\tabdiv\textbf{3.8\tabdiv\underline{4.4}\tabdiv4.9}\tabdiv5.9 & 
4.1\tabdiv\textbf{4.9\tabdiv\underline{5.9}\tabdiv7.0}\tabdiv8.7 \\
\hline
\multicolumn{3}{l}{\textbf{Area Imaged}\tablenotemark{a}} \\
\hline
Pixels across cube major axis & 
120\tabdiv\textbf{240\tabdiv\underline{288}\tabdiv384}\tabdiv512 & 
720\tabdiv\textbf{1152\tabdiv\underline{1536}\tabdiv2304}\tabdiv4608 \\
Area mapped [arcmin$^2$] & 
1.2\tabdiv\textbf{4.0\tabdiv\underline{8.2}\tabdiv22.2}\tabdiv22.8 & 
0.7\tabdiv\textbf{2.8\tabdiv\underline{6.2}\tabdiv7.8}\tabdiv15.2 \\
Spatial dynamic range [$\sqrt{{\rm area/beam}}$] & 
11\tabdiv\textbf{22\tabdiv\underline{28}\tabdiv42}\tabdiv50 & 
54\tabdiv\textbf{81\tabdiv\underline{112}\tabdiv150}\tabdiv264 \\
\hline
\multicolumn{3}{l}{\textbf{Noise} per 2.54~km~s$^{-1}$ channel after imaging} \\
\hline
Noise in residuals [mJy~beam$^{-1}$] & 
5.2\tabdiv\textbf{16\tabdiv\underline{22}\tabdiv67}\tabdiv117 & 
0.8\tabdiv\textbf{3.7\tabdiv\underline{5.5}\tabdiv7.1}\tabdiv10.6 \\
Peak intensity [Jy~beam$^{-1}$] & 
0.11\tabdiv\textbf{0.42\tabdiv\underline{1.4}\tabdiv3.2}\tabdiv27 & 
0.04\tabdiv\textbf{0.10\tabdiv\underline{0.29}\tabdiv0.61}\tabdiv1.1 \\
Peak dynamic range & 
5.4\tabdiv\textbf{16\tabdiv\underline{51}\tabdiv116}\tabdiv264 & 
7.1\tabdiv\textbf{21\tabdiv\underline{51}\tabdiv94}\tabdiv189 \\
\enddata
\tablenotetext{a}{Refers to individual mosaics galaxy parts. These are imaged separately and then linearly mosaicked in the image plane (Section~\ref{sec:postprocess}).}
\tablecomments{These numbers refer to internal release ``version 4'' constructed with the PHANGS-ALMA pipeline ``version 2.0.'' This corresponds to the first PHANGS-ALMA public release. We report numbers for the full set of processed data, though some of these are not part of the initial public release because they are archival or still proprietary. See Figures \ref{fig:beamstats}, \ref{fig:areastats}, and~\ref{fig:noisestats}. Note that some galaxies have been imaged with only the \mbox{7-m} array, so the samples contributing to the two columns differ.}
\end{deluxetable*}

\section{Imaging and Deconvolution of Interferometric Data}
\label{sec:imaging}

We use \texttt{CASA}'s \texttt{tclean} task to image the calibrated measurement sets and to deconvolve the emission into a ``clean'' cube or image.

We adopt a two-stage approach to deconvolution that appears well-suited to complex line emission data. First, we run a multi-scale deconvolution with a high threshold, corresponding to signal-to-noise ratio of~$4$, and little or no constraint on where the deconvolution can place components, i.e., little or no ``clean masking.'' Then, we construct a new, more restrictive clean mask based on the signal in the current cleaned cube. Applying this clean mask, we shift to a standard single-scale deconvolution approach and clean down to a lower threshold, corresponding to signal-to-noise ratio of~$1$. This deep cleaning ensures a good deconvolution of the numerous small angular scale sources seen in our observations of nearby galaxies.

Throughout this process, we force frequent major cycles. During a major cycle, the model is projected into $u{-}v$ space and subtracted from the visibility data. The residual $u{-}v$ data are then imaged and used in the next deconvolution step. By comparing the model and data in the $u{-}v$ plane, we minimize the impact of \texttt{CASA}'s assumption that the synthesized beam, i.e., the interferometric response to a point source, does not vary as a function of position on the sky. In actuality, the synthesized beam can vary across a large mosaic. This variation can introduce minor inconsistencies during the image-plane deconvolution performed in the minor cycles. By frequently projecting back into $u{-}v$ space, these inconsistencies are mostly corrected. More generally, the major cycle represents a direct comparison between data and model. Frequent major cycles also improve the accuracy of the deconvolution and can help overcome, e.g., limited sampling of the $u{-}v$ plane due to the lack of significant rotation synthesis in a single $\sim 1$~hour ALMA observing block.

Periodically stopping and restarting the clean procedure also allows us to check convergence of the deconvolution. We stop the procedure when the fractional change in the model flux with each new clean call drops below 1\%. In PHANGS-ALMA this condition always coincides with the peak residual inside the clean mask approaching the specified threshold, either $4$~times the rms noise for the multi-scale case or $1$~times the rms noise for the single-scale case. When setting these thresholds, the noise level is estimated from the data based on the median absolute value of the residual image. The noise estimated in this way changes relatively little during the course of deconvolution.

This procedure has proven robust. It runs with minimal human intervention across all of PHANGS-ALMA and many other line emission maps. It also works well with many VLA \mbox{21-cm} \hi\ data sets, though we note a few caveats below. In our view, the key choices were:

\begin{itemize}
\item Use multi-scale clean with no clean mask or a very non-restrictive clean mask and a relatively high signal-to-noise threshold.
\item Force many major cycles.
\item Clean deep with a carefully directed single-scale clean, adopting a low signal-to-noise threshold.
\item Direct this single-scale clean by applying automated masking to the current deconvolved image, rather than, e.g., the residuals.
\end{itemize}

The pipeline allows user-input clean masks, but these are not necessary for good performance. When we use clean masks at the multi-scale stage, they must be very broad in order to avoid divergence due to interactions between the clean algorithm and the mask boundary. Any user-supplied mask is then used as a prior during the automated creation of the single-scale mask. At this stage, the user-supplied masks help avoid cleaning noise spikes in the often large, signal-free regions of the cube. Avoiding these noise spikes will have a mild impact on the final noise properties, but the main gain is to save computing time during the single-scale clean. Thus while we do use input clean masks for PHANGS-ALMA, these masks are not crucial to the overall performance of the pipeline deconvolution. Indeed, our first-pass imaging for the PHANGS-ALMA targets without any clean masks yielded almost the same results as the final imaging run. By contrast, supplying an over-restrictive mask often biases the deconvolution and can lead to divergence during multi-scale cleaning.

We illustrate the procedure for one galaxy in Figures \ref{fig:deconv_7m} and~\ref{fig:deconv_12m+7m}. Figure~\ref{fig:deconv_7m} shows the deconvolution of the \mbox{7-m} data for that galaxy. Figure~\ref{fig:deconv_12m+7m} shows the combined deconvolution of the \mbox{12-m} and \mbox{7-m} data. Both figures show snapshots of a 20-channel ``slab'', i.e., an integral across twenty velocity channels, in one PHANGS-ALMA galaxy. Because the integral extends across the slab, the signal-to-noise of these images is improved by a factor of ${\sim}5$ compared to the individual channel maps themselves. Thus, these visualizations show a very aggressive stretch that could bring out artifacts not necessarily visible in individual channel maps.

\medskip

\noindent \textbf{\texttt{CASA} version:} \texttt{tclean} refers to the latest CLEAN algorithm implementation available in \texttt{CASA}. This task evolved significantly over the course of the PHANGS-ALMA project. For the ``v2 pipeline'' and v4 PHANGS-ALMA data release associated with this paper, we imaged the data using \texttt{tclean} in its serial (i.e., non-parallel) mode in \texttt{CASA} version $5.4.0$. 

\medskip 

\noindent \textbf{PHANGS-ALMA CO(2--1) imaging summary:} Table~\ref{tab:phangsimaging} and Figures \ref{fig:beamstats}, \ref{fig:areastats}, and~\ref{fig:noisestats} summarize our application of this procedure to image the PHANGS-ALMA \cotwo\ data. They report the minimum, maximum, median, and $16^{\rm th}{-}84^{\rm th}$ percentile range of key quantities including the properties of the synthesized beam, the area imaged, and the noise and dynamic range achieved in the cubes. For PHANGS-ALMA we imaged both the \mbox{7-m} array data and the combined \mbox{12-m+7-m} array data. We report numbers for both array combinations, though we emphasize that when both arrays are available we strongly prefer the combined \mbox{12-m}+\mbox{7-m} result to that from the \mbox{7-m} alone (see Appendix~\ref{sec:arrays}).

\subsection{Imaging}

\begin{figure*}[ht!]
\gridline{
\fig{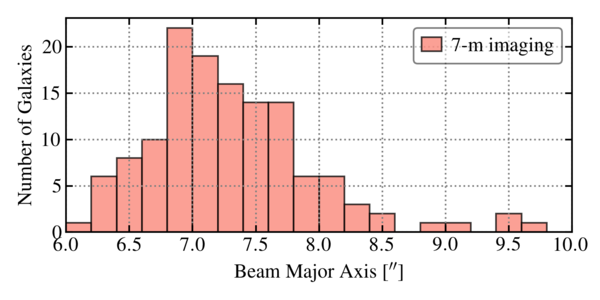}{0.48\textwidth}{}
\fig{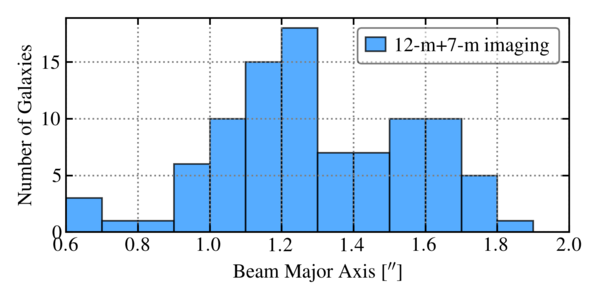}{0.48\textwidth}{}
}
\vspace{-24pt} % reduce excessive gaps between figure rows
\gridline{
\fig{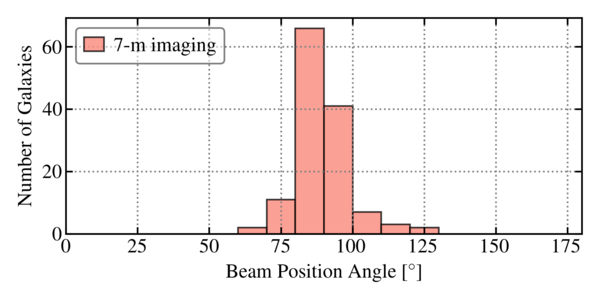}{0.48\textwidth}{}
\fig{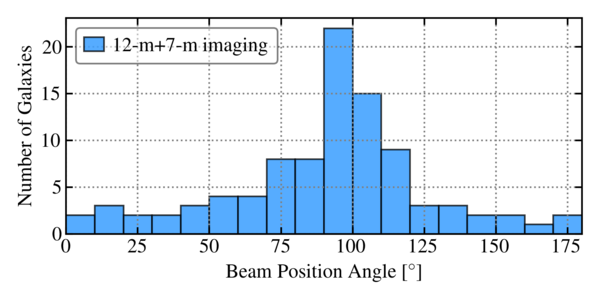}{0.48\textwidth}{}
}
\vspace{-24pt} % reduce excessive gaps between figure rows
\gridline{
\fig{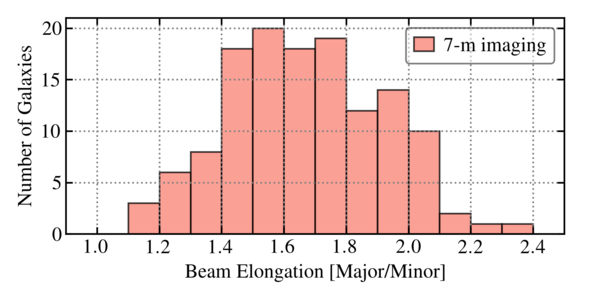}{0.48\textwidth}{}
\fig{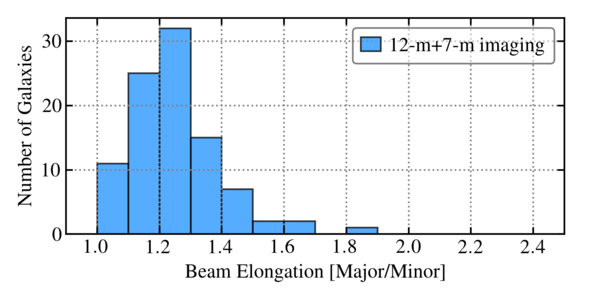}{0.48\textwidth}{}
}
\vspace{-24pt} % reduce excessive gaps between figure rows
\gridline{
\fig{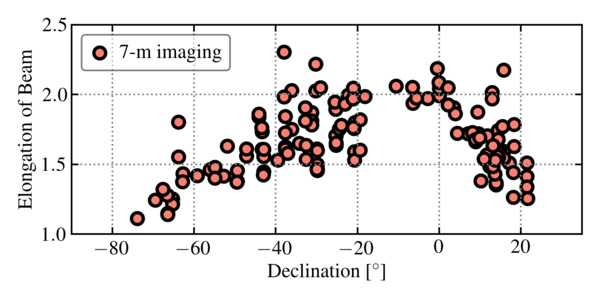}{0.48\textwidth}{}
\fig{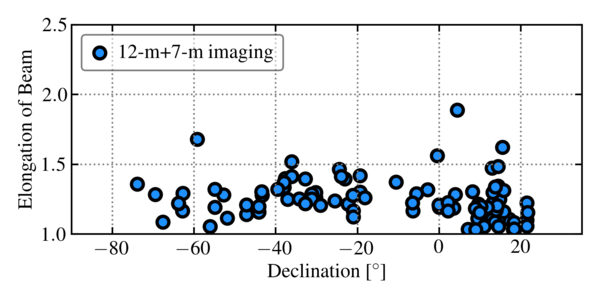}{0.48\textwidth}{}
}
\vspace{-24pt} % reduce excessive gaps between figure rows
\caption{
\textbf{Imaging properties related to the beam.} Properties of the imaged \cotwo\ PHANGS-ALMA cubes for the ACA \mbox{7-m} data only (\textit{left} column) and the combined \mbox{12-m} and \mbox{7-m} data (\textit{right} column). From top to bottom, we show the FWHM major axis of the synthesized beam, the position angle (measured North through East) of the major axis of the synthesized beam, the beam elongation (defined as major over minor axis), and the elongation as a function of declination of the source. See Table~\ref{tab:phangsimaging}. Note that some galaxies have been imaged with only the \mbox{7-m} array, so the samples contributing to the two columns differ.
\label{fig:beamstats}}
\end{figure*}

Most of the inputs to \texttt{tclean} are tunable parameters in the pipeline. By default, PHANGS-ALMA uses the following imaging parameters:

\begin{enumerate}

\item \textbf{Cell size.} We use the ALMA observatory-developed \texttt{analysisutils} package to estimate the size of the synthesized beam based on the $u{-}v$ coverage of the data. Then the pipeline picks a cell size that is both a round number, e.g., $0.05\arcsec$ or $0.2\arcsec$, and oversamples the synthesized beam by a factor of $\gtrsim 4$ along the minor axis and more along the major axis. 

As shown in Table~\ref{tab:phangsimaging} and Figure~\ref{fig:beamstats}, for PHANGS-ALMA we place $4{-}7$ pixels along the beam minor axis and $6{-}10$ pixels across the beam major axis.

\item \textbf{Image size.} The pipeline chooses an image size with a linear extent $>20\%$ larger than the field of view of the data themselves. We choose an image size in pixels that matches the recommendations for best performance using \texttt{CASA}'s Fast Fourier Transform (FFT) algorithm, i.e., that is even and can be factorized to 2, 3, 5, and 7 only.

As Table~\ref{tab:phangsimaging} and Figure~\ref{fig:beamstats} show, for PHANGS-ALMA this translates to typically $240{-}384$ pixels across the ACA \mbox{7-m} data cubes and $1{,}152{-}2{,}304$ pixels across the combined \mbox{12-m}+\mbox{7-m} data cubes. The $>20\%$ buffer to the image size can be seen as white space in Figures \ref{fig:deconv_7m} and~\ref{fig:deconv_12m+7m}. 

\item \textbf{Frequency grid.} For line cubes, the pipeline adopts the frequency grid set during the $u{-}v$ data processing described above. For the delivered PHANGS-ALMA \cotwo\ imaging this translates to ${\sim}2.54$~km~s$^{-1}$ channel width with minor variations from target to target. 

\item \textbf{Gridding algorithm, weighting, and primary beam cutoff.} By default, the pipeline uses \texttt{CASA}'s ``mosaic'' gridding algorithm and weights the $u{-}v$ data according to the ``Briggs'' scheme. It defaults to robustness parameter $r = 0.5$, which offers a good compromise between noise and resolution. By default, it images out to a primary beam cutoff of $0.25$. We adopt all of these parameters when imaging PHANGS-ALMA.

Following the observatory recommendations, we set \texttt{mosweight} to True and calculate the $u{-}v$ weighting for each field separately. Following the documentation, this can improve imaging performance for mosaics at the expense of a slightly larger beam. Because we imaged in \texttt{CASA} $5.4$, before the \texttt{perchanweightdensity} parameter was introduced, our imaging effectively sets \texttt{perchanweightdensity} to False. This parameter instructs \texttt{tclean} to weight each channel individually. Similar to \texttt{mosweight} it should lead to better imaging performance at the expense of a slightly larger beam size. In the future, runs of the PHANGS-ALMA imaging pipeline using \texttt{CASA} version $5.5$ the user can choose whether to adopt per-channel weighting by setting the \texttt{perchanweightdensity} in the clean call.

\item \textbf{Independently image mosaics observed separately.} For PHANGS-ALMA, we observed some galaxies in multiple parts. Each part corresponds to a ${\sim}150$ field mosaic and the parts were observed separately. We imaged each separate part independently.
\end{enumerate}

The choice to independently image each separately observed mosaic is important for PHANGS-ALMA. When we observed a galaxy using several adjacent mosaics, these mosaics were sometimes observed at different times and even different array configurations. This implies a spatially variable synthesized beam across the field, and \texttt{CASA} cannot currently account for position-dependent synthesized beams. Our initial attempts to jointly image multiple large mosaics frequently resulted in divergence. This problem was resolved when we shifted our strategy to image each part separately and then linearly mosaic the parts together.

\medskip

\noindent \textbf{Dirty image and clean mask alignment:} As the first step in imaging, we constructed a ``dirty'' cube.  This cube used our adopted imaging parameters but we performed no deconvolution. 

If the user supplied a clean mask, as was the case for PHANGS-ALMA, then at this stage we used \texttt{CASA}'s \texttt{importfits} and \texttt{imregrid} tasks to align the clean mask to the astrometric grid and axis order of the dirty cube.

Slabs, i.e., integrals over 20 successive channels, in PHANGS-ALMA dirty cubes appear as the top row in Figures \ref{fig:deconv_7m} and~\ref{fig:deconv_12m+7m}. As expected, these dirty images look highly distorted due to spatial filtering through the incomplete $u{-}v$ coverage of the interferometer. The imprint of the user-supplied clean mask for PHANGS-ALMA appears as a contour in the second row.

\subsection{Deconvolution}

The pipeline  uses \texttt{tclean} to deconvolve emission and create a clean cube or image. As described above, this has two main stages: a ``wide'' multi-scale clean and a ``directed, deep'' single-scale clean. We follow a few general principles in both stages:

\begin{enumerate}

\item \textbf{Force frequent major cycles.} The pipeline requires ``major cycles'' to happen frequently. During a major cycle, the approximate image-plane deconvolution is projected back into visibility (Fourier) space and the model is properly subtracted from the data. While computationally expensive, this process produces a more correct residual image, allowing for a more stable, precise deconvolution.

In practice, the pipeline enforces major cycles within each \texttt{tclean} call in two ways. First, it limits the number of ``iterations'' allowed before forcing a major cycle using the \texttt{cycleniter} keyword. Second, it uses a combination of \texttt{cyclefactor} and \texttt{minpsffraction} to set an aggressive threshold for triggering a major cycle. Once the data are cleaned so that the maximum residual approaches this threshold level, \texttt{tclean} triggers a major cycle. For PHANGS-ALMA our default values for these parameters were $\texttt{cyclefactor} = 3.0$ and $\texttt{minpsffraction}=0.5$. These imply that the threshold is never lower than $0.5$ times the peak residual or three times the maximum sidelobe level times the peak residual. By default, the pipeline also uses $\texttt{maxpsffraction}=0.8$ to ensure that some emission is deconvolved in each cycle.

\item \textbf{Multiple \texttt{tclean} calls with more components deconvolved in later calls.} The deconvolution involved many repeated calls to \texttt{tclean}. When the pipeline initially calls \texttt{tclean}, it allows only for a small number of clean components, with the number set via the \texttt{niter} keyword. It also allows for only a limited number of components to be cleaned per channel before enforcing a major cycle. This is set via the \texttt{cycleniter} keyword. Once the overall number of allocated clean components is exceeded, \texttt{tclean} stops. Stopping and resuming \texttt{tclean} forces a major cycle.

Over the course of the first five \texttt{tclean} calls, the pipeline increases \texttt{niter} and \texttt{cycleniter}. By default, the pipeline increases \texttt{niter} by a factor of $2$ each step. It linearly increases \texttt{cycleniter}, starting at 100 and increasing it by 100 at each step in the loop. The choice to limit the number of components in any individual call to \texttt{tclean} is part of our strategy to trigger frequent major cycles.

This gradual increase in allocated clean components resembles the approach used to create the PdBI CO image of M51 by \citet{PETY13}. The numerical choice of how to progressively increase the number of iterations is \textit{ad~hoc}.

\item \textbf{Check for convergence between clean calls.} These repeated calls to \texttt{tclean} allow us to check for convergence in the deconvolution. After each call and before the next one, we calculate the sum of flux in the model (i.e., the clean components). We compare this flux to the previous model flux to calculate the fractional change in flux and the gain in flux per allocated clean component. When the fractional change in the model flux drops below some threshold, usually $1\%$, we terminate that stage of the deconvolution and move to the next one. In the case of the multi-scale clean, we move to automated masking and single-scale cleaning. In the case of single-scale cleaning, we finish the deconvolution and move to postprocessing. 

\item \textbf{Common restoring beam.} By default, we use a common restoring beam, meaning that \texttt{tclean} restores deconvolved emission with a single elliptical, Gaussian beam across all planes of the cube. The alternative offered by \texttt{CASA} is to track the beam per plane, reflecting differences in how the $u{-}v$ coverage maps to angular scale as the frequency changes. For PHANGS-ALMA, the fractional bandwidth, $\delta \nu / \nu$, across our cubes is modest, always $< 0.0035$. As a result, the synthesized beam does not change much with frequency and we do not keep track of a beam per plane. This choice can be changed by the user. For example, a change may be required when many data are flagged in a few channels, which would otherwise result in a large common beam.
\end{enumerate}

\subsection{Multi-scale clean}

In the first stage of deconvolution, we employ the \texttt{CASA} implementation of the ``multi-scale'' deconvolution algorithm \citep[][]{CORNWELL08}. For this stage, PHANGS-ALMA uses a broad clean mask supplied by the user, but the operation also works well with no mask. The scales to be cleaned are also specified by the user as part of defining the configurations. We follow the \texttt{CASA} recommendation regarding choice of scales and use scales from the beam size to within a factor of ${\sim}2$ of the largest recoverable scale. 

Multi-scale clean includes a tuning parameter, \texttt{smallscalebias}, that can be used to bias the results toward small or large scales. We set \texttt{smallscalebias} to $0.9$ by default, indicating a preference for small scales. During development, we experimented with scales from $0.4$ to $0.9$. We found higher values less likely to yield divergence. Note that these tests used earlier versions of \texttt{CASA}, mostly $4.5$ and $4.7$. This may reflect the common presence of a few bright, clumpy structures in our CO maps.

For PHANGS-ALMA, when deconvolving only \mbox{7-m} data, we employed scales of $0\arcsec$ (i.e., a point source), $5\arcsec$, and~$10\arcsec$. When deconvolving the combined \mbox{12-m}+\mbox{7-m} data, we considered scales of $0\arcsec$, $1\arcsec$, $2.5\arcsec$, $5\arcsec$, and~$10\arcsec$. When deconvolving only \mbox{12-m} data, we used scales of $0\arcsec$, $1\arcsec$, $2.5\arcsec$, and~$5\arcsec$. These deconvolution scales correspond to the size of round Gaussian clean components before convolution with the dirty beam.

We impose a threshold of $4$ times the rms noise on the multi-scale cleaning process. For this purpose, we take a single robustly estimated noise value to describe the whole cube (but see \S \ref{sec:noise}). When the peak value in the residual map for each channel falls below this level, cleaning stops in that channel. We estimate the noise from the residual cube, and update this noise estimate between calls to \texttt{tclean}. Because we use a robust noise estimator and the cubes contain a large amount of empty volume, the estimated value of the noise changes little between calls. We found that adopting lower S/N thresholds for the multi-scale clean led to divergence in the deconvolution \citep[for similar conclusions using VLA data see][]{KOCH18}.

As described above, after each call to \texttt{tclean} we sum the total flux in the model image, i.e., the sum of deconvolved flux. When this flux changes by $< 1\%$ between subsequent calls to \texttt{tclean}, we move to the next stage of the deconvolution. Usually this convergence coincides with the peak residual approaching the S/N-based threshold. If the deconvolution has not converged, then we increased the \texttt{niter} and \texttt{cycleniter} and we continue the multi-scale deconvolution with a new call to \texttt{tclean}.

\subsection{Masking and single-scale clean}
\label{sec:masking_and_clean}

After the multi-scale deconvolution converged, there were often still significant residuals around the brightest sources. At this stage, we proceed deconvolving with the classic, single-scale CLEAN algorithm \citep{HOGBOM74} and use it to clean down to a threshold equivalent to signal-to-noise of~$1$. We also generate and apply a much more restrictive clean mask at this step. This masking avoids spending large amounts of effort cleaning signal-free regions of the data cube and makes it possible for the deconvolution to clean very deeply in regions with signal. The shift to the single-scale clean avoids potential pathological interactions between this more restrictive clean mask and large cleaning scales.

We use the resultant multi-scale deconvolved image to construct a signal-to-noise based mask. To do this, we estimate a characteristic rms noise in the cube based on the median absolute deviation of the whole residual cube. Then, we create a mask that includes all regions that have ${\rm S/N}>4$. We then expand this mask to adjacent regions with ${\rm S/N}>2$. Finally, we extend the mask by one channel in each velocity direction. If the user supplied a clean mask, then during this step we only include pixels in the mask that also lie inside the original clean mask.

In this way, we focus the single-scale clean on regions where signal is already evident in the cleaned maps after the multi-scale clean. We note that this approach differs from the automated masking within the tclean task in \texttt{CASA} ``auto-multithresh''. \texttt{CASA}'s algorithm builds a clean mask based on the current residual emission as part of the major cycle \citep{KEPLEY20}, while we construct a clean mask based on the deconvolved emission outside the deconvolution process. Based on experimentation, we found by eye that our approach did a good job of identifying the regions of the residual image where one would want to clean deeper. Put another way, we use the single-scale clean to ``dig deeper'' to ensure a full deconvolution of already-visible bright regions.

During this single-scale deconvolution, we impose a S/N threshold of~$1$, again using a single robustly-estimated noise value to describe the whole cube. This threshold means that we stop the deconvolution in each channel when the maximum residual in that channel reaches a value equal to the noise level. This limit is much lower than the threshold that we adopted for the multi-scale clean. This change causes the single-scale clean to deconvolve a large network of filamentary ${\rm S/N} \lesssim 4$ residuals commonly remaining after the shallow multi-scale clean.

As with the multi-scale deconvolution, during this step we allocate only a limited number of iterations to each \texttt{tclean} call. Between calls we check for convergence. Again we define this as the flux in the model changing by $<1\%$ between successive clean calls. We begin these convergence checks after three calls to single-scale \texttt{tclean}. This delay allows us time to allocate enough iterations to allow some expectation of convergence. 

Our peak residual threshold in individual calls to \texttt{tclean} interacts with our fractional-change-in-flux criteria. In practice, the fractional change in flux drops below $1\%$ when the peak residuals inside the clean mask approach the threshold. For PHANGS-ALMA, the single-scale clean thus effectively cleans down to a peak ${\rm S/N} = 1$ in the residuals within the clean mask.

\subsection{Input or iterative clean masks}

\begin{figure}[t!]
\gridline{\fig{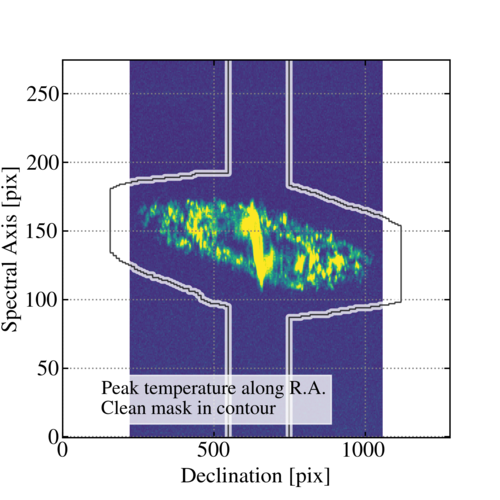}{0.5\textwidth}{}
}
\vspace{-30pt} % reduce excessive gaps between figure rows
\caption{
\textbf{PHANGS-ALMA clean mask in projection.} A two dimensional projection of one PHANGS-ALMA ``user supplied'' clean mask is shown. The background image shows peak intensity along the R.A.\ axis in the \mbox{12-m}+\mbox{7-m} imaging for NGC~4303. The black-and-white contours show the locations along the line of sight where the mask is ``True'' for at least one pixel along the R.A.\ axis. The clean masks are created based on previous rounds of imaging. The figure illustrates how the clean masks broadly circle emission in the cube, rather than applying any substantial restriction. In order to avoid any edge effects, this mask even reaches slightly beyond the ALMA coverage, hence the extension into the white region. Their main effect is to save computing time by avoiding processing the signal-free regions of the cube. The central rectangle shows a region extending over the full velocity width of the cube centered on the galaxy center. We include a feature like this for all galaxies with bright centers.
\label{fig:cleanmask}}
\end{figure}

As discussed above, user-input clean masks are optional in our approach. Indeed, they mostly do not appear necessary. We imaged every PHANGS-ALMA target without a user-supplied mask before imaging them with masks. These initial images generally appeared similar to the final ones. 

The procedure works without an input clean mask because the high threshold adopted for the multi-scale clean makes heavy cleaning of noise spikes unlikely. After this, the pipeline creates a clean mask and our automated masking procedure appears to generally work well. The main gains in using a mask appear to be related to performance. Our single-masking approach will still produce some false positives when applied to large signal-free regions. When we supplied broad clean masks that restricted clean to the general area of the galaxy, we avoided time cleaning spurious ``islands'' of emission during both clean stages.

When provided, input clean masks need to encompass all real emission and be extended compared to the scales used by the multi-scale deconvolution. 
In PHANGS-ALMA our general procedure is to adopt an iterative approach. We image a target without any prior clean mask. Then we convolve the initial deconvolved cube to coarser resolution. Then we adopt a masking approach similar to that used in product creation below. Finally, we dilate the mask by several channels in the velocity dimension and by about the largest recoverable scale in the spatial dimension.

Specifically, we created our clean masks by convolving the initial \mbox{7-m} imaging to coarser angular and spectral resolution, $33'' \times 20$~km~s$^{-1}$. We constructed a mask at this low resolution via sigma-clipping. For any galaxy deemed to have a bright central region, we extended the mask over the inner $40''$ diameter to cover the full velocity range of the cube. We found that this was necessary to ensure complete coverage of any compact, high-velocity material associated with the inner disk or outflows. We inspect each mask on a high stretch in all projections of position-position-velocity space to ensure that the mask includes all emission with enough room for the multi-scale deconvolution to place large components.

\subsection{Comments on PHANGS-ALMA imaging}

\begin{figure*}[ht!]
\gridline{
\fig{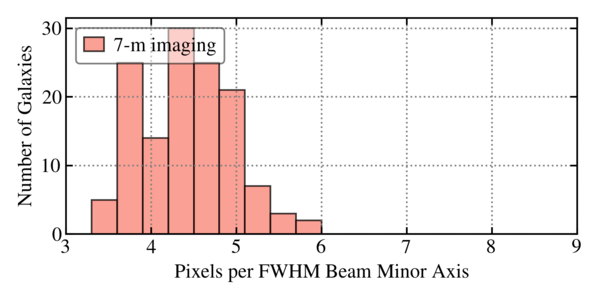}{0.48\textwidth}{}
\fig{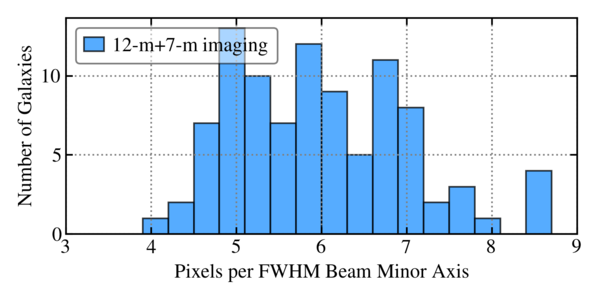}{0.48\textwidth}{}
}
\vspace{-24pt} % reduce excessive gaps between figure rows
\gridline{
\fig{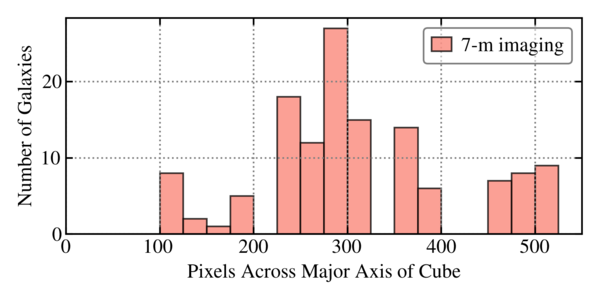}{0.48\textwidth}{}
\fig{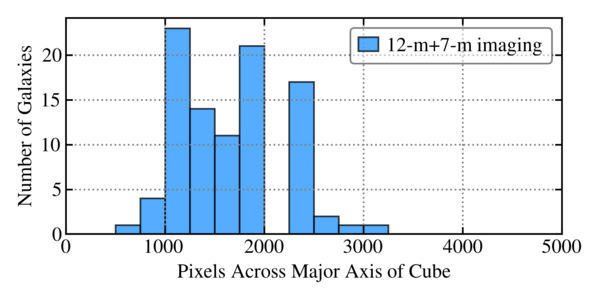}{0.48\textwidth}{}
}
\vspace{-24pt} % reduce excessive gaps between figure rows
\gridline{
\fig{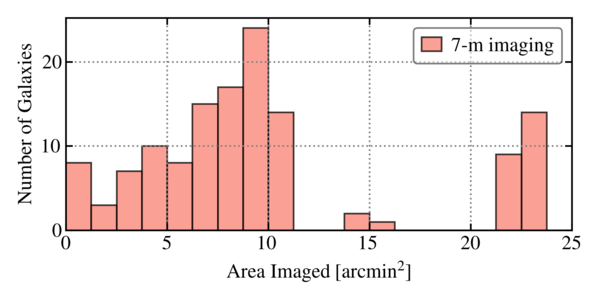}{0.48\textwidth}{}
\fig{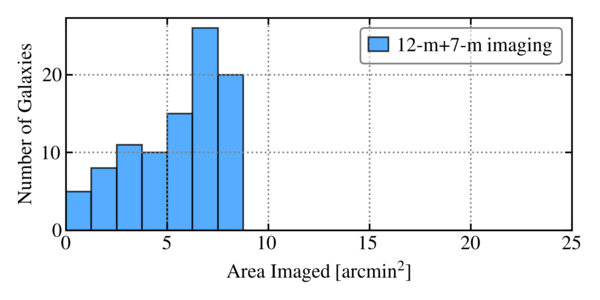}{0.48\textwidth}{}
}
\vspace{-24pt} % reduce excessive gaps between figure rows
\gridline{
\fig{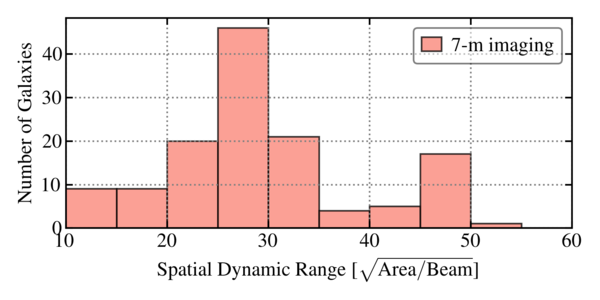}{0.48\textwidth}{}
\fig{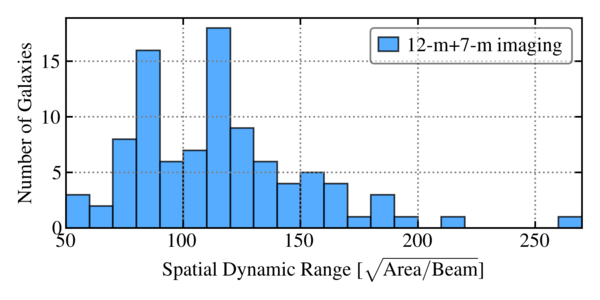}{0.48\textwidth}{}
}
\vspace{-24pt} % reduce excessive gaps between figure rows
\caption{
\textbf{Imaging properties related to the mapping area and image size.} Properties of the imaged \cotwo\ PHANGS-ALMA cubes for the ACA \mbox{7-m} data only (\textit{left} column) and the combined \mbox{12-m} and \mbox{7-m} data (\textit{right} column). From top to bottom, we show the number of pixels across the FWHM of the beam minor axis, the number of pixels across the major axis of the cube, the area of sky imaged, and the spatial dynamic range of the imaged region. See Table \ref{tab:phangsimaging}. Note that some galaxies have been imaged with only the \mbox{7-m} array, so the samples contributing to the two columns differ. Also note that individual mosaic parts are imaged separately. The final spatial dynamic range of those images will be higher than shown here.
\label{fig:areastats}}
\end{figure*}

\begin{figure*}[ht!]
\gridline{
\fig{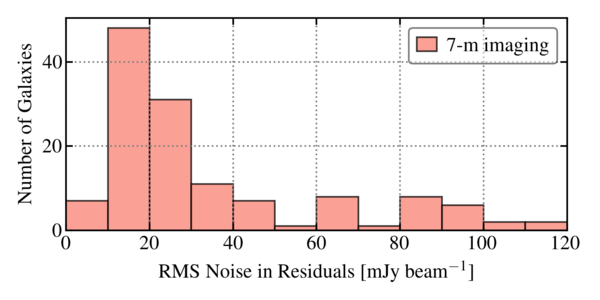}{0.48\textwidth}{}
\fig{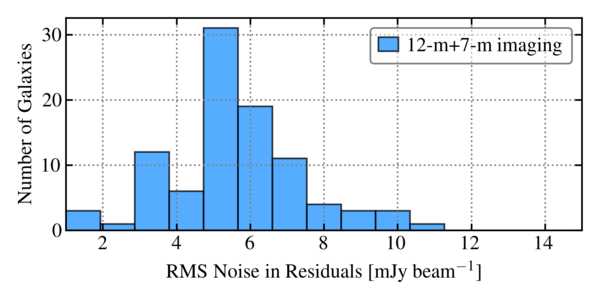}{0.48\textwidth}{}
}
\vspace{-24pt} % reduce excessive gaps between figure rows
\gridline{
\fig{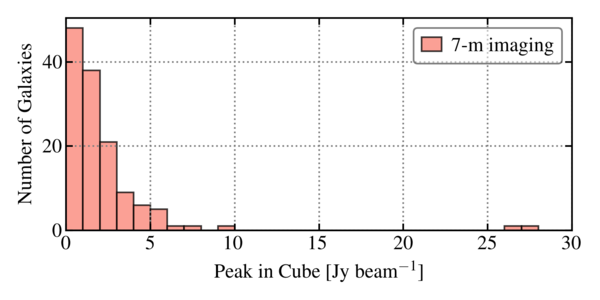}{0.48\textwidth}{}
\fig{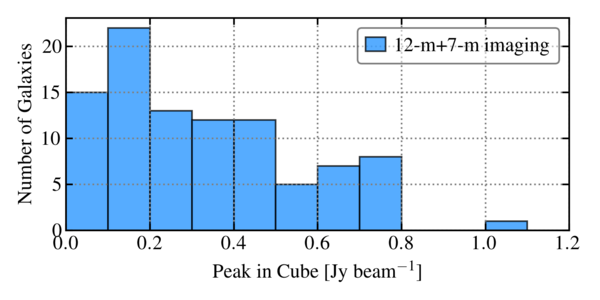}{0.48\textwidth}{}
}
\vspace{-24pt} % reduce excessive gaps between figure rows
\gridline{
\fig{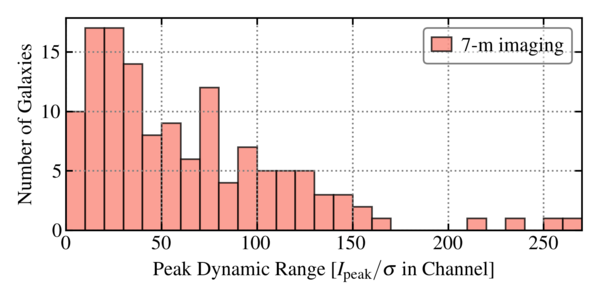}{0.48\textwidth}{}
\fig{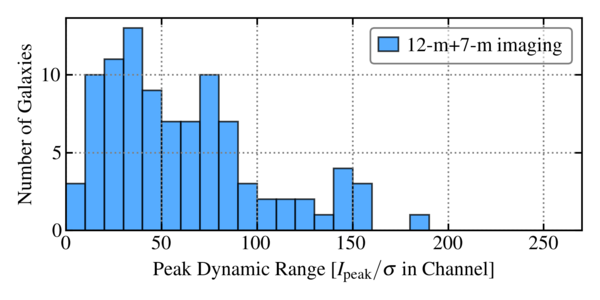}{0.48\textwidth}{}
}
\vspace{-24pt} % reduce excessive gaps between figure rows
\caption{
\textbf{Imaging properties related to noise and dynamic range.} Properties of the imaged \cotwo\ PHANGS-ALMA cubes for the ACA \mbox{7-m} data only (\textit{left} column) and the combined \mbox{12-m} and \mbox{7-m} data (\textit{right} column). From top to bottom, we show the rms noise in the residuals of each cube, the peak intensity in each cube, and the maximum dynamic range (peak intensity over rms noise) in any channel of the cube. See Table~\ref{tab:phangsimaging}. Note that some galaxies have been imaged with only the \mbox{7-m} array, so the samples contributing to the two columns differ.
\label{fig:noisestats}}
\end{figure*}

Table~\ref{tab:phangsimaging} and Figures \ref{fig:beamstats}, \ref{fig:areastats}, and~\ref{fig:noisestats} summarize the application of these algorithms to the PHANGS-ALMA \cotwo\ data.

Imaging only the ACA \mbox{7-m} data yields synthesized beam sizes mostly in the range of $6.8''{-}7.9''$. The beams for the ACA \mbox{7-m} data tend to be significantly elongated, with the major-to-minor axis ratio typically in the range of $1.4{-}2.0$. The elongation is mostly along the East-West direction and worst at intermediate declination as expected based on the information provided in the ALMA Technical Handbook, which reports large beam elongations for $\rm -40^{\circ} < Dec. < 0^{\circ}$.

For the \mbox{7-m} imaging, we typically place $7$~pixels across the major axis of the beam, $4.4$ pixels across the minor axis of the beam, and image a cube $240{-}384$ pixels across. On average, the maps are a few square arcminutes in size, with a median $8.2$~arcmin$^{2}$. Across the entire \mbox{7-m} portion of the survey and including archival data, we mapped about $0.4$~square degrees. The typical spatial dynamic range of an individual \mbox{7-m} image, defined as the number of resolution elements along one dimension of the image, is about~$28$.

For the \mbox{7-m} imaging, we achieve a typical rms noise of $22$~mJy~beam$^{-1}$ per $2.54$~km~s$^{-1}$ channel. The peak dynamic range, meaning peak intensity in a channel divided by rms scatter in that channel, varies across the sample but is mostly in the range of $16{-}116$. Note that this is the dynamic range in an individual channel. The ${\sim}2.54$~km~s$^{-1}$ channel width places about $5{-}10$ elements, and sometimes many more, across a typical emission line. As a result, the line-integrated signal-to-noise is even higher.

The combined \mbox{12-m} and \mbox{7-m} imaging typically yields a beam size of $1.0''{-}1.6''$, with median of $1.2''$. These beams tend to be less elongated, with a median major-to-minor axis ratio of~$1.2$ (consistent with the expected beam shape based on the ALMA configurations). As with the \mbox{7-m} data, the elongation tends to place the major axis in the East-West direction. Here, we place about $6$~pixels along the minor axis of the beam when imaging.

The \mbox{12-m}+\mbox{7-m} cubes are much larger in pixel units, typically $1{,}152{-}2{,}304$ pixels across. Again, the cubes tend to cover a few square arcminutes, usually $2.8{-}7.8$~arcmin$^2$ and $6.2$~arcmin$^{2}$ on average. The slightly smaller mapped area reflects the larger primary beam of the \mbox{7-m} antennas and that the \mbox{7-m} sample includes several very large, nearby galaxies, e.g., NGC~0253, that we did not map with the \mbox{12-m} array. The total area mapped by the \mbox{12-m} survey is  about $0.15$ square degrees, about half the area covered by the \mbox{7-m} survey.

The spatial dynamic range of the combined images is much higher than for the \mbox{7-m}-only data. The typical spatial dynamic range of $112$ corresponds to $>10{,}000$ independent spectra per image.

The typical noise in the residuals of the combined data is $5.5$~mJy~beam$^{-1}$ per 2.54~km~s$^{-1}$ channel and the peak dynamic range is similar to that in the \mbox{7-m}-only images: ${\sim}50$ on average. Again the integrated signal-to-noise in the maps will be even higher.

We consistently deconvolve more flux when imaging the combined \mbox{12-m}+\mbox{7-m} array data than when imaging the \mbox{7-m} array data for the same target. In Appendix~\ref{sec:arrays}, we analyze this effect using both our full data set and simulated data, in which the correct sky image is know \textit{a priori} (see Section~\ref{sec:endtoend}). Our analysis suggests that this discrepancy is a general feature of ALMA observations of nearby galaxies: compact \mbox{12-m} array observations play an important role in achieving a complete deconvolution of emission, even when \mbox{7-m} array observations are present.

\subsection{Limitations of the imaging approach}

Overall, this imaging scheme has proven robust and we have successfully applied it to a variety of ALMA and VLA line and continuum data. However, we have encountered a few cases where the approach does not work or needs modification, and we note these here. First, when imaging sources with bright, not-yet-subtracted continuum emission, our convergence tests need modification. The convergence test focuses on the fractional change in flux. Including one or more high-flux point sources can skew the imaging to converge before any surrounding faint emission has been imaged. More generally, our convergence criteria need to be refined to reflect the desired dynamic range. Our adopted criteria work well for the dynamic range of $\sim 10{-}1{,}000$ expected for PHANGS-ALMA and VLA \mbox{21-cm} imaging of nearby galaxies. 

Second, when imaging structures with extended, highly asymmetric structure, the use of large, symmetric multi-scale clean components can lead to oversubtraction. To some degree, \texttt{tclean} can make up for this by adding negative components to the model. However in some cases, either adjusting the \texttt{smallscalebias} tuning parameter to emphasize small scales or adopting a more restrictive clean mask can improve performance. We have mainly encountered this issue in applying the algorithm to \mbox{21-cm} imaging of Local Group galaxies, where extended, asymmetric emission extends across very large scales.

Third, we made several choices in constructing the imaging algorithm. We chose the signal to noise threshold for the single- and multi-scale clean, as well as various gridding parameters, the set of scales for multi-scale clean, and details of masking. In principle, the PHANGS--ALMA pipeline can be used to conduct a full regression analysis, exploring the uncertainty associated with changing each parameter within a reasonable range. In practice, because it takes roughly a full day for a server with 24 CPUs and 256 GB of memory to process a typical target, we are only able to carry out a limited number of these tests. In Section~\ref{sec:endtoend}, we describe how we run two targets at multiple signal-to-noise levels through complete end-to-end tests of the pipeline. In Appendix~\ref{sec:ssc}, we carry out a similar test to investigate approaches to short spacing correction. These tests are already helpful, but due to practical considerations we have delayed a comprehensive assessment of the uncertainties associated with the choice of imaging parameters to the future.

%% file: totalpower.tex
\section{Calibration and Imaging of Total Power Data}
\label{sec:totalpower}

We process total power data in parallel with the interferometer data using a separate pipeline. For this, we use the modified version of the ALMA total power pipeline presented by \citet{HERRERA20}. We give an overview of the procedure here and refer to \citet{HERRERA20} and the publicly available scripts for more details. We also highlight one specific issue important to the PHANGS-ALMA total power data, the contamination of a subset of our data by a telluric ozone line at $229.575$~GHz. 

This total power pipeline employs a combination of the \texttt{CASA}, GILDAS, and R software packages. Unless otherwise noted, we carry out these steps in \texttt{CASA} version $4.7.2$.

\subsection{Calibration} 

We import the single dish data from the observatory-provided ASDM format to the ``measurement set'' format. Then, we split the data by antenna and write them into the ASAP data format as ``scantables.'' Next, we compute and apply the ``chopper wheel''-based temperature scale using \texttt{CASA}'s \texttt{sdcal2} task. This task calculates the temperature scale from the hot and cold loads plus sky observations \citep{PENZIAS1973}. We use this same task, \texttt{sdcal2}, to subtract the ``OFF'' spectrum from each on-source spectrum. These ``OFF'' spectra are obtained by integrating on empty sky near the source, so the result is a set of calibrated, sky-subtracted spectra.

\subsection{Baseline fitting} 

After calibration, we convert the frequency and velocity scales of the spectra from the observatory into the LSRK frame, around the systemic velocity of each galaxy. This step suffers from the same issues regarding \texttt{CASA} regridding described in Section~\ref{sec:spectral_regridding}. These currently represent an unavoidable limitation of the software.

We start by extracting a wide part of each spectrum centered on the systemic velocity of the galaxy. Then we fit first-order baselines to the line-free regions of each of these calibrated, sky-subtracted spectra. Baselines offsets and frequency-dependent baseline fluctuations are a common feature of single dish data. They reflect imperfect matches between the ``ON'' and ``OFF'' spectra and instabilities in the receiver, sky, or other parts of the signal path. The fitted baselines will also include any genuine continuum emission from the galaxy. As a result, this step removes any sensitivity to the total power data to continuum.

For simplicity, we define the ``line-free'' region to be fit by excluding a fixed velocity range from all spectra in each data set for the baseline fitting procedure. We choose the excluded velocity range to be large so that it easily encompasses all emission from the galaxy. The excluded velocity interval ranges between $200$ and $500$~km~s$^{-1}$, depending on the target. After fitting, we subtract the fitted baselines from the calibrated spectrum. This procedure is carried out independently for each ALMA execution block and for each antenna.
 
\subsection{Unit conversion and combination} 

After calibration and baseline subtraction, we first apply the antenna efficiency factor provided by the observatory as part of the delivery to convert the intensity scale from units of antenna temperature, in Kelvin, to Jy~beam$^{-1}$.

The observatory regularly measures these efficiencies by combining the total power antennas with interferometric observations by the \mbox{7-m} antennas to provide time-dependent conversion factors. This is done on a per antenna and per observation basis. This ensures a highly reliable flux calibration of the total power data. 

In Appendix~\ref{sec:tp-cal} we verify that the individual PHANGS-ALMA total power observations for the same galaxy show only 3\% rms scatter in amplitude scale from observation-to-observation. This is consistent with high quality overall calibration of the ALMA total power data. We make no additional corrections here, nor do we scale the data during combination with the interferometer data.

Last, we merge the data from all observations and antennas into a single \texttt{CASA} measurement set using the \texttt{CASA} task {\tt concat}.

\subsection{Imaging}

We grid the calibrated, sky-subtracted, baseline-corrected spectra into a data cube. To do this, we use \texttt{CASA}'s {\tt sdimaging} task, which convolves the irregularly sampled spectra onto a regular grid \citep[e.g.,][]{MANGUM07}. For the \cotwo\ data, this convolution uses a spheroidal gridding kernel with a support diameter of $12$ pixels and a pixel size of~${\sim}2.8\arcsec$.

\subsection{Inspection and quality assurance}

We perform basic inspection by examining the integrated spectra for each scan and each antenna. Occasionally individual scans or antennas reveal isolated artifacts and are flagged. Less than 1\% of our total data were removed in this manner. We also check the ``line-free'' region from which to calculate the baseline fit and adjust it if it overlaps with any galaxy emission. If any flagging or baseline region selection changes, we re-run the entire pipeline for the target.

After gridding, we also visually inspect the cubes. Except for the telluric contamination discussed below, they showed no signs of residual pathological spectra or artifacts.

\subsection{Telluric ozone contamination of CO(2--1) data} 

The \cotwo\ total power data for six PHANGS-ALMA targets is contaminated by a spurious line feature of ${\sim}50$~km~s$^{-1}$ width that peaks near $V_\text{LSRK}  \approx 1250$~km~s$^{-1}$. We ascribe the observed contamination to a relatively weak telluric ozone line at $229.575$~GHz rest frequency (i.e., offset by ${+}1253$~km~s$^{-1}$ from the rest frequency of the \cotwo\ line).

In our original total power observations, the OFF position was fixed in the equatorial reference frame, and the contamination affected even more galaxies. In these cases, the feature typically appeared either positive or negative throughout each entire image. Most of the affected targets were then reobserved, this time using an OFF position at the same elevation as the target, i.e., using a fixed offset in azimuth rather than a fixed offset in right ascension and declination.

These reobservations improved the situation, reducing the strength of the feature or even suppressing it entirely. In cases where the feature persisted, the reobservations tended to shift the nature of the contamination. Rather than having a fixed sign across the whole data set, in sets observed with a fixed-elevation OFF, the contamination shifted from positive to negative on opposite sides of the target.

This behavior can be naturally expected from the calibration procedure.\footnote{For the relevant information for \texttt{CASA} $4.7.2$, see equation~8.1 in section~8.5.2 of \newline
\href{https://casa.nrao.edu/docs/cookbook/index.html}{https://casa.nrao.edu/docs/cookbook/index.html}.} The ON-OFF subtraction used to remove atmospheric emission from the source will leave a remnant contribution proportional to the difference in airmass between the ON and the OFF spectra. To first order, this difference will be proportional to the offset in elevation. This also explains why the interferometric data are not affected by the contamination, beyond a potential mild increase in noise at these frequencies, as they do not reference to a displaced OFF position.

The subsequent baseline subtraction typically uses a first-order polynomial fit. This fit can remove any residual continuum emission, which will vary smoothly and slowly as a function of frequency. However, the baseline fit cannot remove a narrow line feature like the ozone line. The situation becomes even worse when the ozone line overlaps the velocity range covered by the galaxy. Then the line emission from the sky and the source can become confused.

We tested this scenario by looking for the primary direction along which the ozone feature varied. We found that, as expected, the strength of the feature tended to vary almost linearly along a direction close to the elevation axis at the time of the observations.

We measured the gradient of that linear trend and found that the peak of the ozone feature typically varied by ${\sim}0.02{-}0.03$~mK~arcsec$^{-1}$, with the calculation done in ${\sim}10$~km~s$^{-1}$ channels. This measured gradient is consistent with an order-of-magnitude estimate using the {\tt ATM} atmospheric model \citep{PARDO01} distributed along with the GILDAS software. The exact value of the gradient appears to depend on elevation and atmospheric conditions. In the most extreme case, it reached four times this typical value.

To the best of our knowledge, contamination by the $229.575$~GHz ozone line has not been reported in previous extragalactic, single dish \cotwo\ surveys, even large mapping surveys covering comparable area to PHANGS-ALMA \citep[e.g., the IRAM \mbox{30-m} HERACLES survey,][]{LEROY09}. Our best estimate is that this contamination simply reflects the much better sensitivity in the PHANGS-ALMA data compared to previous mapping surveys. The rms noise of our total power maps is typically $2.5{-}3.0$~mK per $2.5$ km~s$^{-1}$ compared to ${\sim}25$~mK per $5.2$~km~s$^{-1}$ channel in the HERACLES maps.

Results on the telluric contamination have been reported back to the ALMA observatory in a memo by A.~Usero et~al. This memo recommends observing strategies that can mitigate the effect of the ozone line. In general, the contamination is stronger at lower elevation and the linear trend becomes significantly steeper below ${\sim}45^\circ$. 

\subsection{Strategy for fitting and removing telluric ozone contamination}

In six galaxies the telluric ozone line overlaps the \cotwo\ line velocity range and re-observations using a fixed-elevation OFF did not solve the problem. We thus developed a custom procedure to remove the ozone contamination. This procedure, which is currently implemented in the \texttt{R} programming language \citep{RMANUAL}, works as follows:

\begin{enumerate}
\item The procedure operates at the level of individual ``execution blocks'' (EBs), i.e., individual observing sessions. Because these sessions are relatively short, ${\sim}1{-}1.5$~h, we approximate the transformation from the azimuth-elevation frame to the celestial frame as constant and work with the post-gridding cube data for individual EBs. We also assume that atmospheric conditions are stable over this short time so the signature of the telluric contamination is constant.

We model the strength of the ozone line at a sky position $\mathbf{x}$ and velocity $v$ as
\begin{equation}
\label{eq-o3}
T_\mathrm{O_3}(\mathbf{x},v) = L(\mathbf{x})\times P(v)~,
\end{equation}
where $L$ is a linear gradient that models the amplitude as a function of elevation, and $P$ is the spectral profile of the line. Because we work in the LSRK velocity frame, the peak velocity of the ozone profile $P$ will be typically offset from ${+}1253$~km~s$^{-1}$ by the difference between the topocentric velocity and LSRK rest frame. We calculate the expected offset using the {\tt ASTRO} program of the GILDAS software. The velocity difference between the frames varies across EBs by as much as $30$~km~s$^{-1}$. 

\item For each EB, we generate a {\em contaminated} \cotwo\ cube with our standard total power pipeline. The only modification is to exclude an additional velocity range around the ozone line during baseline fitting.

\item To determine $L$ in Eq.~\eqref{eq-o3}, we build a map of the mean intensity in the cube within a ${\pm}25$~km~s$^{-1}$ range centered on the expected peak velocity of $P$ for the ozone line. Then we manually define a two-dimensional mask that encompasses the real CO emission from the galaxy in this velocity range. The signal outside this spatial mask will mostly represent telluric emission/absorption. We fit the unmasked position data as a linear function of right ascension and declination using a noise-weighted least-squares method to generate an estimate of~$L$.

\item Our model (Equation~\ref{eq-o3}) assumes that the spectral profile of the line, $P$, does not vary across the map. To determine $P(v)$ in each velocity channel, $v$, we first build a mask that encompasses all real CO emission, using a dilated mask technique (similar to the masks discussed in Section~\ref{sec:masking_and_clean}). Then we calculate the average $(T/L)$ outside the mask, weighting it by $(L/\sigma)^2$. Here $T$ and $\sigma$ are the measured intensity and the rms noise at the corresponding position and velocity channel, respectively. We smooth the initial estimate of $P$ with a 5-channel boxcar kernel to reduce uncertainty due to noise. We also set $P(v)=0$ beyond $\pm50$~km~s$^{-1}$ from the expected peak velocity of the ozone line. This limit ensures the correction does not create any artifacts at velocities where contamination would be negligible even at our high sensitivity.

\item We build a contamination cube from $L\times P$ and subtract it from the original CO cube to get its \textit{contamination-corrected} version. 

\item Finally, we co-add the \textit{contamination-corrected} cubes from all EBs, weighted by their average rms noise to produce the final total power cube for that galaxy.

\end{enumerate}

\begin{figure*}
\includegraphics[width=\hsize]{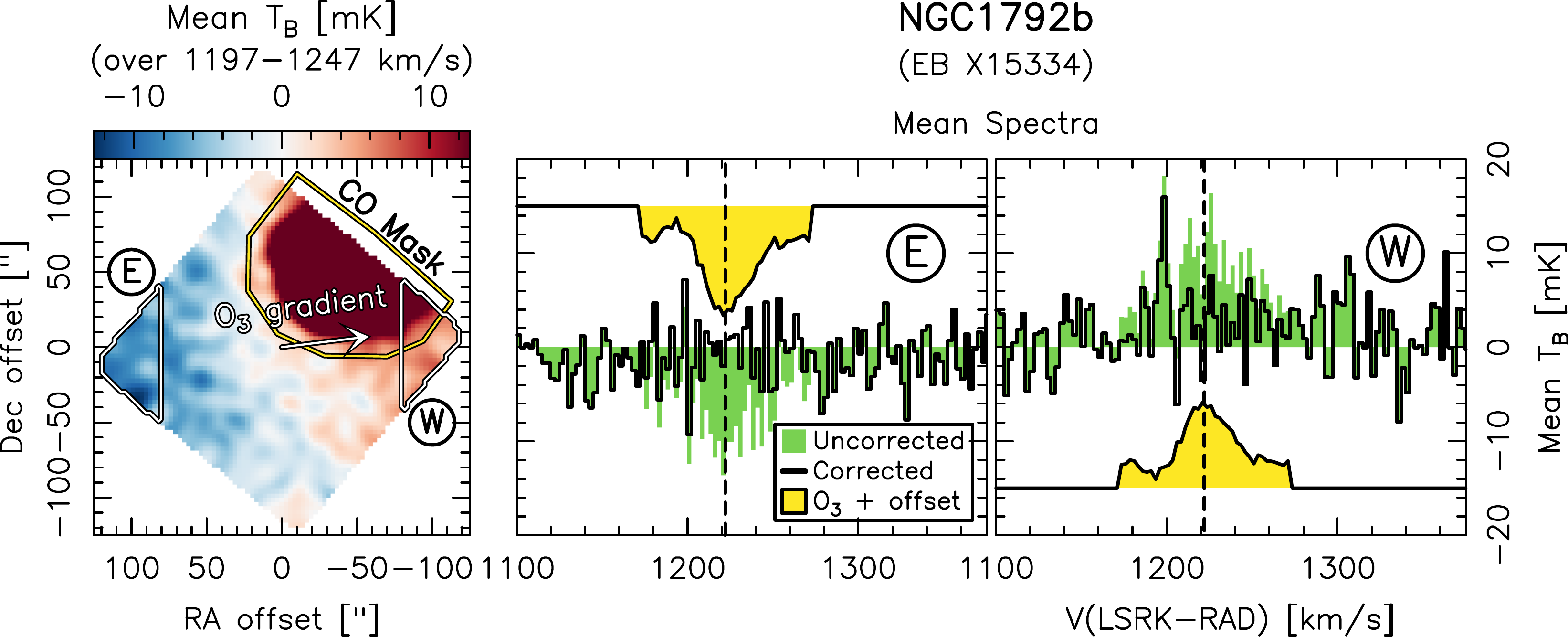}
\caption{\textbf{An example of telluric ozone contamination in the PHANGS-ALMA total power observations and the effects of the correction algorithm.} The left-hand panel shows a mean brightness temperature map of \cotwo\ from the southern half of NGC~1792. This target is one of six still affected by telluric ozone contamination after reobservations using a fixed-elevation OFF position. The map here was derived from a single execution block, i.e., a single observing session. The mean brightness temperature is calculated within $\pm25$~km~s$^{-1}$ from the Doppler-shifted velocity of the ozone feature expected during that session ($V_\text{LSRK}={+}1222$~km~s$^{-1}$). The arrow indicates the orientation of the contamination gradient derived from our fit (see text for details). The yellow polygonal line is the manual mask used to exclude the real CO emission of the target (the color-saturated red patch in the map) from the fit. We highlight the results of our algorithm in two regions at opposite ends of the gradient, labeled E and W. The middle and right-hand panels show \cotwo\ spectra averaged over the E and W regions: before correcting for any contamination (green, filled), after applying our method (black), and the difference between them (yellow, filled, offset vertically for the sake of clarity). The latter is the average spectrum of the ozone signal in either region. As expected, the ozone signal peaks at a velocity of ${+}1222$~km~s$^{-1}$, indicated with a vertical dashed line. It shifts in sign between E and W as the airmass difference between the source and the OFF shifts sign.
\label{fig:ozone}
}
\end{figure*}

As illustrated in Figure~\ref{fig:ozone}, this procedure effectively removed signatures of ozone contamination in the six remaining affected PHANGS-ALMA targets. This approach should be useful for sensitive on-the-fly observations of external galaxies with source velocities in the range ${\sim}1100{-}1400$~km~s$^{-1}$. The procedure could also be generalized to deal with any telluric contamination of on-the-fly mapping maps.

%% file: postprocess.tex
\begin{figure*}[ht!]
\gridline{
\fig{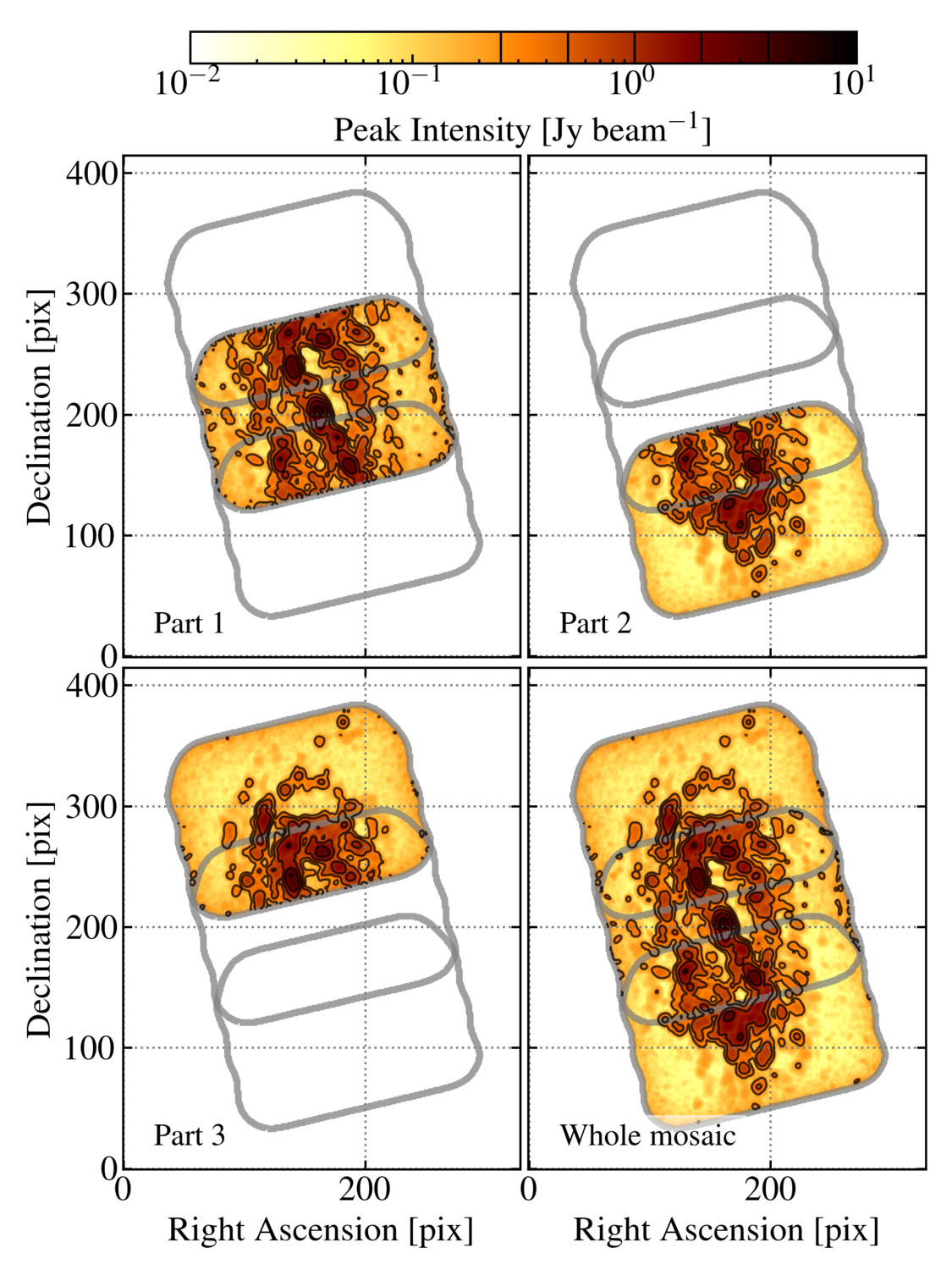}{0.7\textwidth}{}
}
\vspace{-24pt} % reduce excessive gaps between figure rows
\caption{
\textbf{Example of linear mosaicking.} Example of linear mosaicking for PHANGS-ALMA \cotwo\ data from the ACA \mbox{7-m} antennas for NGC~2903. The galaxy was independently observed using three maximum-sized 150-field mosaics (for the \mbox{12-m} array), labeled parts 1, 2, and~3. Each part is imaged separately. Then the individual parts are convolved to a common beam and aligned on a shared astrometric grid. The first three panels show peak intensity images of these three aligned, beam-matched parts. Gray contours show the footprint of the individual parts. The three cubes are then combined, weighting by the local combined primary beam response and the overall noise level in the cube (Equation~\eqref{eq:mosaic}). The final panel shows the peak intensity map from the resulting combined mosaic. We apply a similar procedure to the single dish data and then combine the interferometric and total power data after this mosaicking step.
\label{fig:linmos}}
\end{figure*}

\section{Cube Post-Processing}
\label{sec:postprocess}

After imaging and deconvolution, we process the interferometric cubes into a final ``science-ready'' form. This has six main steps: primary beam correction, convolution to a round synthesized beam, linear mosaicking to combine multi-part galaxies, feathering to combine interferometric and total power data, downsampling of cubes, and conversion to a Kelvin intensity scale.

\subsection{Primary beam correction} 

First, we created a version of each cube that was corrected by the combined primary beam response of all mosaic pointings,~$B$, in each channel. To do this, we use the \texttt{CASA} task \texttt{impbcor}, which divides the image cube by the combined primary beam response map output by \texttt{tclean}. A byproduct of this correction is to increase the noise near the mosaic edges, where $B$ is low.

\subsection{Convolution to a round beam} 

The imaging yields elliptical synthesized beams, e.g., as shown in Table~\ref{tab:phangsimaging} and Figure~\ref{fig:beamstats} for PHANGS-ALMA. The beam does not tend to align in any useful way with galactic structure and only makes analysis more complex. Therefore, as part of post-processing, we used the \texttt{CASA} task \texttt{imsmooth} to convolve each cube to a final round, Gaussian-shaped beam.

In addition to increasing the minor axis of the beam, \texttt{imsmooth} required us to slightly pad the major axis of the beam to find a viable convolution kernel. This increased the major axis beam size by a small amount, $\lesssim 10\%$. In principle, this resolution loss can be avoided by constructing an appropriate kernel in the Fourier domain. This kernel would have infinite width (in Fourier space) along the major axis in order to avoid convolution in that direction. \texttt{CASA} currently lacks this capability, so we instead pad the major axis. This allows a kernel to be constructed in the image domain and transformed into the Fourier domain.

Our final images combine deconvolved emission (i.e., the sum of all clean components) and residuals, which are mostly noise. The deconvolved emission has been convolved with a Gaussian ``clean'' beam calculated from fitting the core synthesized beam. The residual emission, including any actual emission too faint to be deconvolved, still incorporates the dirty beam. Note that by convolving the cube to have a round beam, we also change the ``dirty'' beam associated with this residual emission. The synthesized dirty beam can have a complex shape, but the core is similar to the elliptical Gaussian restoring beam. Therefore the convolution to a round beam will also ``round'' the core of the dirty beam, thus keeping the shape of the dirty beam and restoring beam approximately similar and making the dirty beam more symmetric.

At this point, the images have a round beam, units of Jy~beam$^{-1}$, and represent deconvolved images of the sky no longer tapered by the combined primary beam response.

\subsection{Stitching multi-part galaxies via linear mosaicking}  

Due to the combination of ALMA's fields-per-scheduling block limitation and its powerful mosaicking capabilities, many science projects now observe multiple large mosaics to be combined into a single large image during processing. In practice, for PHANGS-ALMA many targets were observed using two or three separate mosaics. In one case, NGC~0253, five distinct maximum sized mosaics were used. Figure~\ref{fig:linmos} shows an example of a three-part observation targeting NGC~2903.

As mentioned previously, these mosaics tend to be observed at different times with different $u{-}v$ coverage and weather conditions and so have different synthesized beams. \texttt{CASA} does not currently track positional variations in the synthesized beam, making it challenging to image all of these data simultaneously. Instead, we stitch these images together in the image domain via linear mosaicking.

To do this, we first identify a common spatial resolution for all parts of the mosaic. This common resolution is slightly larger than the coarsest resolution for any single part. As with the convolution to a round beam, this process used the \texttt{CASA} task \texttt{imsmooth} and required a small amount of ``padding,'' i.e., increasing the size of the beam to allow the routine to successfully carry out the convolution. This is another step where we lose a modest amount of resolution, $< 10\%$. This beam matching is not an issue for the single-dish data because the beam shape is constant.

After this convolution, we constructed a new astrometric grid that covered all of the individual mosaic parts. The individual parts already shared the same spectral
axis thanks to the pre-processing of the visibility data (Section~\ref{sec:staging}). We then used the \texttt{CASA} task \texttt{imregrid} to align all of the parts of the mosaic to this common astrometric grid. We also aligned the combined primary beam coverage cubes onto the same grid. Last, we aligned the single dish total power data onto this grid for use in feathering (Section~\ref{sec:feather}).

After this, we combined all mosaic parts into a single image. To do this, we weight each cube by the local value of the primary beam response squared times the inverse of the typical noise in that cube. That is, for each voxel we calculate
\begin{equation}
\label{eq:mosaic}
\left< I \right> = \frac{\sum_{i = 0}^{n} \left( \mathit{B}_i^2 \sigma_i^{-2} \right) \times I_i}{\sum_{i = 0}^{n} \left( \mathit{B}_i^2 \sigma_i^{-2} \right)}~.
\end{equation}
\noindent Here $\left< I \right>$ is the mean intensity, which is placed in the new cube, and the sum over $i$ refers to a sum over all mosaic parts that contribute at that voxel. The combined primary beam response squared, $\mathit{B}_i^2$, should track regional variations of the noise within the primary-beam corrected cube. The quantity $\sigma_i^2$ is the overall noise variance in the cube, so that the product $\sigma_i/\mathit{B}_i$ corresponds to the local rms noise. Weighting the intensities by the inverse of their noise variance should produce the lowest possible noise in the output cube.

We did experiment with joint imaging of the mosaic parts and found that variations in the $u{-}v$ coverage across the mosaic sometimes lead to divergence in the deconvolution. Stitching via linear mosaicking proved to be a much more stable option.

\begin{figure*}[ht!]
\gridline{
\fig{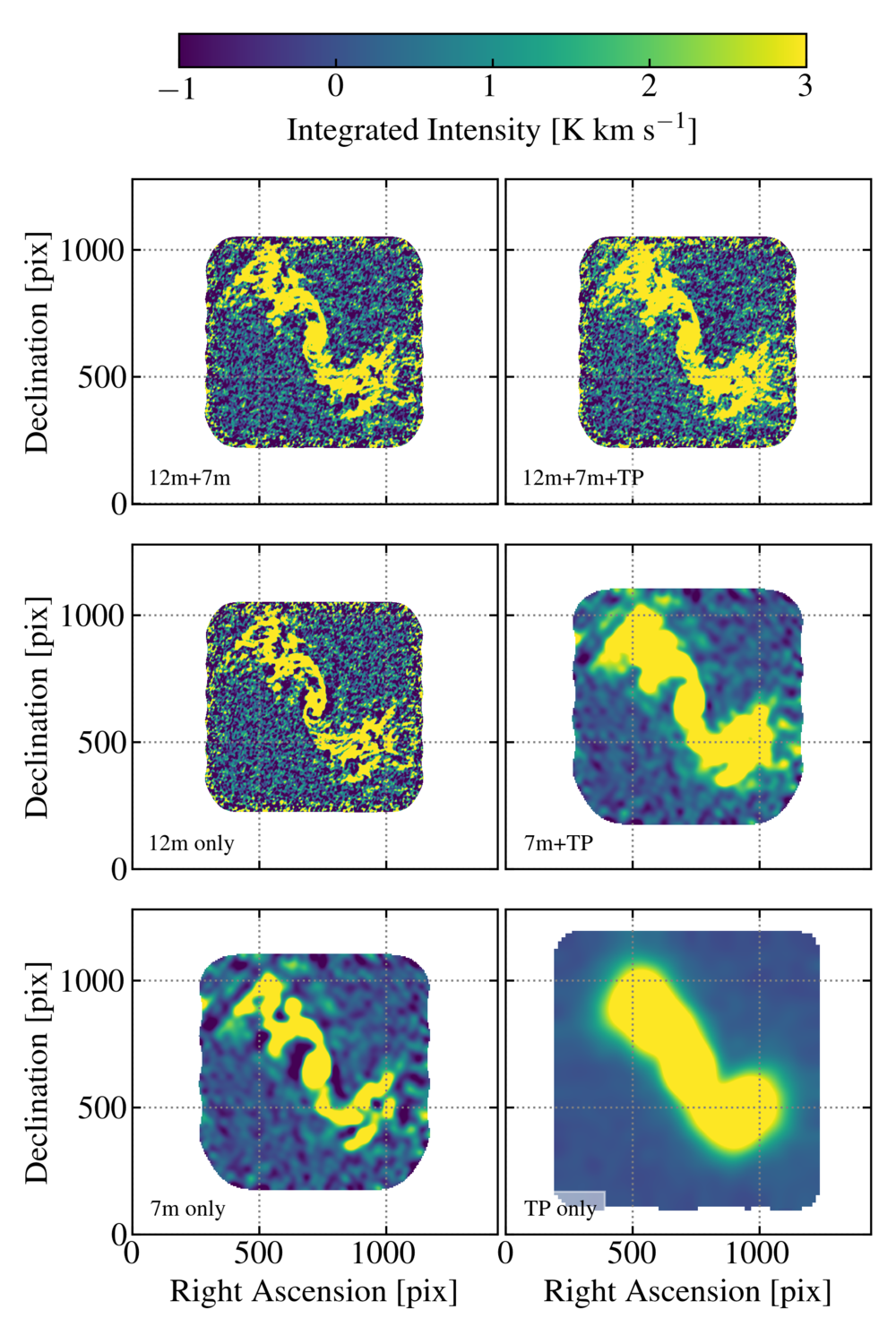}{0.7\textwidth}{}
}
\vspace{-24pt} % reduce excessive gaps between figure rows
\caption{
\textbf{Comparison of images from all array combinations for one galaxy.} Integrated emission from 20-channel-thick slabs in the \cotwo\ data cubes for the same target shown in Figures~\ref{fig:deconv_7m} and~\ref{fig:deconv_12m+7m}. Each panel shows the image for a different array combination: (\textit{top left}) the jointly imaged \mbox{12-m}+\mbox{7-m} data with no total power data, (\textit{top right}) the jointly imaged \mbox{12-m}+\mbox{7-m} data feathered with the total power data, (\textit{middle left}) the \mbox{12-m} data only, (\textit{middle right}) the \mbox{7-m} data feathered with the total power data, (\textit{bottom left}) the \mbox{7-m} data only, and (\textit{bottom right}) the total power data only. All panels show the same high stretch, in units of K~km~s$^{-1}$, and use the same astrometric grid. The image demonstrates the artifacts, especially negative ``bowls,'' due to missing short-spacing data and imperfect deconvolution in \mbox{7-m}-only or \mbox{12-m}-only imaging. Contrasting the images shows how the inclusion of \mbox{7-m} data improves the recovery of extended emission compared to the \mbox{12-m}-only image. In Appendix~\ref{sec:arrays} and Section~\ref{sec:endtoend}, we show and discuss that the combined \mbox{12-m}+\mbox{7-m} deconvolution recovers more flux and yields better overall results than the \mbox{7-m}-only case. Including the total power data almost entirely removes the negative artifacts and adds a faint, extended component in both the \mbox{12-m}+\mbox{7-m} and \mbox{7-m} cases. 
\label{fig:allarrays}}
\end{figure*}

\begin{figure*}[t!]
\begin{center}
\includegraphics[width=0.45\textwidth]{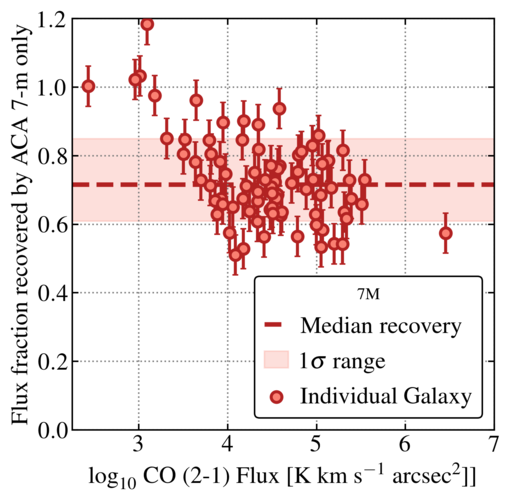}
\includegraphics[width=0.45\textwidth]{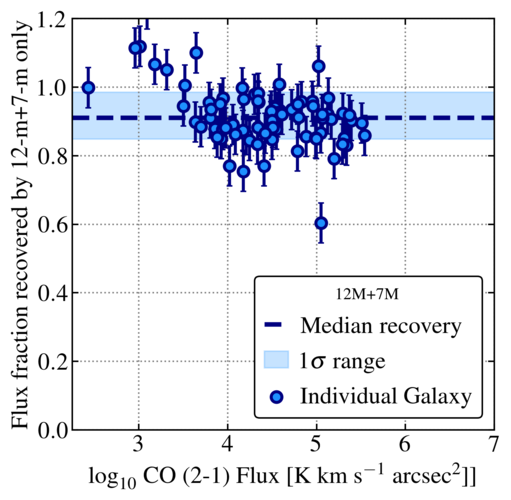}
\end{center}
\caption{\textbf{Fraction of PHANGS-ALMA CO(2--1) flux recovered by the interferometer before short-spacing correction.} Fraction of the total flux recovered by the interferometer only for the PHANGS-ALMA \cotwo\ data, for the \textit{left} ACA \mbox{7-m} array only and \textit{right} the combined \mbox{12-m}+\mbox{7-m} arrays. The $y$-axis shows the ratio between the flux in the interferometer-only integrated intensity map and the integrated intensity map constructed from the feathered data. Specifically, we calculate these fluxes from the integrated intensity maps constructed using the ``broad'' masks described in Section~\ref{sec:products}. The error bars combine the statistical uncertainty from noise with typical uncertainties for ALMA's interferometric and total power calibration, added in quadrature. On average, the ACA \mbox{7-m}-only data recover $72\%$ of the \cotwo\ flux seen by the total power antennas (horizontal dashed line), with $16{-}84^{\rm th}$ percentile range of $61\%{-}85\%$ (shaded region) across the sample. The combined \mbox{12-m}+\mbox{7-m} imaging does much better, recovering a median $91\%$ of the total flux with a $16{-}84^{\rm th}$ percentile range of $85\%{-}99\%$. The high values in low brightness galaxies reflect low signal-to-noise cases where the imaging fails to properly deconvolve the galaxy. We discuss the better performance of the \mbox{12-m}+\mbox{7-m} imaging compared to the \mbox{7-m}-only in Appendix~\ref{sec:arrays}.
\label{fig:recovery}}
\end{figure*}

\subsection{Combination of total power and interferometric data via feathering} \label{sec:feather}

We combined the cleaned \mbox{7-m} and \mbox{12-m}+\mbox{7-m} cubes with the single dish cubes using \texttt{CASA}'s \texttt{feather} task. \texttt{Feather} combines the interferometric and total power cubes in the Fourier domain, using the total power data at low angular frequencies (i.e., to fill in short- and zero-spacings) and the interferometer data for information at high angular frequencies. We used this task with the default options as we already reprojected the different input images to the same grid and we ensured that they all were converted to Jy~beam$^{-1}$ units.

Figures~\ref{fig:allarrays} and ~\ref{fig:recovery} show the impact of the short-spacing correction for the PHANGS--ALMA \cotwo\ data. Figure~\ref{fig:allarrays} illustrates the data before and after feathering and for different arrays. We show the integrated intensity obtained from collapsing the same 20 channel-thick slab of NGC~4303 seen in Figures~\ref{fig:deconv_7m} and~\ref{fig:deconv_12m+7m}. All six panels show the same high stretch. Also note that here we include images made using only the \mbox{12-m} array and total power data. The \mbox{12-m} only and \mbox{7-m} only images illustrate negative artifacts, or ``bowling,'' around bright emission due to missing short-spacing data. The images that include total power data show how the single dish data fills in the bowls and also adds an extended, faint component to the image \citep[see][for a much more detailed demonstration in M51]{PETY13}.

Figure~\ref{fig:recovery} shows results separately for the \mbox{7-m}-only data and the combined \mbox{12-m}+\mbox{7-m} data. On average, the \mbox{7-m}-only cubes recover ${\sim}70\%$ of the emission found in the final, short-spacing corrected cubes. The \mbox{12-m}+\mbox{7-m} cubes do much better, recovering ${\sim}90\%$ of the flux seen in the total power data on average. There is a large scatter in the recovery fraction of the ACA \mbox{7-m} data, with the $16{-}84^{\rm th}$ range spanning from $61\%$ to $85\%$ recovery. For high brightness targets, the \mbox{12-m}+\mbox{7-m} data cluster in the range $85{-}99\%$ recovery. The difference between the two arrays suggests that a large part of the ``missing flux'' in the \mbox{7-m}-only case reflects shortcomings of the deconvolution, not only spatial filtering. We discuss the point more in Appendix~\ref{sec:arrays}.

\medskip

\noindent \textbf{Other approaches:} We experimented with other approaches, including \texttt{tp2vis} \citep{KODA19} and the use of either the total power data or previous rounds of high SNR imaging attempts as a model (for more details see Appendix~\ref{sec:ssc}). To evaluate the competing methods, we created a set of images with known flux based on collapsed versions of our CO data cubes. Then, we simulated interferometric observations of the known source using \texttt{simalma}. We simulated total power observations by convolving the true image to the resolution of the single dish data. Then we applied each method of reconstruction: feathering, \texttt{tp2vis}, and seeding the deconvolution using an input model.

The calculations in Appendix~\ref{sec:ssc} yield \textit{much} more spatial filtering than our real data. Based on comparison to the more realistic simulations carried out in Section~\ref{sec:endtoend}, this appears to reflect that our actual imaging operates in individual velocity channels. As emphasized in that section, the calculations in Appendix~\ref{sec:ssc} should be taken as experiments that consider a ``worst case'' scenario for spatial filtering in nearby galaxies.

These tests showed that \texttt{feather} recovered a known input image with about the same fidelity and flux accuracy as the other approaches. Typically, all of the methods implied $10{-}15\%$ inaccuracies in overall recovery of the input image, but also treat an extreme case. Still this part of the calculation certainly represents one of the dominant uncertainties in high signal-to-noise \mbox{7-m}-only observations. Short-spacing correction is an area where we expect research and development to improve our data products in the coming years.

\medskip

\noindent \textbf{Apodization:} In the current version of the pipeline we do \textit{not} apodize (``taper'') the single dish image before feathering. We feather the best-estimate image of intensity from both the single dish and interferometric data. This approach effectively treats the total power information as zero outside the field of view of the interferometer.

In theory, apodizing the single dish and feathering before primary beam correction may seem preferable, and some \texttt{CASA} and ALMA documentation recommends this approach because it carefully matches the field of view of the two data sets and avoids any sharp edges. We conducted tests using simulated sources and emission near the edge of the field of view and found that apodizing before feathering led to distortions at the edges of the output. The leading hypothesis for this effect is that apodization interacts with the primary beam of the single dish telescope to distort the shape of bright sources in the mosaic during feathering (C.~D.\ Wilson et al.\ private communication).

Given that there is some uncertainty regarding the treatment of edges in \texttt{feather}, we carry out linear mosaicking of both the total power and interferometric data \textit{before} feathering. In PHANGS-ALMA, most cases with bright emission near the edge of the observed field of view are part of a larger, multi-part mosaic. By stitching these parts together before feathering, we minimize the impact of our treatment of the map edges.

\subsection{Downsampling and trimming of data cubes} 

After imaging and convolution to a round beam, our cubes usually have $\gtrsim 7$ pixels across the FWHM of the synthesized beam. The imaging and processing also left the cubes with a large amount of empty space surrounding the data. Both the oversampling and the padding are useful for imaging but unnecessary for scientific analysis. They also substantially inflate the data volume of the cubes. Therefore, at this stage we trim and downsample the cubes to lower their volume without reducing information content.

For any cube with pixel scale fine enough that $> 6$ pixels fit across the (now round) beam FWHM, we rebinned the cube. This rebinning increased the pixel size by a linear factor of two, which corresponded to a factor of four decrease in the number of pixels in the cube. After this rebinning, the pixels still critically sampled the beam. 

Finally, we extracted only the part of each cube that contained data, dropping any extra padding in right ascension and/or declination.

\subsection{Conversion to Kelvin intensity scale} 
Finally, we convert our cubes from units of Jy~beam$^{-1}$ to brightness temperature, $T_{\rm b}$, measured in Kelvin. This removes the beam from the units and recasts the maps onto a straightforward intensity scale, which is ideal for studying complex, resolved \cotwo\ emission.

To convert, we use the current synthesized beam size and the observed frequency in the central channel of the cube to set a constant scaling factor.  Formally, this conversion varies across our bandpass by a factor of $2\Delta \nu / \nu$, which is ${\sim}0.006$ for the maximum $1000$~km~s$^{-1}$ bandwidth of the PHANGS-ALMA \cotwo\ cubes, but we do not include this variation. We recorded the Jansky-to-Kelvin conversion in the header of the final cube. For the PHANGS-ALMA \cotwo\ cubes the values have 16$^{\rm th}$ to 84$^{\rm th}$ percentile range of $0.33{-}0.47$~K~Jy$^{-1}$ for the ACA \mbox{7-m} data and $6.2{-}7.0$~K~Jy$^{-1}$ for the \mbox{12-m}+\mbox{7-m} combined data.

\subsection{Exporting to FITS}

At the end of this process, we export the trimmed, corrected line cubes (and images) to FITS format. During this step, we have ensured that the headers are correct and contain no extraneous information. At this stage, we have primary beam corrected, short-spacing corrected, round beam data cubes in units of brightness temperature.

%% file: products.tex
\section{Data Product Creation}
\label{sec:products}

\begin{deluxetable}{lc}
\tabletypesize{\small}
\tablecaption{
PHANGS-ALMA Resolution and Noise
\label{tab:productprocess1}}
\tablewidth{0pt}
\tablehead{
\colhead{Item} & 
\colhead{Description} 
}
\startdata
\hline
Resolutions\tablenotemark{a} & \\
\hline\hline
7{-}m+TP & 84 galaxies\tablenotemark{b} \\
... native angular [$''$] & 
$ 7.6 ^{+ 0.8 }_{-0.5}$ \\
... native physical\tablenotemark{c} [pc] & 
$550 ^{+140 }_{-150}$ \\
12{-}m+7{-}m+TP & 77 galaxies\tablenotemark{b} \\
... native angular [$''$] & 
$ 1.3 ^{+0.4}_{-0.2}$ \\
... native physical\tablenotemark{c} [pc] & 
$100^{+31}_{-35}$ \\
\multicolumn{2}{l}{Common resolutions (when allowed by data)} \\
... angular & native, $2''$, $7.5''$, $11''$, $15''$ \\
... physical\tablenotemark{c} [pc] & 60, 90, 120, 150, 500, 750, 1000\\
\hline\hline
\multicolumn{2}{l}{Noise in individual 2.54~km~s$^{-1}$ channels} \\
\hline\hline
7{-}m+TP & 84 galaxies\tablenotemark{b} \\
... median noise, native res. [mK] & $12^{+4.5}_{-3.5}$ \\
... median noise, 750~pc [mK] & $7.1^{+3.8}_{-3.6}$ \\
... full fractional spectral variation & 
$0.25^{+0.06}_{-0.06}$
\\
... $\pm 1\sigma$ fractional spatial variation & 
$0.80^{+0.10}_{-0.17}$ 
\\
12{-}m+7{-}m+TP & 77 galaxies\tablenotemark{b} \\
... median value, native res. [mK] & 
$85^{+40}_{-40}$
\\
... median value, 150~pc [mK] & 
$ 53^{+25}_{-20}$
\\
... full fractional spectral variation & 
$0.23^{+0.05}_{-0.09}$
\\
... $\pm 1\sigma$ fractional spatial variation & 
$1.0^{+0.2}_{-0.28}$ \\
\hline
\enddata
\tablenotetext{a}{Data are convolved to each of these resolutions whenever the native resolution is fine enough to allow this. Target resolutions are the same for configurations with and without total power. The quoted value refers to the FWHM of a Gaussian beam.}
\tablenotetext{b}{When this table was compiled, $6$ galaxies were still missing total power data due to the telluric contamination described in Section \ref{sec:totalpower}. Since then, these data have been corrected and all $90$ galaxies have total power data. The median properties of the data are essentially unchanged. In total in the public data release, $81$ galaxies have 12{-}m+7{-}m+TP data and and $90$ have 7{-}m+TP.}
\tablenotetext{c}{When convolving to a fixed physical resolution, we adopt the current best estimate of the galaxy's distance \citep[][for PHANGS-ALMA]{ANAND21}.}
\tablecomments{These numbers refer to the first public data release, internal ``version~4'', constructed with ``version 2.0'' of the PHANGS-ALMA pipeline. They refer to the products created for the \cotwo\ survey. The number of galaxies indicates the number of targets with these array combinations processed by this release. The $\pm$ values refer to the 16th and 84th percentiles of the sample distribution.}
\end{deluxetable}

We create a series of data products from the science-ready data cubes. First we convolve the cubes to a set of fixed angular and physical resolutions. Table~\ref{tab:productprocess1} lists the target resolutions and other details of the product creation process. These fixed-resolution cubes are intended to allow rigorous comparison among targets at different distances \citep[e.g.,][]{HUGHES13A,ROSOLOWSKY21}. The processing described in Section~\ref{sec:staging} already places the cubes at nearly matched velocity resolution.

For each cube, we estimate the noise at each location in the data cube. We combine this noise estimate with the data themselves to create two kinds of masks (Table \ref{tab:productprocess2}), a ``broad'' mask focusing on high completeness and a ``signal'' mask focusing on including only emission detected at high confidence.

We apply these masks and collapse the cubes along the spectral axis to produce a variety of ``moment'' maps (Table \ref{tab:derivedproducts}): integrated intensity, peak intensity, intensity-weighted mean velocity, line width, and so on. Whenever feasible, we also calculate corresponding uncertainty maps.

In the PHANGS-ALMA pipeline, these operations occur outside of \texttt{CASA} in a \texttt{python} environment. We use routines built around the \texttt{numpy}, \texttt{scipy}, \texttt{astropy}, \texttt{radio-beam}, and \texttt{spectral-cube} packages.

For continuum products, we carry out the convolution and estimate a single noise value from the signal free-region of the image. Most of this section describes processing of data cubes.

\subsection{Convolution to Fixed Resolutions}

\begin{figure*}[ht!]
\begin{center}
\includegraphics[width=0.4\textwidth]{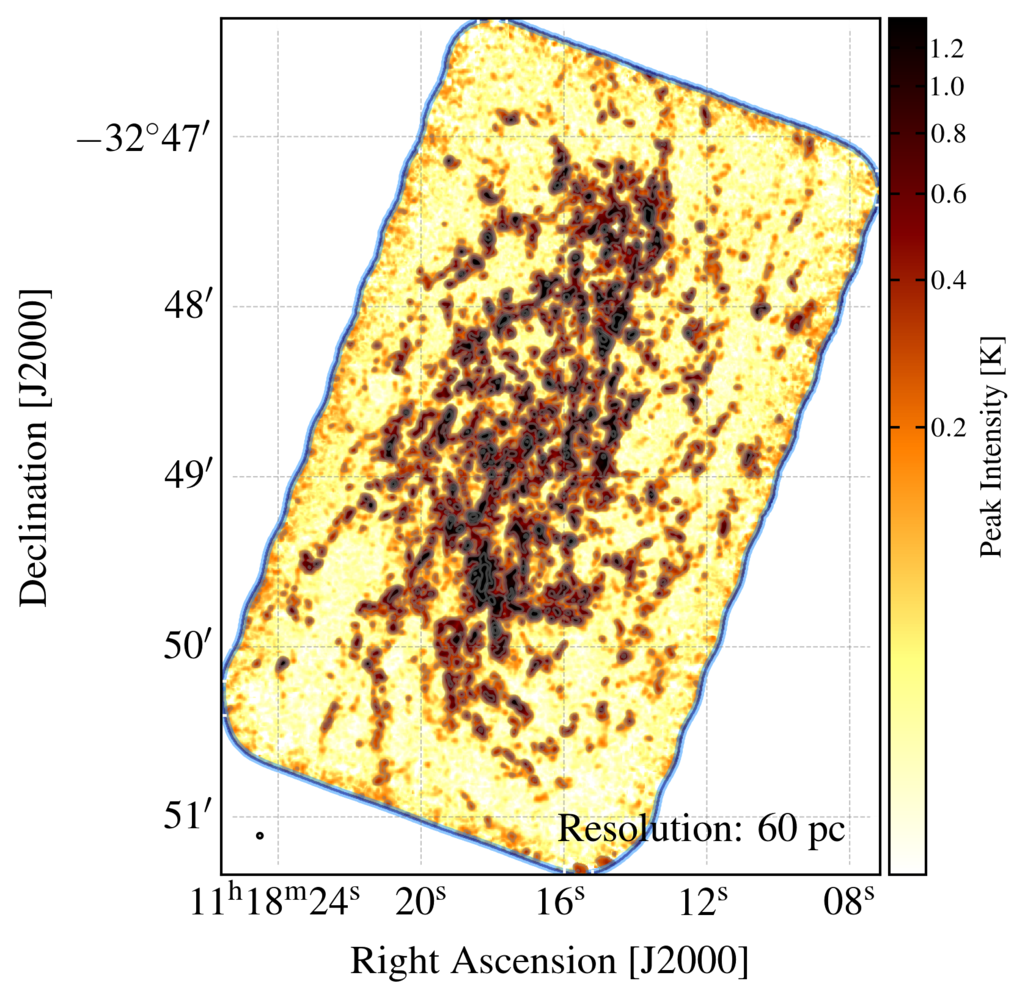}
\includegraphics[width=0.4\textwidth]{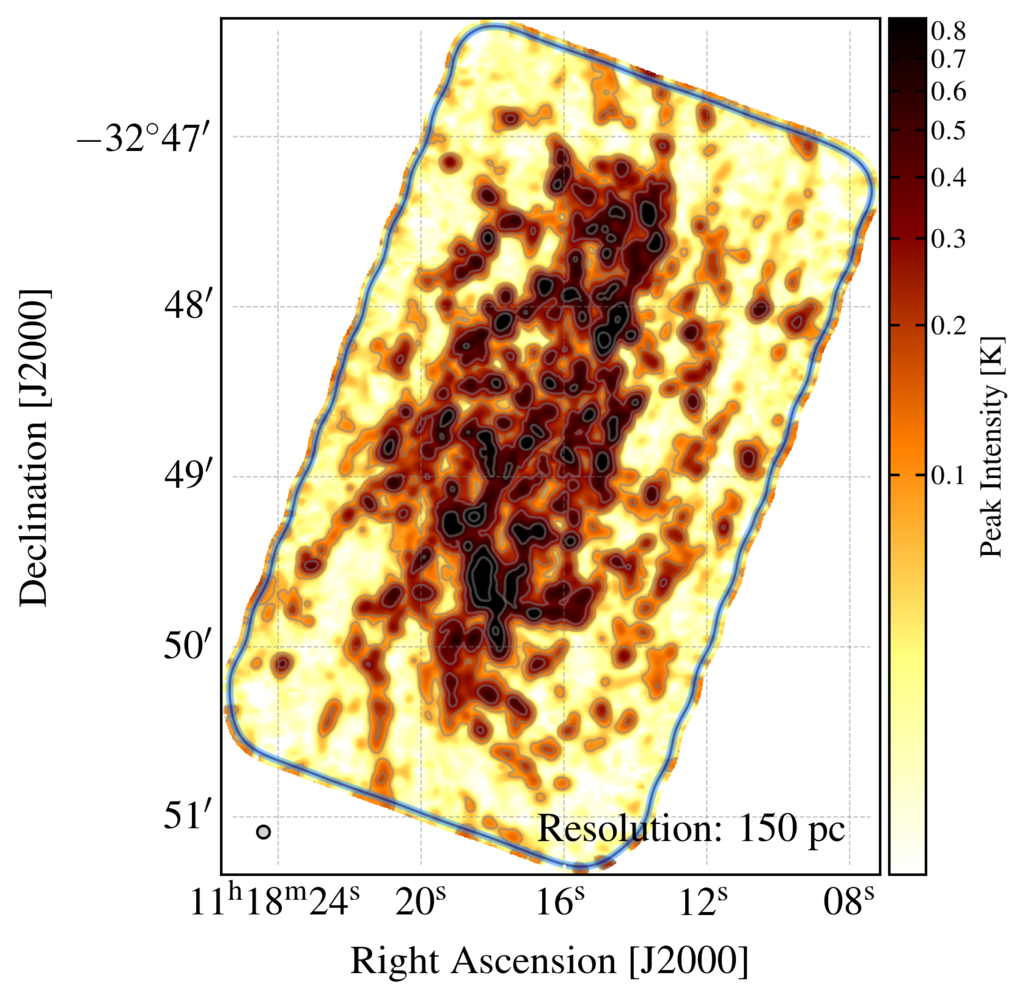}
\vspace{-24pt} % reduce excessive gaps between figure rows
\includegraphics[width=0.4\textwidth]{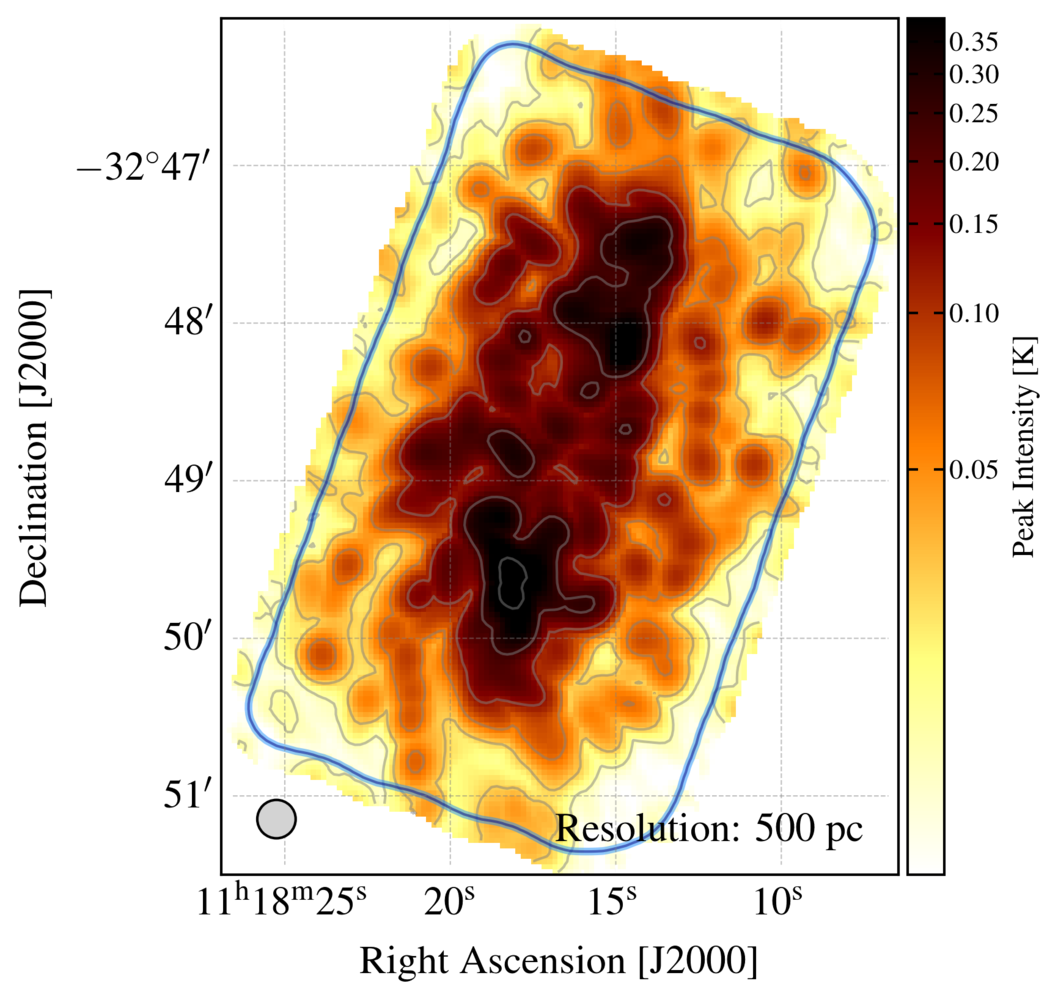}
\includegraphics[width=0.4\textwidth]{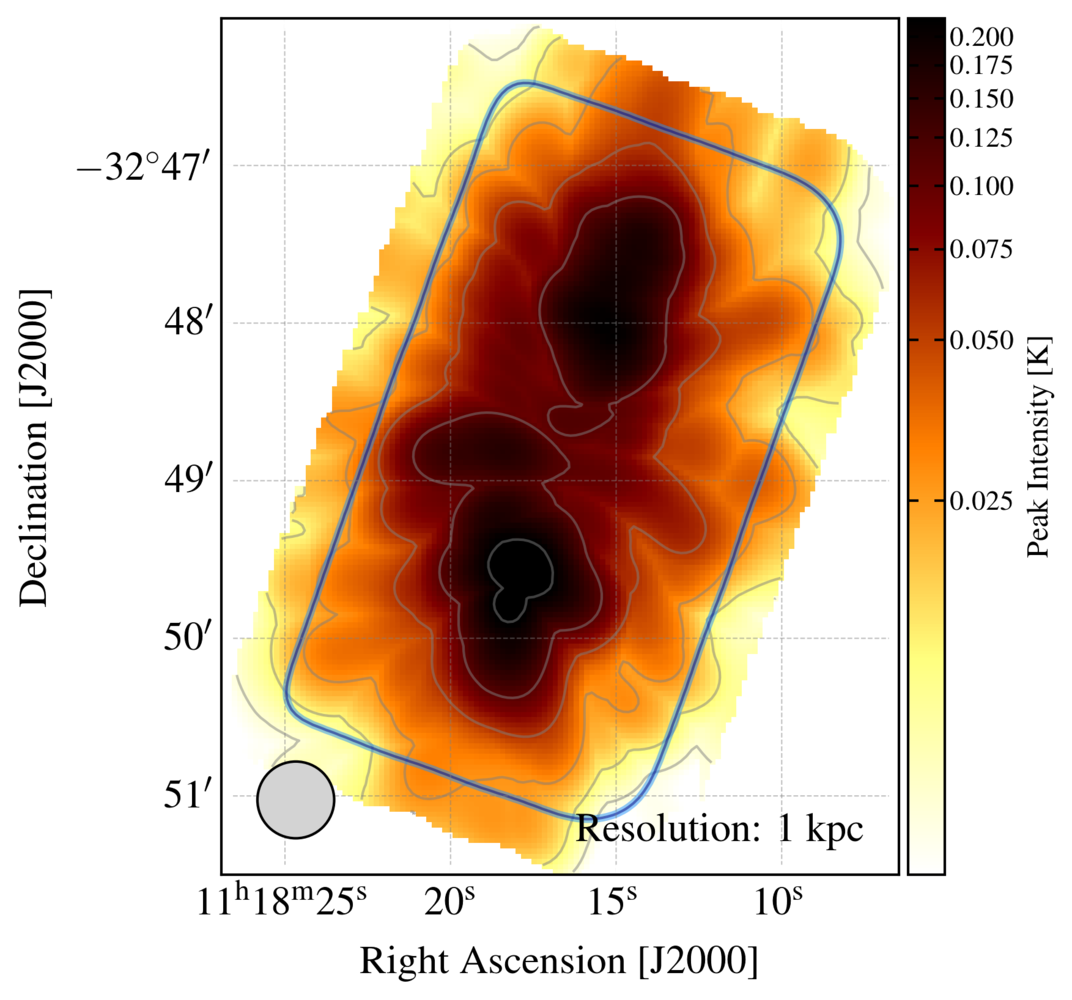}
\end{center}
\vspace{+0.2in}
\caption{
\textbf{Example of data products and coverage derived from the convolved cubes.} Peak temperature maps using a $12.5$~km~s$^{-1}$ spectral window at four resolutions for NGC~3621. Each panel shows the product derived after convolution to a different physical resolution (from top left to bottom right): 60~pc, 150~pc, 500~pc, and $1{,}000$~pc. The image is set to cover the $1{-}99$\% range of the data on an arcsinh stretch. The contours show 5, 10, 20, 40, 80, and 160 times the noise level. Circles in the lower left of each panel show the smoothed beam FWHM sizes. The images show that as the resolution degrades, the sensitivity and extent of detections increases but fine details are washed out. The blue lines show the area of 95\% coverage after the convolution. Regions outside this contour include fewer data than those inside, and so suffer from increasing edge effects as one approaches the map edge.
\label{fig:convexample}
}
\end{figure*}

\begin{figure*}[ht!]
\begin{center}
\includegraphics[width=0.45\textwidth]{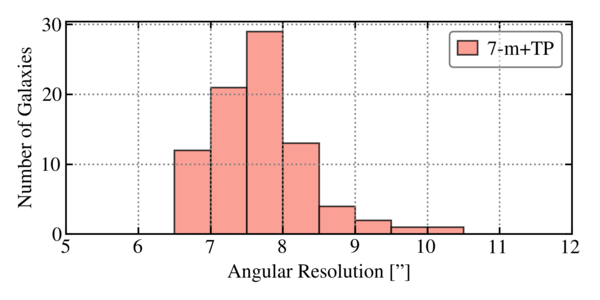}
\includegraphics[width=0.45\textwidth]{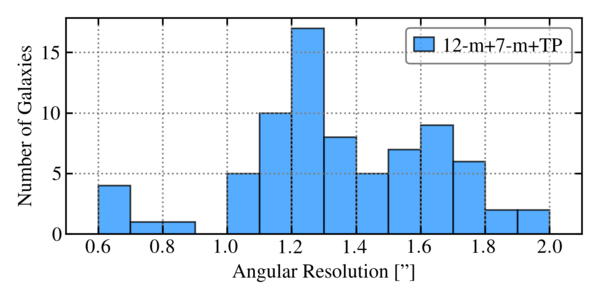}
\vspace{-18pt} % reduce excessive gaps between figure rows
\includegraphics[width=0.45\textwidth]{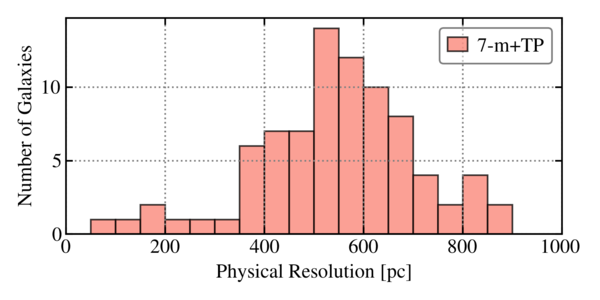}
\includegraphics[width=0.45\textwidth]{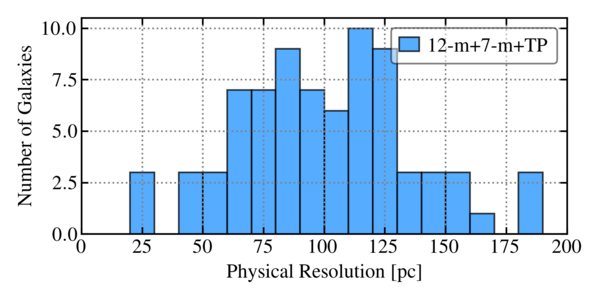}
\end{center}
\caption{
\textbf{Native physical and angular resolutions of the PHANGS-ALMA cubes.} Histograms of the native angular (\textit{top row}) and physical (\textit{bottom row}) resolutions of the PHANGS-ALMA \mbox{7-m}+TP (\textit{left}) and \mbox{12-m}+\mbox{7-m}+TP (\textit{right}) cubes after processing (Table~\ref{tab:derivedproducts}). To calculate the physical resolutions, we adopt the distance compilation from \cite{ANAND21}. The wider range of angular resolutions for the \mbox{12-m}+\mbox{7-m} data reflects the fact that the \mbox{12-m} array configuration used to take these data varied somewhat, consistent with standard ALMA observing strategies. The extreme outliers at fine physical resolution for the \mbox{7-m}+TP data reflect cases where we have used the ACA \mbox{7-m}+TP only to target very nearby, extended systems.
\label{fig:reshist}
}
\end{figure*}

We convolve each data cube to a series of fixed angular and physical resolutions. This has two purposes. First, convolving to coarser angular resolution improves the surface brightness sensitivity and increases the fraction of the flux detected at good signal-to-noise (see Figure~\ref{fig:convexample}). This allows us to use coarser resolution versions of the cube to create high-completeness masks to be applied to the sharper resolution data. Second, convolving to multiple fixed physical resolutions plays a crucial role in testing scientific hypotheses. At the most basic level, this allows for rigorous comparison among galaxies observed with different beams and lying at different distances \citep[e.g.,][]{HUGHES13A, ROSOLOWSKY21}. Increasingly, spatial scale is also by itself viewed as an important variable when studying stochastic processes and the hierarchical structure of the interstellar medium \citep[e.g.,][]{SCHRUBA10,SCHINNERER19,CHEVANCE20}.

For PHANGS-ALMA, we convolved the data to a series of fixed angular and physical scales. The target angular scales have FWHM beam sizes of $2''$, $7.5''$, $11''$, and $15''$ and the fixed physical resolutions are $60$, $90$, $120$, $150$, $500$, $750$, and $1{,}000$~pc. When convolving to a fixed physical resolution, we adopt a distance to the galaxy, which the user supplies as an input to the pipeline. Then we calculate the angular scale corresponding to the target physical resolution at the adopted distance of the galaxy. We use Euclidean geometry for this calculation. For PHANGS-ALMA, we adopt the distances derived and compiled by \citet{ANAND21}. For these target angular resolutions, the \mbox{12-m}+\mbox{7-m} data can typically be convolved to all scales. The \mbox{7-m} data can all be convolved to $11''$and $15''$, but less than half can be convolved to $7.5''$. Roughly, the target resolutions of  60, 90, 120, and 150~pc correspond to the quartiles of the distribution of \mbox{12-m}+\mbox{7-m} physical resolutions  (see Figure~\ref{fig:reshist} and Table~\ref{tab:productprocess1}). For a typical target, the 7{-}m data can only reach $\gtrsim 500$~pc resolution, but a key extension of PHANGS--ALMA targets targets galaxies with $d < 5$~Mpc. In these targets, the 7{-}m data can also reach physical resolutions $\lesssim 165$~pc.

When we convolve the cubes, we treat any area outside the map as missing. This means that near the edges of the map, comparatively fewer data contribute to the final map. As a result, the noise will be higher near the map edges. We create a ``coverage cube'' to track the amount of data contributing to each sight line. To create this cube, we replace all locations with data in the original cube by $1.0$ and all locations without data by $0.0$. Then we also convolve this cube. In the resulting coverage cube, a value of $0.95$ indicates that during the convolution, $95\%$ of the effective area of the convolving beam contained data, i.e., had values of $1.0$, while $5\%$ of the convolved area did not, i.e., had values of $0.0$. We use this coverage cube to clip some final data products to avoid strong edge effects.

\medskip

\textbf{Application to PHANGS-ALMA:} Table~\ref{tab:productprocess1} and Figures \ref{fig:convexample} and \ref{fig:reshist} illustrate some details of the resolution and convolution for PHANGS-ALMA. Table~\ref{tab:productprocess1} and Figure~\ref{fig:reshist} report the native angular and physical resolutions of the \mbox{12-m}+\mbox{7-m}+TP and \mbox{7-m}+TP data after postprocessing. The \mbox{7-m}+TP data show a narrow range of angular resolutions, consistent with the almost fixed configuration used to observe them. The distances to nearby galaxies, including the PHANGS-ALMA targets, are almost always uncertain by $5{-}30\%$ \citep[e.g.,][among many others]{TULLY16,MCQUINN17,ANAND21}. After accounting for the current best-estimate distances to the targets \citep{ANAND21}, the \mbox{7-m}+TP data show a wide range of physical resolutions, typically ${\sim}550$~pc but with outliers down to $<200$~pc. These high resolutions arise from \mbox{7-m} observations of very nearby systems with $d \lesssim 5$~Mpc, many of which we have so far targeted only with the ACA.

In Figure~\ref{fig:reshist} and Table~\ref{tab:productprocess1} the \mbox{12-m}+\mbox{7-m}+TP data show a wider range of angular resolutions. This mostly reflects that ALMA delivers data within some tolerance of the nominal angular resolution and that the \mbox{12-m} array cycles between array configurations. As a result the exact $u{-}v$ coverage differs from galaxy to galaxy. After accounting for distance, the typical physical resolution of the \mbox{12-m}+\mbox{7-m}+TP data is 100~pc, with the highest resolution ${\sim}25$~pc, all galaxies better than $200$~pc, and 90\% of galaxies having physical resolution better than $150$~pc.

Note that the resolutions in these final cubes have been inflated by several postprocessing steps (Section~\ref{sec:postprocess}). We convolved to a round synthesized beam and also degraded to the coarsest common resolution when linearly mosaicking individual ``parts'' of multi-part mosaic galaxies. Each of these steps involves a convolution to a moderately coarser resolution, and in the case of multi-part galaxies one part may have much higher resolution than the other \citep[a prominent example of this in PHANGS-ALMA is NGC~4321, M100, where one half of the galaxy has much higher resolution ($1.0''$) than the other ($1.6''$), e.g., see][]{HENSHAW20}.

Figure~\ref{fig:convexample} shows an example of the convolution to fixed physical resolution applied to one PHANGS-ALMA galaxy, NGC~3621. Each panel shows the galaxy at a fixed physical resolution, from 60~pc to 1~kpc. The contour shows the area of high ($>95\%$) coverage as defined above. The figure shows increased surface brightness sensitivity, increased filling fraction of emission, and decreased detail as the resolution degrades. The bottom right panel also illustrates how at coarser resolutions, edge effects become important. As the beam becomes larger, a larger fraction of the original flux sits near the edge of the field of view, where the sensitivity is reduced due to incomplete sampling.

\begin{figure}
\centering
\includegraphics{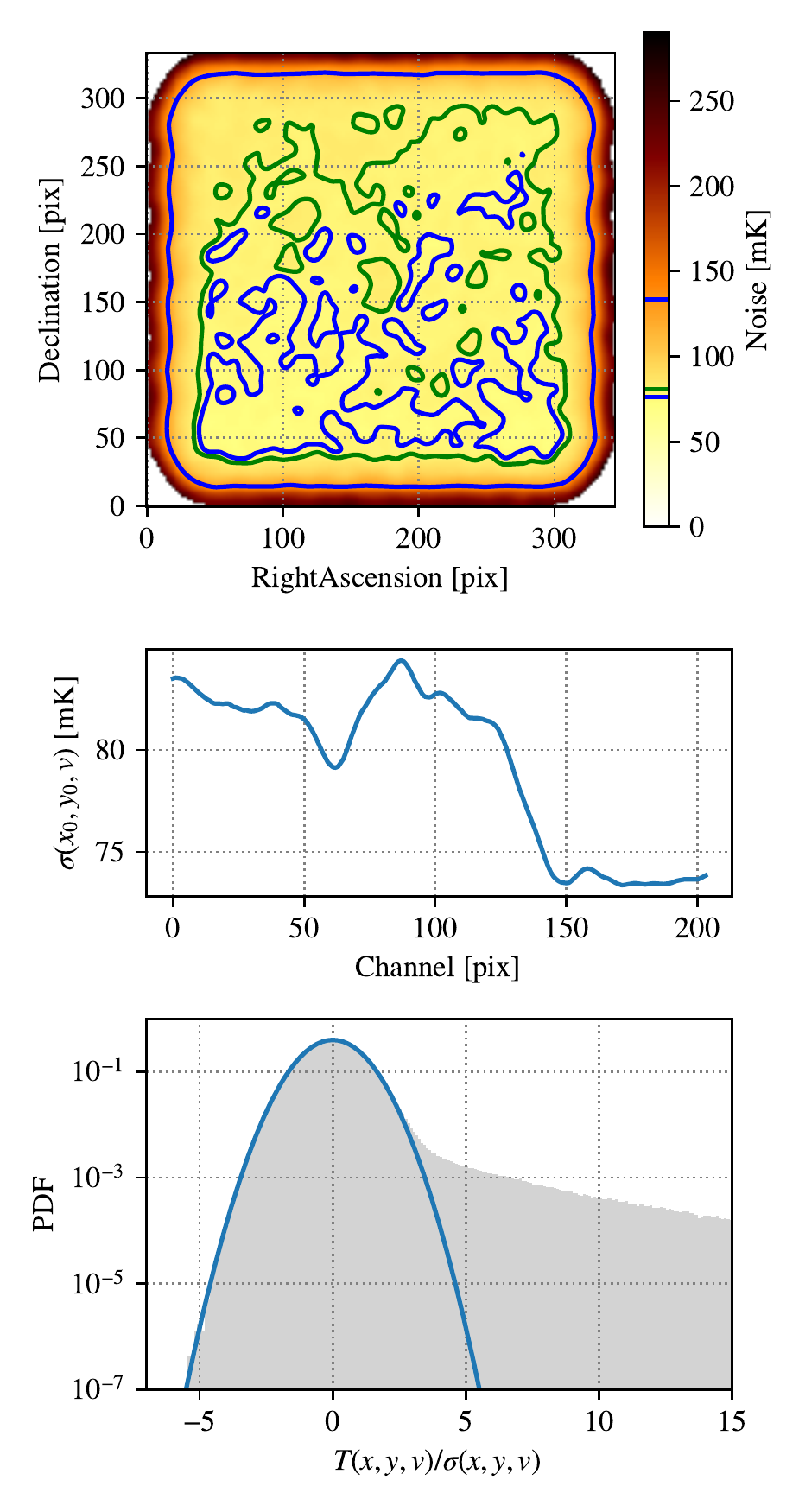}
\vspace{-24pt} % reduce excessive gaps between figure rows
\caption{{\bf Examples of empirically generated noise cubes.} The top and middle panels show the noise map in the spatial (top) and spectral (middle) coordinates for the \mbox{12-m}+\mbox{7-m} imaging of NGC~4303. The contours in the top panel show the 16th (blue), 50th (green) and 84th (blue) percentiles of the noise values. The noise profile along the spectral axis is extracted from the center of the map. The bottom panel shows the probability density function of the signal-to-noise implied by this noise cube. The blue parabola shows the PDF of a normal distribution with mean of zero and variance of one. The normal distribution is an excellent description of the signal-to-noise values except for the strong positive tail of values arising from signal in the cube.}
    \label{fig:noise_figure}
\end{figure}

\subsection{Noise Estimation}
\label{sec:noise}

\begin{figure*}[ht!]
\begin{center}
\includegraphics[width=0.45\textwidth]{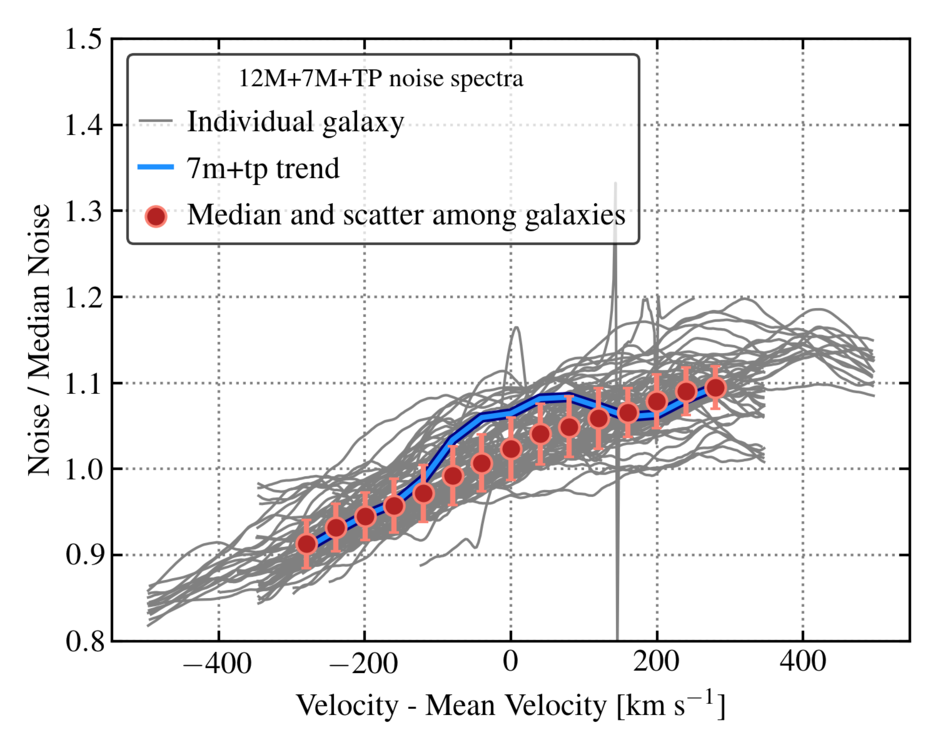}
\includegraphics[width=0.45\textwidth]{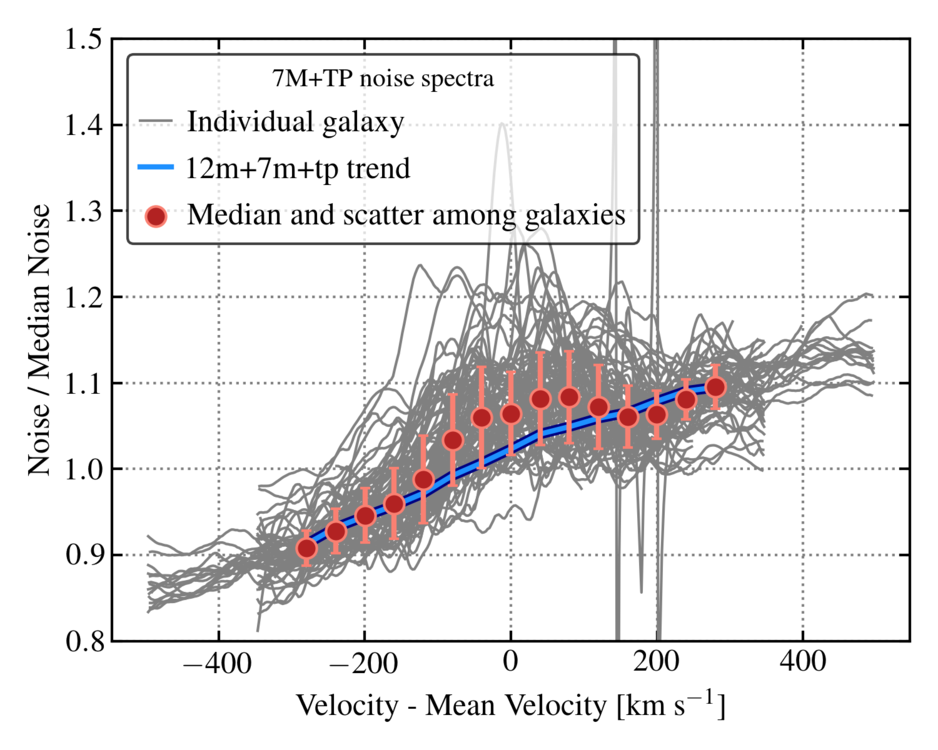}
\vspace{-18pt} % reduce excessive gaps between figure rows
\end{center}
\caption{
\textbf{Normalized noise spectra for PHANGS-ALMA data cubes.} Normalized noise, calculated by our three-dimensional noise estimator (Section~\ref{sec:noise}) as a function of Doppler shift velocity for PHANGS-ALMA galaxies. We plot noise divided by the median noise in the cube and velocity offset from the mean velocity in the cube. We calculated the median noise in the cube from a $100$~km~s$^{-1}$-wide window at each edge of the spectrum. Most galaxies and both array combinations show the same overall trend in noise as a function of velocity. We attribute this to a mixture of the gridding effects discussed in Section~\ref{sec:spectral_regridding} and the noise response of the Band~6 receiver used for the survey \citep[e.g.,][Figure 22 and C. Brogan priv. communication]{KERR14}. The smooth trend in the \mbox{12-m}+\mbox{7-m}+TP data (\textit{left} panel) suggests that the iterative signal rejection works well for these data. The \mbox{7-m}+TP data (\textit{right}) show the imprint of the galaxy emission superimposed on the background trend near the mean (systemic) velocity. This modest (${\sim}10\%$) effect reflects that our rejection of signal from the noise estimate works well but not perfectly in these lower resolution cases.
\label{fig:noisespec}
}
\end{figure*}

\begin{figure*}[ht!]
\begin{center}
\includegraphics[width=0.45\textwidth]{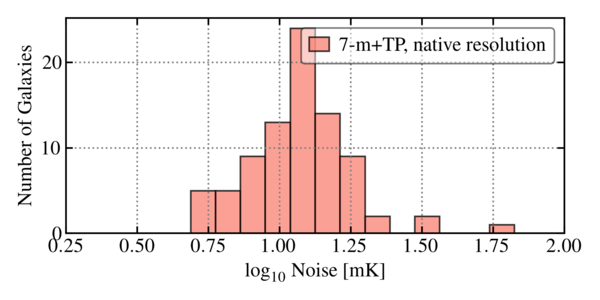}
\includegraphics[width=0.45\textwidth]{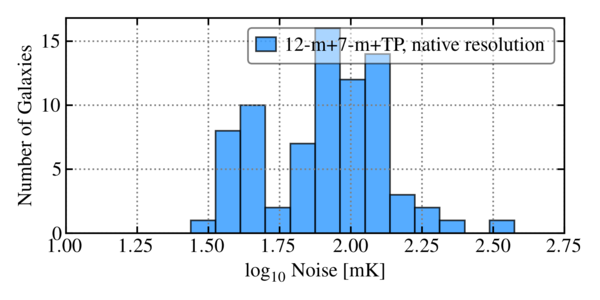}
\vspace{-18pt} % reduce excessive gaps between figure rows
\includegraphics[width=0.45\textwidth]{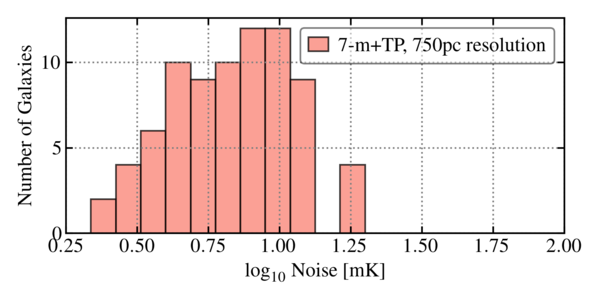}
\includegraphics[width=0.45\textwidth]{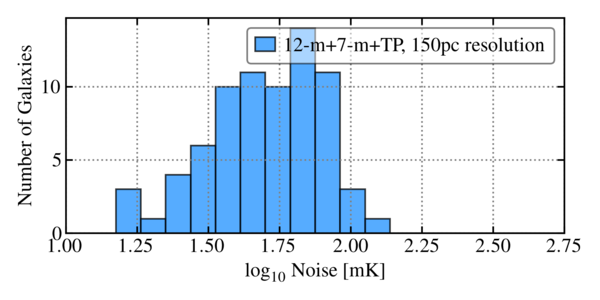}
\end{center}
\caption{
\textbf{Characteristic noise in the PHANGS-ALMA cubes.} Characteristic noise values, in units of milli-Kelvin, for each PHANGS-ALMA \mbox{7-m}+TP and \mbox{12-m}+\mbox{7-m}+TP cube at their native resolution (\textit{top row}). These ``characteristic'' values are drawn from the center of the three dimensional noise cube. The spatial (Figure~\ref{fig:noise_figure}) and spectral (Figure~\ref{fig:noisespec}) noise estimates represent deviations about this. The lower panels show the noise after convolving the \mbox{7-m}+TP data to 750~pc resolution and the \mbox{12-m}+\mbox{7-m}+TP data to 150~pc resolution, omitting galaxies that cannot be convolved to this resolution because of their distance and angular resolution. The convolution lowers the characteristic noise (Table~\ref{tab:derivedproducts}).
\label{fig:noisehist}
}
\end{figure*}

For each cube at each spatial scale we produce a three-dimensional estimate of the rms noise. We treat this as a separable problem. First we construct a noise map that captures spatial variations of the noise, $R(x,y)$. Then we measure a normalized noise spectrum that captures the relative spectral variations, $s(v)$.  The noise in the data cube is then:
\begin{equation}
    \sigma(x,y,v) = R(x,y) s(v)~.
\end{equation}

We determine $R(x,y)$ and $s(v)$ empirically, determining the values from the data themselves using an iterative procedure.  First, we use a robust noise estimator, the median absolute deviation of the data around zero, to characterize the noise in a rolling spatial box. In estimating the noise, we exclude positive data that have high significance with respect to the noise level. These are likely to be associated with real emission and not noise. We do not exclude high significance negative data because this has not been a concern for PHANGS-ALMA, but we might modify the calculation to do this in the future.

To save computation time and increase the sample size of the data used for noise estimation, we calculate the noise in boxes centered on a sparsely-sampled square grid rather than at every pixel in the cube.  The size of the box and the grid spacing are tunable parameters. Larger boxes yield more robust noise estimates, thanks to the large sample size, at the expense of washing out small-scale variations in the noise. For the PHANGS-ALMA public release we used a box size of ${\sim}3\times$ the FWHM beam size in width.  We calculate the noise on a rectilinear grid of positions with spacing of $1.2$ beam FWHM.  We then smooth the empirical noise estimates with a Gaussian kernel with size equal to the box size yielding an estimate of $R(x,y)$.

We then estimate $s(v)$ by normalizing the cube by the spatial response, $R$, and estimating the median factor that each channel is different from the spatial noise estimate derived for the cube as a whole. We then smooth these estimates with a third order Savitsky-Golay filter to estimate $s(v)$. In PHANGS-ALMA, both the performance of the receiver and the regridding effects described in Section~\ref{sec:staging} lead to spectral variations of the noise.

This process is iterative. We generate an estimate of $\sigma(x,y,v)$, divide the cube by this estimate, and then repeat the noise estimation process. Variations in the noise estimate are accumulated to form a final estimate of $\sigma(x,y,v)$. The iterative process drives $I(x,y,v) / \sigma(x,y,v)$ in the signal-free regions to a zero-centred normal distribution with standard deviation of~$1$. In practice, we find that three iterations are sufficient to arrive at a stable estimate of the noise.  Figure~\ref{fig:noise_figure} shows the variations seen in a typical noise map and that the resulting noise cube characterizes the spatial and spectral variations of noise in the cube.

\medskip

\textbf{Application to PHANGS-ALMA:} Table~\ref{tab:productprocess1} and Figures \ref{fig:noise_figure}, \ref{fig:noisespec}, and \ref{fig:noisehist} report some results of applying this algorithm to the PHANGS-ALMA \cotwo\ data. Table~\ref{tab:productprocess1}, Figure \ref{fig:noisespec}, and Figure~\ref{fig:noisehist} report typical noise values and typical spatial and spectral variations in the cubes. We show the normalized noise spectra of each galaxy in Figure~\ref{fig:noisespec}.

Table \ref{tab:productprocess1} and Figure \ref{fig:noisehist} show median noise levels of 12~mK for the native resolution \mbox{7-m}+TP data, $7$~mK for the 750~pc resolution \mbox{7-m}+TP data, $85$~mK for the native-resolution \mbox{12-m}+\mbox{7-m}+TP data, and $53$~mK for the 150~pc resolution \mbox{12-m}+\mbox{7-m}+TP data. In each case individual galaxies scatter by $\sim \pm 50\%$ about these median values. As expected, the convolution lowers the overall noise level, but the fractional scatter in the data set remains about the same at each resolution. 

Table~\ref{tab:derivedproducts} notes the magnitude of spatial and spectral noise variation across the final PHANGS-ALMA cubes. We typically find ${\sim}25\%$ variation in the spectral dimension. We observe much larger spatial variations, with $\pm 1\sigma$ variations of $80{-}100\%$ on average (i.e., the 84$^{\rm th}$-16$^{\rm th}$ percentile value divided by the median is ${\sim}0.8{-}1.0$). As shown in Figure~\ref{fig:noise_figure}, this mostly reflects the large variation of noise near the map edge due to the changing primary beam response. For galaxies observed in multiple parts (e.g., Figure~\ref{fig:linmos}) the different parts often have different surface brightness sensitivities. This also contributes to this spatial variation.

For most galaxies, the noise spectra shown in Figure~\ref{fig:noisespec} exhibit a common behavior. The two arrays also show similar behavior to one another. The noise tends to increase from low to high recessional velocity, i.e., with decreasing frequency. The decrease has a coherent shape across most galaxies, with median variation magnitude of ${\sim}25\%$ in both arrays. We understand this overall gradient as a combined result of the spectral regridding effects discussed in Section~\ref{sec:spectral_regridding} and the behavior of the Band~6 receiver used to make the measurement. The spectral gridding introduces a gradual gradient across the bandpass as slightly different amounts of independent data contribute to different channels. The receiver effect refers to the fact that we place the CO line relatively close to the lower edge of the upper sideband of the ALMA Band~6 receiver. The receiver temperature rises with decreasing frequency in this regime (C.~Brogan et al.\ private communication). One notable outlier in the \mbox{12-m}+\mbox{7-m}+TP plot is NGC~0628, where we placed the line in the middle of the lower sideband.

We see the same trend in noise as a function of velocity in the \mbox{7-m}+TP data, but we also find enhanced noise near the systemic velocity of the galaxy. This reflects the fact that our iterative noise rejection does not do a perfect job of filtering out the emission from the galaxy in this case. The emission in the \mbox{7-m}+TP maps tends to be more extended with a larger filling factor and higher median S/N compared to the \mbox{12-m}+\mbox{7-m}+TP maps. As a result, it appears to bias our noise estimates high by about 10\% over the velocity range of the galaxy. Other than this effect, the average spectral variation of the noise matches well between the \mbox{12-m}+\mbox{7-m}+TP and \mbox{7-m}+TP data. That is, the blue and red lines overlap away from the systemic velocity in the two panels of Figure~\ref{fig:noisespec}.

Finally, recall that at several steps during the imaging (Section~\ref{sec:imaging}), we use a single robustly-determined noise value to describe the data, rather than the three-dimensional estimate here. Note that this processing happens before any primary beam correction and treats individual mosaics separately. Therefore, most of the spatial noise variations will be suppressed. We expect these estimates to be accurate to ${\sim}30\%$.

\subsection{Masking}

\begin{figure*}[ht!]
\begin{center}
\includegraphics[width=0.45\textwidth]{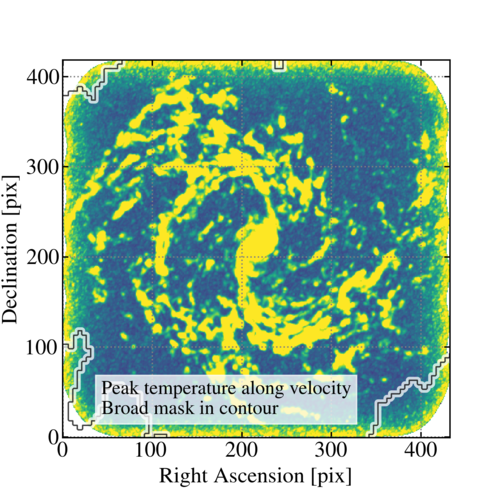}
\includegraphics[width=0.45\textwidth]{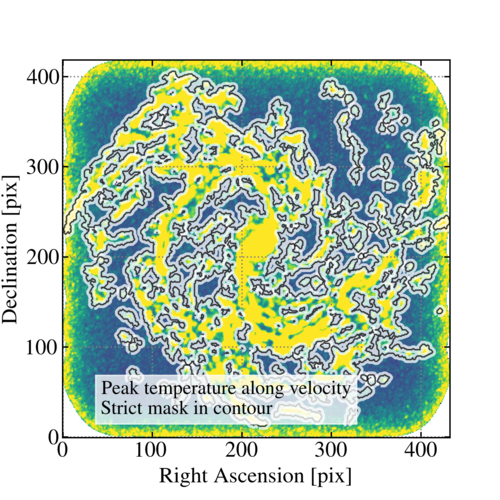}
\vspace{-18pt} % reduce excessive gaps between figure rows
\includegraphics[width=0.45\textwidth]{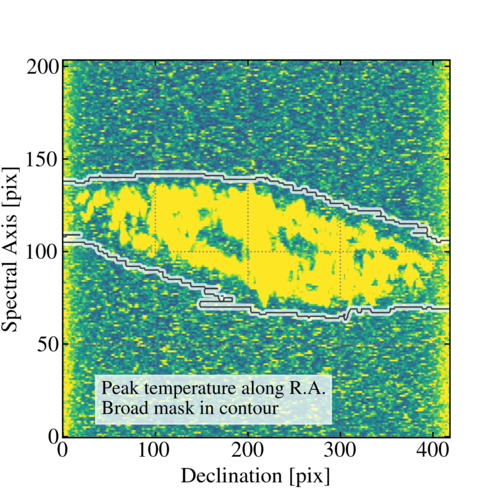}
\includegraphics[width=0.45\textwidth]{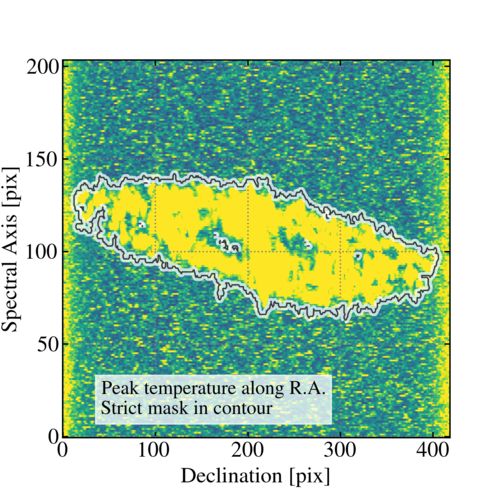}
\end{center}
\caption{
\textbf{Illustration of our two masking schemes.} The \textit{left} column shows the ``broad mask'' in black-and-white contour over a peak intensity map collapsed along the spectral dimension (\textit{top}) and the right ascension direction (\textit{bottom}) for NGC~4303. The contour shows all lines of sight where the mask has at least one ``True'' (1) value along the line of sight. The \textit{right} panels show the same images but contours now indicate the ``strict mask.'' The strict mask identified emission that can be distinguished from noise with high confidence. The figure shows that the strict mask roughly includes all of the emission that can be seen in the peak intensity maps on a high stretch. We construct the broad mask from the union of strict masks made at all resolutions. The broad mask has high completeness, meaning that it contains most of the emission in the cube. As illustrated here, the broad mask often extends across the whole map and tracks the circular rotation of the galaxy.
\label{fig:maskillustrated}
}
\end{figure*}

\begin{figure*}[ht!]
\begin{center}
\includegraphics[width=0.45\textwidth]{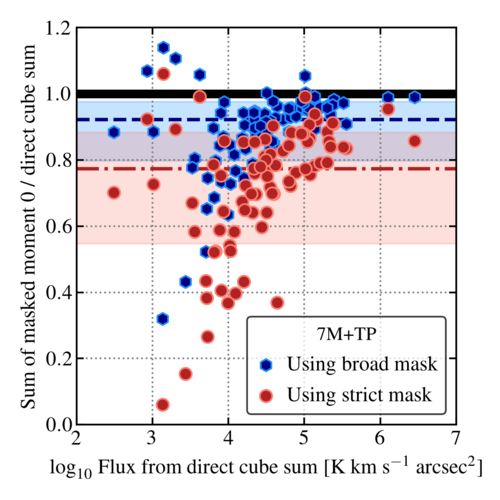}
\includegraphics[width=0.45\textwidth]{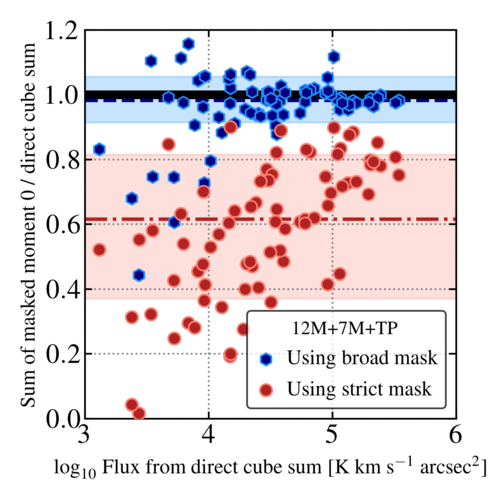}
\end{center}
\caption{
\textbf{Flux recovery in our two masking schemes.} Fraction of flux recovered using the moment~0, i.e., integrated intensity, maps created using the ``strict'' (red) and ``broad'' (blue) masks. In both panels, we show the ratio of flux in the masked moment maps to the total flux calculated by summing the entire cube. The shaded regions and lines show the $16{-}84$\% range and median for each type of mask (see also Table~\ref{tab:productprocess2}). The high confidence ``strict'' masks produce moment maps that include less of the overall flux: ${\sim}60\%$ on average for the \mbox{12-m}+\mbox{7-m}+TP data and ${\sim}80\%$ on average for the \mbox{7-m}+TP data. However, each sight line in a strictly masked moment map is highly likely to contain real emission (see Figure~\ref{fig:maskillustrated} and \citealt{SUN18,SUN20B}). Maps constructed using the high completeness broad masks include almost all flux for the \mbox{12-m}+\mbox{7-m}+TP data and are $\gtrsim 90\%$ complete for the \mbox{7-m}+TP data. A few cases show ratios above $1$, which could result from mild calibration differences between the total power and interferometer data or, more likely, failure of the direct integral of the cube to yield an accurate flux, e.g., due to mild baseline issues or field of view clipping effects in small maps.
\label{fig:completeness}
}
\end{figure*}

\begin{deluxetable}{lc}
\tabletypesize{\small}
\tablecaption{
PHANGS-ALMA Masking
\label{tab:productprocess2}}
\tablewidth{0pt}
\tablehead{
\colhead{Item} & 
\colhead{Description} 
}
\startdata
\hline
\multicolumn{2}{l}{Mask summary} \\
\hline\hline
``Strict mask'' & \\
\multicolumn{2}{l}{... low false positive rate} \\
\multicolumn{2}{l}{... based on S/N threshold} \\
\multicolumn{2}{l}{... constructed for every resolution} \\
``Broad mask'' & \\
\multicolumn{2}{l}{... high completeness} \\
\multicolumn{2}{l}{... union of signal masks for all resolutions} \\
\multicolumn{2}{l}{... one mask per configuration and target} \\
\hline
\hline
\multicolumn{2}{l}{Completeness\tablenotemark{a}} \\
\hline\hline
7{-}m+TP & 84 galaxies \\
... strict mask-to-direct sum & $0.77^{+0.11}_{-0.12}$ \\
... broad mask-to-direct sum & $0.92^{+0.06}_{-0.12}$ \\
... strict mask-to-broad mask & $0.84^{+0.08}_{-0.14}$ \\
12{-}m+7{-}m+TP & 77 galaxies \\
... strict mask-to-direct sum & $0.62^{+0.20}_{-0.25}$ \\
... broad mask-to-direct sum & $0.98^{+0.08}_{-0.07}$ \\
... strict mask-to-broad mask & $0.65^{+0.17}_{-0.25}$ \\
\enddata
\tablenotetext{a}{Completeness here refers to the fraction of flux included in each mask at the native resolution of the cube. For both masks, we reference this to a direct sum of the cube, also at the native resolution. We also calculate the ratio of flux between the two masks. Quoted values are medians and the error bars refer to the $16^{\rm th}$ and $84^{\rm th}$ percentile. These measurements are visualized in Figure~\ref{fig:completeness}.}
\tablecomments{These numbers refer to the first public data release, internal ``version~4'', constructed with ``version 2.0'' of the PHANGS-ALMA pipeline. They refer to the products created for the \cotwo\ survey. The number of galaxies indicates the number of targets with these array combinations processed by this release.}
\end{deluxetable}

Since the data cubes include large, signal-free volumes, we create masks to identify the regions of the cube containing signal. We then apply these when creating higher level data products.

We create two types of masks, which we illustrate in Figure~\ref{fig:maskillustrated}. First, we create a high-confidence ``strict mask'' that includes only voxels highly likely to contain real signal. Second, we create a high-completeness ``broad mask,'' which contains most known signal in the cube. Though there are many approaches to masking, these two cases cover most common applications. The strict mask, which is illustrated in the right column of Figure~\ref{fig:maskillustrated}, includes only bright emission and few or no noise-dominated sight lines. It should be used when running calculations sensitive to noise, e.g., many types of kinematic analysis. The broad mask, illustrated in the left column of Figure~\ref{fig:maskillustrated}, should include almost all regions with real emission. This comes at the expense of including more noise-dominated sight lines. The broad mask should be used for any analysis aimed at a complete characterization of the emission.

We create strict masks for each cube at each resolution. These mostly follow the standard recipes defined for CPROPS \citep{ROSOLOWSKY06}. They begin with a core mask that includes all voxels with signal-to-noise ratio above $4$ over two successive velocity channels\footnote{In principle, the appropriate number of channels required for joint detection depends on the expected line width and the line spread function. In PHANGS-ALMA, two channels corresponds to a line width of ${\sim}5$~km~s$^{-1}$, about the narrowest full line width that we might expect for a GMC \citep[e.g.,][]{BOLATTO08,HEYER09}. The channels are also mostly independent \citep[when tested following the method in][]{LEROY16} so that the line spread function is about one channel.}. We also create a lower signal-to-noise outer mask that includes all voxels with signal-to-noise above $2$ in two successive velocity channels. We then construct a final mask that consists of all contiguous regions in the outer mask that contain any pixels from the higher-significance core mask. As long as the channel width is a few times narrower than the typical line width, this algorithm does an excellent job of identifying all significant features in the cube. The example in the right column of Figure~\ref{fig:maskillustrated} shows that for NGC~4303 the strict mask indeed does a good job of highlighting all of the real emission one would pick out from a peak temperature map.

We offer the user the option to trim small-volume or small-area regions from the core mask, with both specified in units of the beam area. In order to maintain a relatively clean, easily modeled criteria for inclusion in the mask that applies for individual lines of sight \citep[][]{SUN18,SUN20B}, we do not use these volume options for the main PHANGS-ALMA data products. We do apply one additional condition on the strict mask, however. When masking data cubes that have been created by convolution, we restrict the core mask to only include regions that had high coverage in the original map. Specifically, we only allow regions that have a value greater than $0.95$ in the ``coverage cube'' (see above) to contribute to the core mask. This avoided spurious contributions from map edges where the noise estimate can become slightly inaccurate. 

We create broad masks by taking the union of all strict masks from all resolutions. Both high resolution and low resolution masks contribute to the final result. The masks at the coarse resolution tend to do an excellent job of capturing extended, faint emission. These tend to be most important for overall recovery of flux in PHANGS-ALMA targets. The masks at high resolution tend to capture bright compact features, e.g., these high resolution masks do a better job of recovering the broad line wings associated with galactic nuclei than the low resolution ones. By combining these masks, we construct a best estimate of where we have detected any signal in the cube at any resolution. As illustrated in the left column of Figure~\ref{fig:maskillustrated}, the broad masks do a good job of encompassing all emission from the galaxy at the expense of including a moderate amount of ``empty'' noise-dominated volume. In many PHANGS-ALMA cases, including the one illustrated, the broad mask captures the overall rotation of the galaxy and extends across most of the area of the map. 

Note that the broad masks resemble the clean masks described in Section~\ref{sec:imaging}. For PHANGS-ALMA these are not identical. We create external clean masks and supply them. If one wanted to use the PHANGS-ALMA pipeline to create clean masks in an automated way, one could process the data, create broad masks, then feed them back in as clean masks. In practice, the main differences between our broad and clean masks are that the clean masks had an additional dilation in all three dimensions (i.e., they have been slightly ``inflated'') and that the clean masks for galaxies with bright centers include a wide velocity region near the center of the galaxy (compare Figures \ref{fig:cleanmask} and~\ref{fig:maskillustrated}).

\medskip

\textbf{Application to PHANGS-ALMA:} Table~\ref{tab:productprocess2} and Figures \ref{fig:maskillustrated} and~\ref{fig:completeness} show some outcomes of applying this masking to PHANGS-ALMA. Figure~\ref{fig:maskillustrated} illustrates the differences between the strict and broad masks for a typical bright galaxy, NGC~4303. Table~\ref{tab:productprocess2} and Figure~\ref{fig:completeness} report the fraction of flux captured by each mask for our data.

For the \mbox{12-m}+\mbox{7-m}+TP data the strict masks have ${\sim}60\%$ completeness on average, meaning that they include about 60\% of the emission found via a direct sum of the cube. As in \citet{SUN18,SUN20B}, we find a wide range of completeness among the PHANGS-ALMA data, with the $16{-}84\%$ range spanning about $40{-}80\%$. There is not a perfect mapping between integrated CO flux and completeness in the strict maps, but our lowest completeness galaxies do tend to have lower overall flux. These are often, but not always, lower mass, more \hi-dominated systems. For the \mbox{12-m}+\mbox{7-m}+TP data, with only a few exceptions, the broad masks do a good job of achieving nearly $100\%$ completeness. The outliers tend to be the lowest flux galaxies. This reflects that if we fail to detect diffuse signal in any mask, even at low resolution, the broad mask will underestimate the true flux.  Masked flux fractions larger than unity can occur because the masked regions do not include the negative noise fluctuations that are included in the sum over the cube.

For \mbox{7-m}+TP data, the strict masks have higher overall completeness, almost $80\%$ on average with a range of about $70{-}90\%$. The completeness of the broad mask compared to direct integration of the cube is actually moderately lower for the \mbox{7-m}+TP data than the \mbox{12-m}+\mbox{7-m}+TP data, only $92\%$ on average for the \mbox{7-m}+TP data. This likely reflects the fact that our set of spatial scales only reaches to $15\arcsec$, which is still somewhat compact compared to the \mbox{7-m} native resolution of ${\sim}7.5\arcsec$. Still, the completeness of the broad masks for the \mbox{7-m}+TP data is quite high for all high-flux targets. As with the \mbox{12-m}+\mbox{7-m}+TP data, the completeness drops in faint, lower surface brightness targets. Because there is less difference between the broad and the strict masks for the \mbox{7-m}+TP data, the completeness of the two track one another closely as a function of total flux, but with the strict masks mildly offset to lower completeness.

Overall, masks perform as intended in PHANGS-ALMA. The broad masks achieve near-100\% completeness in many cases, while the strict masks have lower completeness but higher confidence. The difference is much less marked in the \mbox{7-m}+TP data compared to the \mbox{12-m}+\mbox{7-m}+TP data because the strict masks already have high completeness due to the high surface brightness sensitivity of the \mbox{7-m}+TP data.

\subsection{Map Creation}

\begin{deluxetable*}{lccc}
\tabletypesize{\small}
\tablecaption{PHANGS-ALMA Derived Product Summary \label{tab:derivedproducts}}
\tablewidth{0pt}
\tablehead{
\colhead{Map} & 
\colhead{Expression} &
\colhead{Unit} &
\colhead{Uncertainty Method}
}
\startdata
Integrated Intensity &  $W(x, y) = \sum_i I(x,y,v_i) M(x,y,v_i) \delta v $ & K~km~s$^{-1}$ & Gaussian \\
$\ldots$ (\texttt{mom0}) & & & \\
Peak Intensity
&  $I_\mathrm{peak}(x,y) = \underset{v_i}{\max} \left[I(x,y,v_i)\right]$ & K & None \\
$\ldots$ (\texttt{tpeak}) & & & \\
Peak Intensity (Smoothed)\tablenotemark{a} 
                            &  $I_\mathrm{peak,\Delta V}(x,y) = \underset{v_i}{\max} \left[I(x,y,v_i) * K(\Delta V) \right]$ & K & None\\
$\ldots$ (\texttt{tpeak1p5}) & & & \\
Mean velocity
&  $\bar{v}(x, y) = \frac{1}{W(x, y)} \sum_i v_i I(x,y,v_i) M(x,y,v_i) \delta v $ & km~s$^{-1}$ & Gaussian \\
$\ldots$ (\texttt{mom1}) & & & \\
Velocity at Peak Intensity
&  $v_\mathrm{peak}(x, y) = \underset{v_i}{\mathrm{argmax}}[I(x,y,v_i)]$ & km~s$^{-1}$ & None \\
$\ldots$ (\texttt{vpeak}) & & & \\
Interpolated Peak Velocity
& 
$v_\mathrm{quad}(x, y) = v_\mathrm{peak}(x,y) - \frac{A}{B}$~for 
& km~s$^{-1}$ & Gaussian \\
$\ldots$ (\texttt{vquad})
&
$A = I(x,y,v_\mathrm{peak}+1) - I(x,y,v_\mathrm{peak}-1)~{\mathrm{\,}}$
& &
\\
&
$B=I(x,y,v_\mathrm{peak}-1) + I(x,y,v_\mathrm{peak}+1) -2 I(x,y,v_\mathrm{peak})$
& &
\\
RMS line width
&  $\sigma_v(x,y) = \left[\frac{1}{W(x,y)} \sum_i (v_i-\bar{v})^2 I(x,y,v_i) M(x,y,v_i) \delta v\right]^{1/2} $ & km~s$^{-1}$ 
& Gaussian \\
$\ldots$ (\texttt{mom2}) & & & \\
Equivalent/effective width
&  $\mathrm{EW}(x,y) = [\sum_i I(x, y, v_i) \delta v]/[\sqrt{2\pi} I_{\rm peak} (x, y)]$ & km~s$^{-1}$ & Gaussian \\
$\ldots$ (\texttt{ew}) & & & \\
\enddata
\tablecomments{For all entries, $I(x,y,v)$ is the position-position-velocity data cube produced by the pipeline, $M(x,y,v)$ is a Boolean mask indicating where CO emission is found, $\delta v$ is the channel width, taken as a constant.}
\tablenotetext{a}{Here $K(\Delta V)$ is a boxcar smoothing kernel of full width $\Delta V = 12.5~\mathrm{km~s}^{-1}$ and $*$ is the convolution operator.}
\end{deluxetable*}

\begin{figure*}[ht!]
\begin{center}
\includegraphics[width=0.45\textwidth]{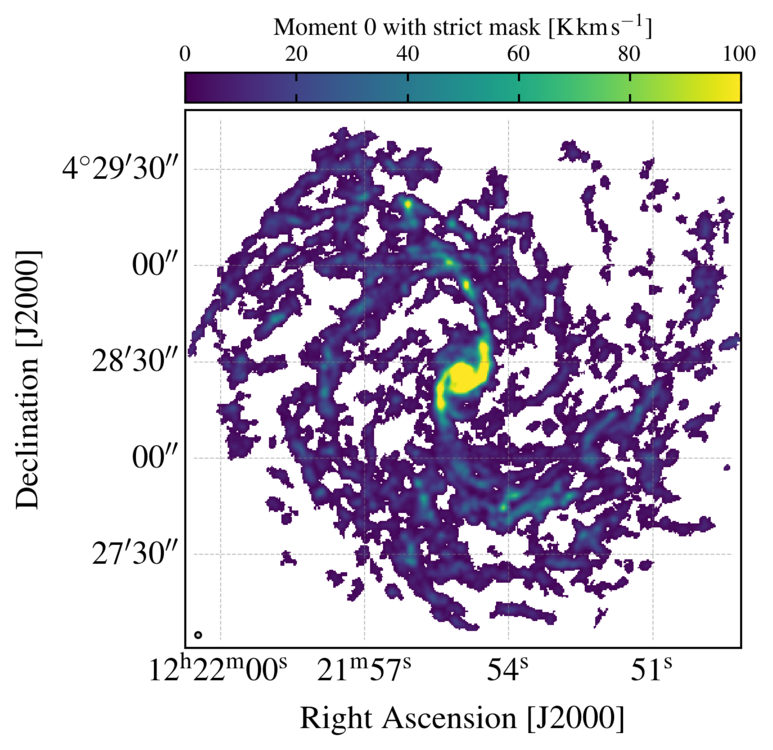}
\includegraphics[width=0.45\textwidth]{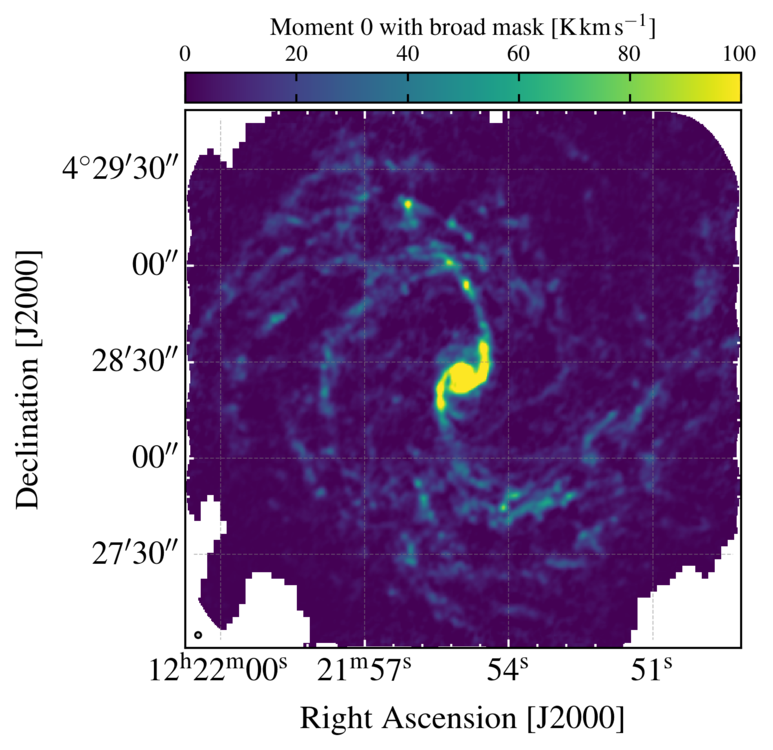}
\includegraphics[width=0.45\textwidth]{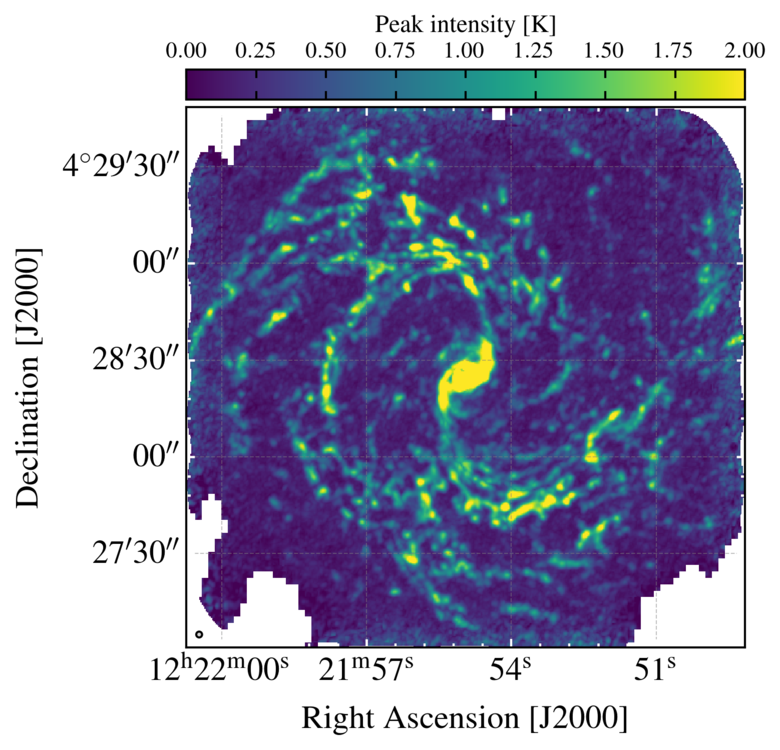}
\includegraphics[width=0.45\textwidth]{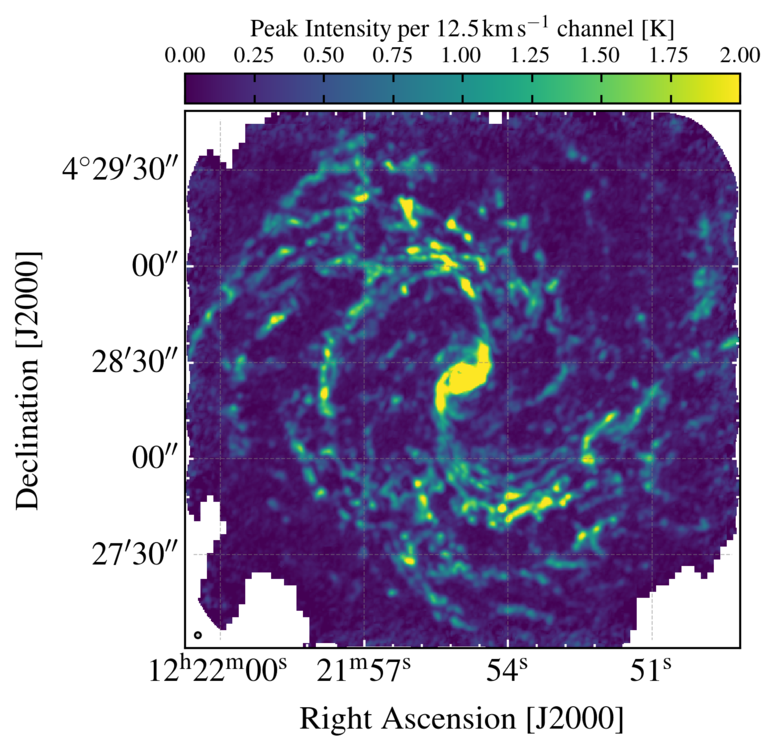}
\end{center}
\caption{
\textbf{Products showing integrated and peak intensity.} Four views of intensity for NGC~4303. The \textit{top} row shows the line-integrated intensity, or ``moment 0,'' calculated after applying the strict (\textit{left}) or broad (\textit{right}) mask to the data. The two distributions appear very similar, but the broad mask includes more area, including many sightlines with little or no emission. The \textit{bottom} row shows peak intensity calculated using either a $2.54$~km~s$^{-1}$ spectral window (\textit{left}) or a $12.5$~km~s$^{-1}$ spectral window (\textit{right}). Both do an excellent job of highlighting faint structure, e.g., in the interarm regions. The $12.5$~km~s$^{-1}$ filter used in the right-hand map approximately matches the typical line width of emission. As a result it has moderately lower noise and higher signal-to-noise.
\label{fig:mom0tpeak}
}
\end{figure*}

We combine the cubes, noise estimates, and masks to produce a suite of high level data products and associated uncertainties. In general, we deliver each product at each possible resolution. Figures \ref{fig:mom0tpeak}, \ref{fig:mom1}, and~\ref{fig:mom2} illustrate these products for one PHANGS-ALMA galaxy.

We produce associated uncertainty maps via Gaussian error propagation. For a map that estimates a two-dimensional product over a spectrum using a function~$f$, the variance in our estimate of $f$ will be 
\begin{equation}
    \sigma_f^2 = \sum_{i,j} \left(\frac{\partial f}{\partial v_i}\right)\left(\frac{\partial f}{\partial v_j}\right)\sigma_{ij}^2~,
\end{equation}
where $\sigma_{ij}^2$ is the variance-covariance matrix with $\sigma_{ii}^2 = \sigma_i^2$, and the sum runs over channels $i$ and $j$, with $v_i$ and $v_j$ the intensity in each channel.  We use the three-dimensional error estimates to determine the uncertainties from Section~\ref{sec:noise}. We also include the effects of channel-to-channel correlation in our uncertainties.  We model the covariance between channels in terms of a correlation coefficient,~$r$, measured as a function of channel separation. The covariance is then
\begin{equation}
    \sigma_{ij}^2 = r(|i-j|) \sigma_i \sigma_j~.
\end{equation}
We measure the channel correlation empirically from our imaging products \citep[e.g.,][]{LEROY16,KOCH18B} and find $r(0)=1, r(1)\approx 0.05$, and $r\approx 0$ otherwise. This implies that covariance between channels increases $\sigma_f^2$ by ${\sim}10\%$ relative to the uncorrelated case.

The pipeline produces the following data products as summarized in Table~\ref{tab:derivedproducts}.

\begin{enumerate}
\item \textbf{Integrated intensity (\texttt{mom0}):} We integrate the cube along the spectral dimension to produce the integrated intensity in units of K~km~s$^{-1}$, also referred to as the ``moment~0'' map. We create versions using both the strict and broad maps. These products and the associated uncertainties (\texttt{emom0}) represents our basic assessment of the distribution of line emission on the sky. The ``broad'' versions of these maps should show the location of essentially all emission in the cube. Figure~\ref{fig:mom0tpeak} shows both the broad and strict versions of these maps for NGC~4303 at 150~pc resolution.

\item \textbf{Peak intensity with and without a matched-line width filter (\texttt{tpeak} and \texttt{tpeak12p5kms}):} We calculate the peak intensity along each line of sight, in units of~K. Such ``peak temperature'' maps offer a useful way to see faint signal and highlight structure in the cube with minimal masking.

We also found it very useful to create ``matched-line width'' versions of the peak temperature map. To produce these, we smooth the data cube along the spectral dimension using a tophat kernel with width equal to the expected line width. We used $12.5$~km~s$^{-1}$ for PHANGS-ALMA. These matched-filter peak intensity maps produce some of the cleanest views of faint structure in the cubes.   Figure~\ref{fig:mom0tpeak} shows peak temperature maps of NGC~4303 at 150~pc resolution after applying the broad mask. We show both the single-channel and $12.5$~km~s$^{-1}$-wide versions.

\begin{figure*}[ht!]
\begin{center}
\includegraphics[width=0.45\textwidth]{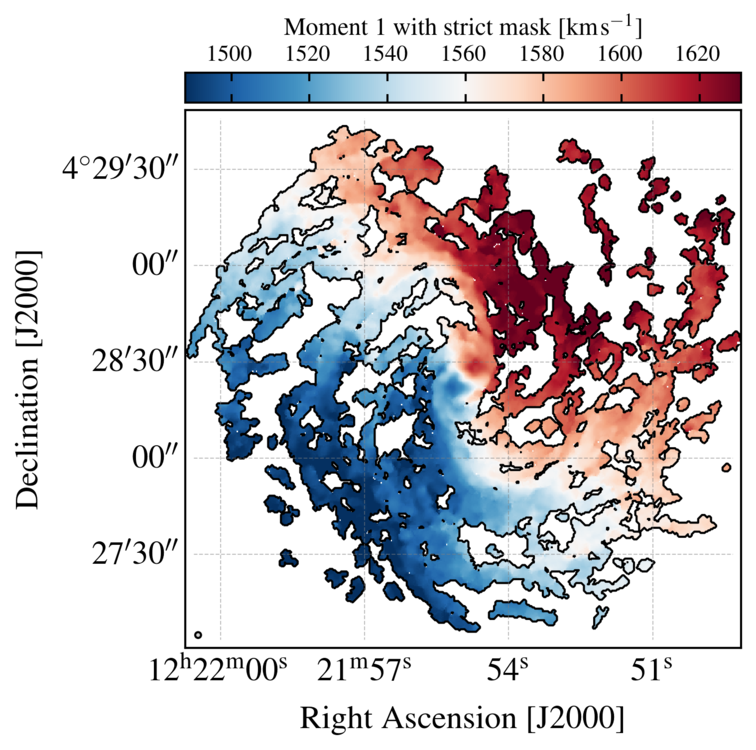}
\includegraphics[width=0.45\textwidth]{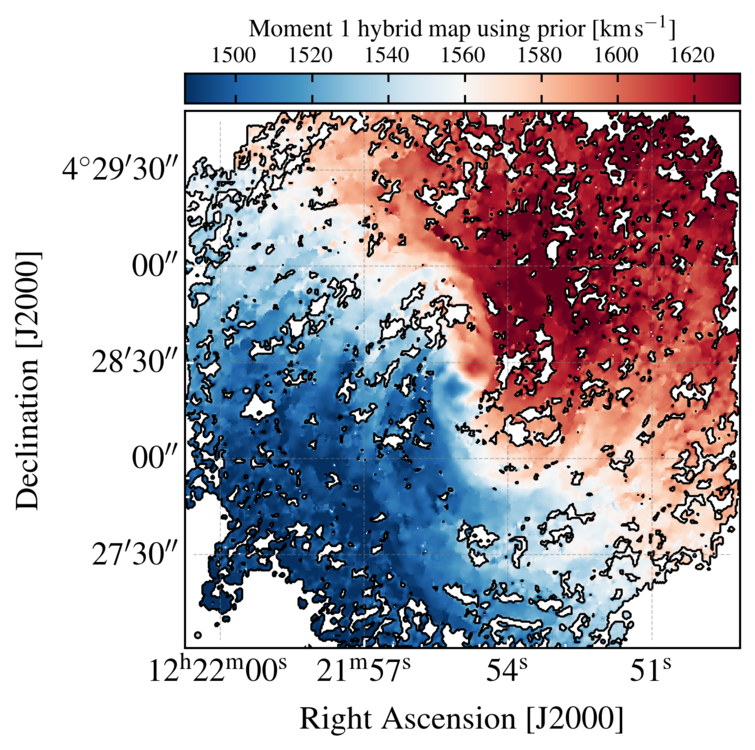}
\end{center}
\caption{
\textbf{Data products showing the velocity field.} The \textit{left} panel shows intensity-weighted mean velocity or ``moment~1'' calculated after applying the strict mask to the 150~pc resolution data cube for NGC~4303. The \textit{right} panel shows the intensity-weighted mean velocity calculated combining the strict and broad moment~1 maps. The broad map is only used where there is no strictly-masked measurement, the integrated intensity exceeds a S/N of $2$, and the velocity field lies within some tolerance ($\pm30$~km~s$^{-1}$) of a prior estimate, in this case a lower resolution strictly-masked velocity field. This ``moment~1 with a prior'' significantly expands the coverage of the velocity field while still yielding coherent structure.
\label{fig:mom1}
}
\end{figure*}

\item \textbf{Intensity-weighted mean velocity, with and without priors, and velocity at peak intensity (\texttt{mom1}, \texttt{mom1wprior}, \texttt{vpeak}, and \texttt{vquad}):} We calculate the intensity-weighted mean velocity of each spectrum, in units of km~s$^{-1}$, also known as the ``moment~1'' map. We also calculate the uncertainty associated with this map (\texttt{emom1}). The left panel of Figure~\ref{fig:mom1} shows the intensity-weighted mean velocity field calculated after applying the strict mask to the 150~pc resolution version of the NGC~4303 PHANGS-ALMA \cotwo\ cube.

We also record the velocity associated with the peak intensity and the centroid velocity near the peak calculated following \citet{TEAGUE18}. This estimator uses a quadratic function to interpolate the spectral coordinate of the local maximum at subchannel resolution.  We also calculate the uncertainty from this estimator.

We also create a version of the velocity field designed to include measurements with lower signal-to-noise ratio than those captured by the strict mask, but reject data likely to represent outliers. This map includes all moment~1 values derived from the strict mask. It also includes all moment~1 values calculated from the broad mask that meet three conditions: (i) there is no strict mask measurement for that line of sight, (ii) the integrated intensity (i.e., moment~0) value along that line of sight calculated after applying the broad mask has signal to noise above some threshold, and (iii) the measured velocity is within some tolerance of a prior guess at the velocity field.

For PHANGS-ALMA this ``moment~1 with prior'' includes lines-of-sight with moment~0 signal-to-noise above $2$, uses the $15''$ resolution moment~1 map as a prior, and allows values within $\pm 30$~km~s$^{-1}$ of that prior. In some cases, this approach can dramatically expand the coverage of the velocity field at high resolution \citep[e.g., see][]{LANG20}. This processing follows the approach applied by \citet{COLOMBO14B} to M51. They used a model rotating disk with the measured M51 rotation curve as the prior. Using the CO rotation curves of \citet{LANG20} might be our approach in a future release.

The right panel of Figure~\ref{fig:mom1} shows the intensity-weighted mean velocity field using this hybridization and prior technique. The figure shows a dramatic expansion in area covered compared to the moment~1 calculated from the strictly masked cube, but still reveals a coherent velocity structure with only modest impact from noise.

\begin{figure*}[ht!]
\begin{center}
\includegraphics[width=0.45\textwidth]{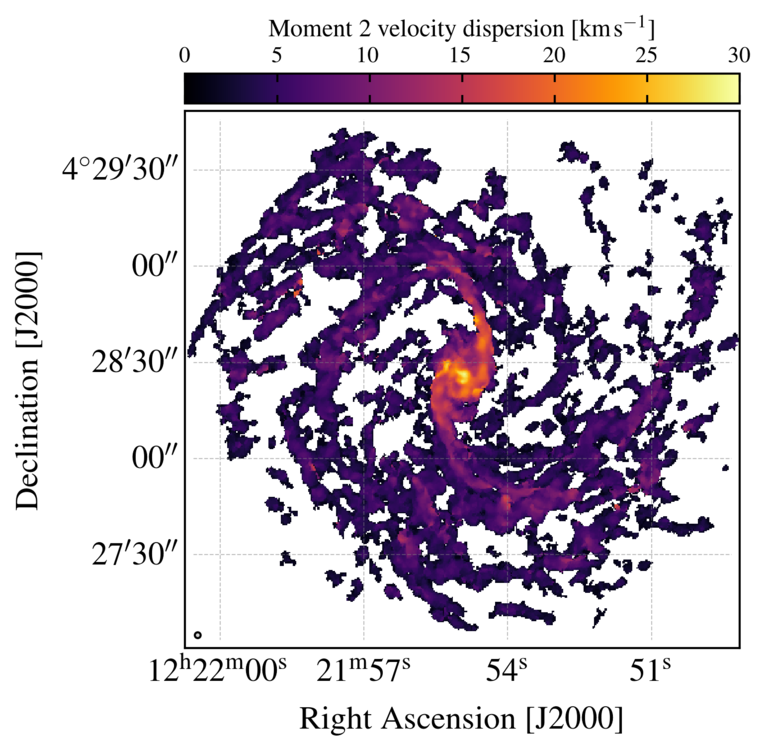}
\includegraphics[width=0.45\textwidth]{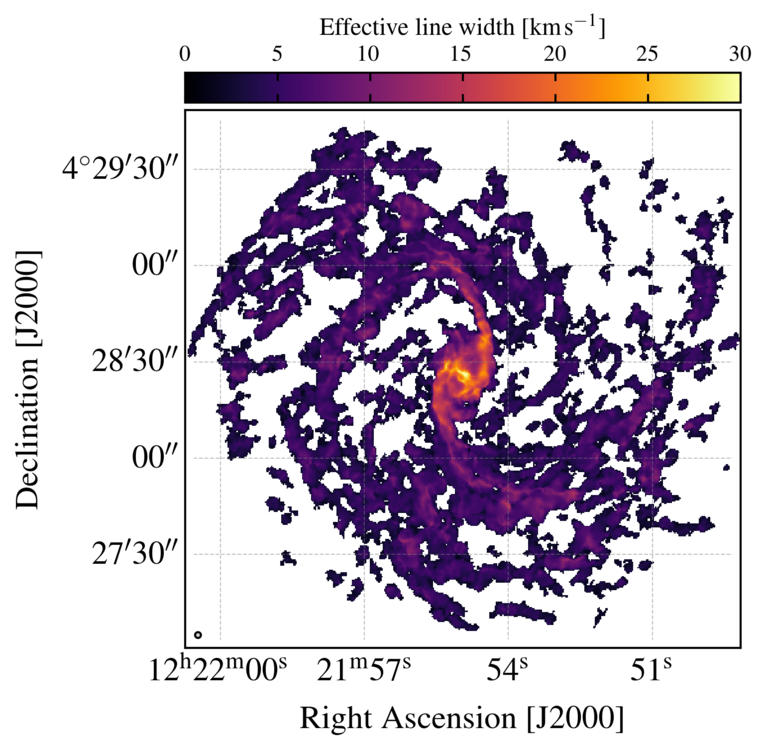}
\end{center}
\caption{
\textbf{Products showing line widths.} Two measures of line width for NGC~4303 calculated as part of our product creation. \textit{Left:} RMS velocity dispersion, also referred to as the ``moment~2'' map, calculated after applying the strict mask to the cube. \textit{Right:} ``Effective width'' or ``equivalent width'' (Equation~\ref{tab:derivedproducts}, \citealt{HEYER01}) also calculated from the strictly masked cubes. The two maps show overall similar distributions, with disagreements arising in cases where the line profile is non-Gaussian and where noise affects the spectrum. 
\label{fig:mom2}
}
\end{figure*}

\item \textbf{Root-mean-square line width and ``effective width'' or ``equivalent width'' (\texttt{mom2} or $\sigma_v$ and \texttt{ew}):} We record the intensity-weighted rms scatter of emission about the intensity-weighted mean velocity, i.e., the second moment, and the associated uncertainty. For a Gaussian line profile, this corresponds to the $1\sigma$ width of the line. Because this estimator becomes unstable in the presence of noise, we only calculate it for the strict maps. This use of the strict maps can, in turn, bias this line width to low values because faint line wings can be missed by the strict mask. In these cases, best practice is to correct $\sigma_v$ using an analytic or data driven extrapolation \citep[e.g., see][]{ROSOLOWSKY06} or to cleanly define a selection function when studying line widths \citep[e.g.,][]{SUN20B}. The pipeline does not currently implement any such clipping or sensitivity correction for $\sigma_v$.

We also record the ``effective width'' or ``equivalent width'' of each line following the definition of \citet{HEYER01}, as well as the uncertainty. This definition of line width is more robust to noise and outliers compared to the second moment, but more sensitive to the velocity resolution of the data and shape of the line profile. In this definition (see Table~\ref{tab:derivedproducts}), the effective width is the integrated intensity divided by the peak intensity. Note that this differs from the optical definition of equivalent width. The name ``effective width'' has been suggested to avoid confusion. Figure~\ref{fig:mom2} shows both the moment~2 and effective width maps for the PHANGS-ALMA map of NGC~4303 after applying the strict mask.

\end{enumerate}

For quantities currently without associated uncertainties, our recommendation is to use a Monte Carlo calculation along with the noise cube to simulate uncertainties.

%% file: quality.tex
\section{Quality Assurance and Regression Tests} 

To ensure that the PHANGS-ALMA data products were science-ready, we implemented a set of quality assurance (QA) and regression procedures. During the initial internal data releases, we built a detailed report for each data cube, which was passed to two experts for a careful by-eye inspection. Later in the project, we constructed automated regression tests, which we benchmarked against previous versions of the imaging. We also carried out an end-to-end check on the staging, imaging, and post-processing using a simulated data set. This check verified the ability of the pipeline to recover a known input image. All of these tests helped to highlight several subtle issues related to the accuracy of the deconvolution, which we discuss in more detail in Appendix~\ref{sec:arrays}.

\label{sec:quality}

\subsection{Manual Data Inspection}

For the initial internal PHANGS-ALMA data releases, we generated a collection of plots and tables that we refer to as a ``QA report'' for each data cube. These reports were distributed among ${\sim}15$ team members with experience analyzing millimeter data. Each report was assigned to at least two reviewers. 

The reviewers were asked to identify potential pathologies and assess the overall quality of the image. Their feedback was used primarily to identify major failure cases, but also to improve our overall deconvolution and data processing strategies. As an example of this feedback, the spectral noise patterns discussed in Section~\ref{sec:spectral_regridding} were first identified during an early round of QA report inspection.

\medskip

\noindent \textbf{Contents of the Inspection Reports:} The QA report aimed to present the data in a digestible form that captured the properties of the emission, characterized the noise, and highlighted any potential problems. For the initial internal PHANGS-ALMA data inspections, each report presented the following diagnostics:

\begin{enumerate}
\item Summary of the beam size, shape, and orientation and the astrometric grid. These parameters were extracted from the FITS header of the cube.

\item Channel maps showing the deconvolved data cube, user-defined clean mask (if any), data cube of residual emission, and the ratio of the data to the residuals.

\item A moment-0 map produced with no masking, that is, generated by summing the full cube over the full imaged bandwidth.

\item Tables reporting the sum of the emission inside and outside the user-supplied clean mask and a mask identifying significant emission. This is not identical to the ``strict mask'' defined in Section~\ref{sec:products}, but it is constructed along similar lines.

\item Histograms of the pixel values in the full cube, and separate histograms for pixels inside and outside the user-supplied clean mask and mask identifying significant emission.

\item Integrated spectra, constructed by summing the full data cube, and separate spectra from summing the cube with the user-supplied clean mask and significant emission mask applied.

%(see example in Figure~\ref{fig:qa_antonio_2dhist_example})
\item Two-dimensional histograms illustrating the distribution of pixel values within each channel for the full data cube, as well as versions for the residual image and the masked and unmasked regions of the cube.

\item Power spectra of emission calculated from the individual channel maps.
\end{enumerate}

\noindent We found that this collection of plots allowed the reviewers to identify pathological data and to evaluate the performance of the deconvolution and masking.

In practice, the reports used for internal QA were generated by an IDL pipeline that ran independently of the PHANGS-ALMA data reduction pipeline. We subsequently developed a \texttt{python} version of the QA report generation tool that can be run independently or integrated into the PHANGS pipeline.%\footnote{The \texttt{python} QA report generation code is available at \href{https://github.com/PhangsTeam/phangs-alma-html-report}{https://github.com/PhangsTeam/phangs-alma-html-report}.}

\subsection{Regression Tests Against Previous Versions}

The PHANGS-ALMA imaging and production creation pipelines have been iterated several times. This included a major code revision and several substantive revisions of the imaging and product creation algorithms. For products created after the initial round of quality assurance, we could use the previous, already quality-assured imaging as a benchmark against which to compare the new products. 

For these newer products, we automatically generated the QA reports as before, but we adopted a different, less time-consuming QA strategy. We created a suite of regression tests that benchmarked each new cube and map against the equivalent, quality-assured product created by a previous version of the pipeline. The statistics used for these regression tests included: beam shape, astrometric grid parameters, key statistics describing the distribution of pixel values, integrated flux, flux above fixed intensity and signal-to-noise thresholds, and outlier-resistant standard deviation estimates. 

Using the regression tests, we checked whether each of the parameters extracted from the new and previous versions of the data were in good agreement. To define acceptable agreement, we imposed typical tolerance levels of 1 up to 20\%, depending on the parameter under consideration and our knowledge of the changes that had been implemented in the pipeline.  Following these regression tests, we then focused our manual QA efforts -- including detailed inspection of the QA reports -- on cases where the regression tests indicated significant differences between the new and old data products. We found that this approach represented an acceptable compromise between rigorously testing the impact of each change to the PHANGS-ALMA pipeline on all PHANGS-ALMA data products and overwhelming our manual QA team.

\subsection{End-to-End Test of the PHANGS Pipeline}
\label{sec:endtoend}

\begin{figure*}[ht!]
\begin{center}
\includegraphics[width=0.4\textwidth]{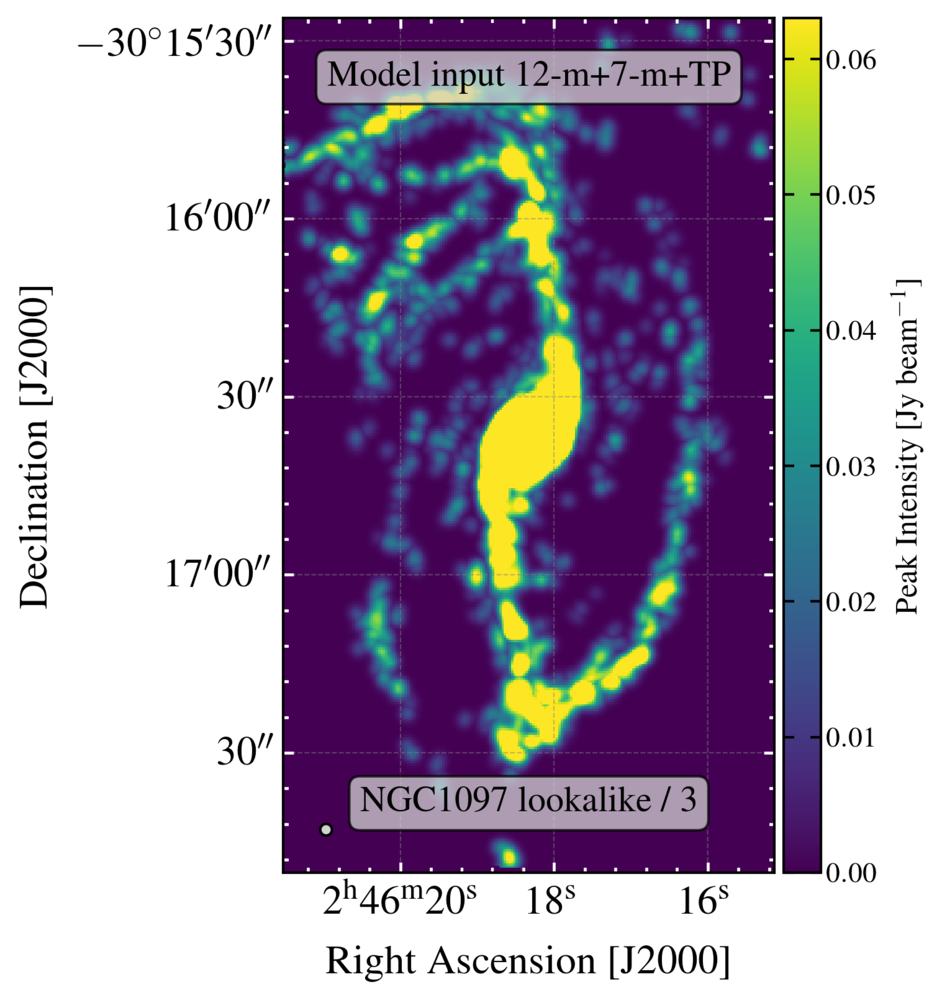}
\includegraphics[width=0.4\textwidth]{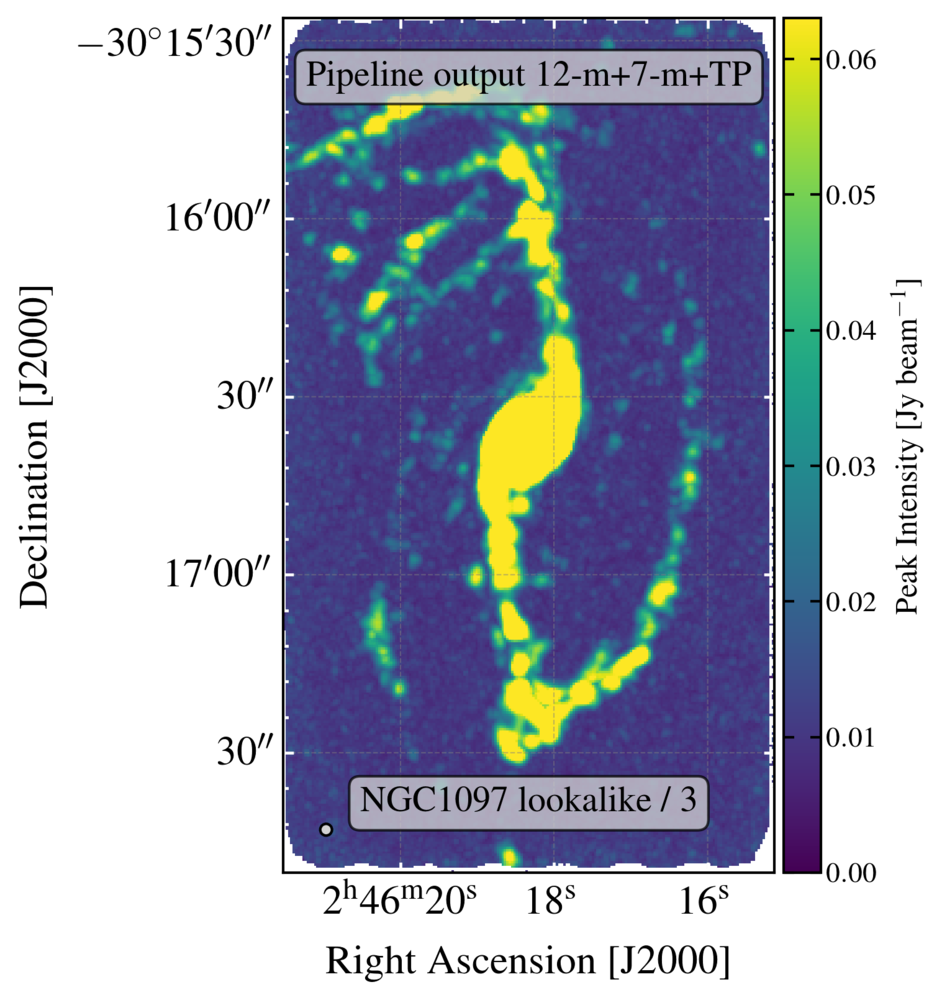}
\vspace{-24pt}
\includegraphics[width=0.4\textwidth]{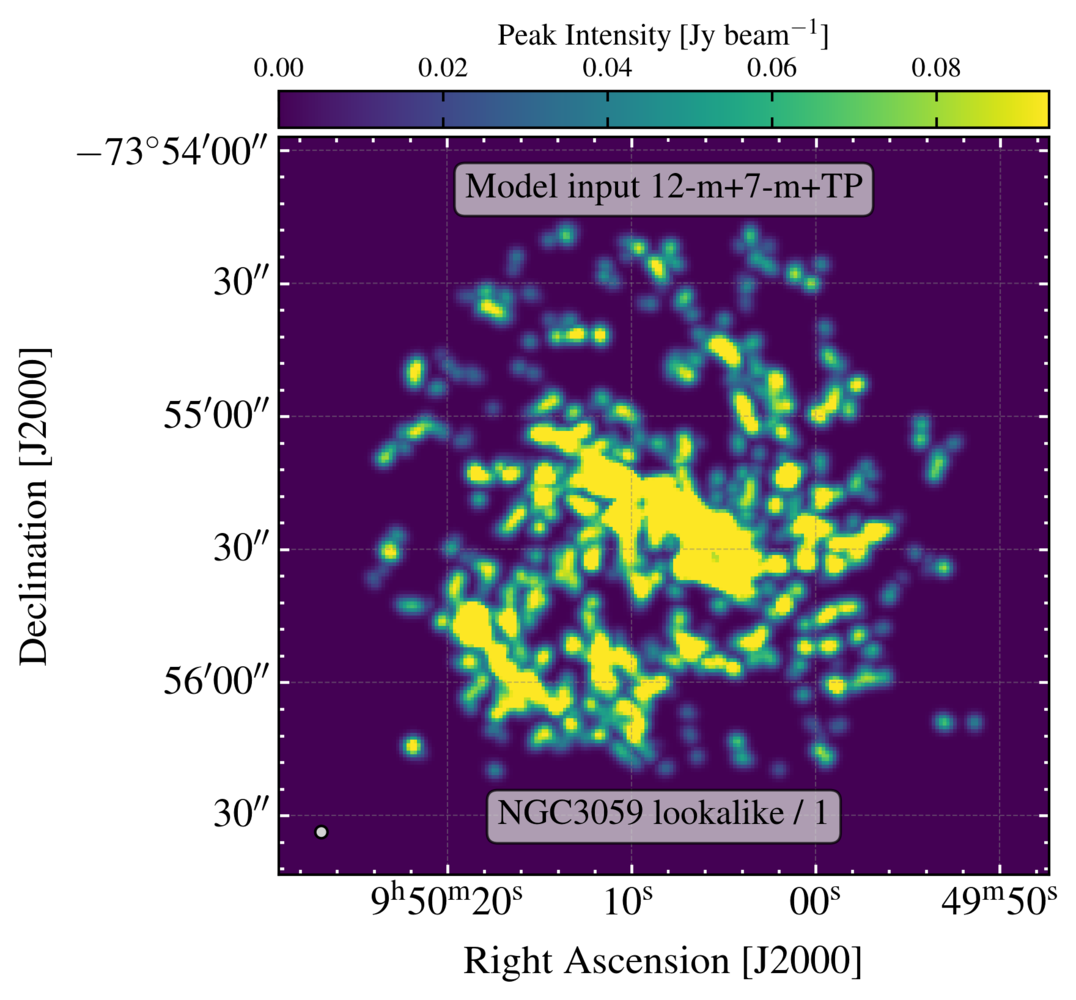}
\includegraphics[width=0.4\textwidth]{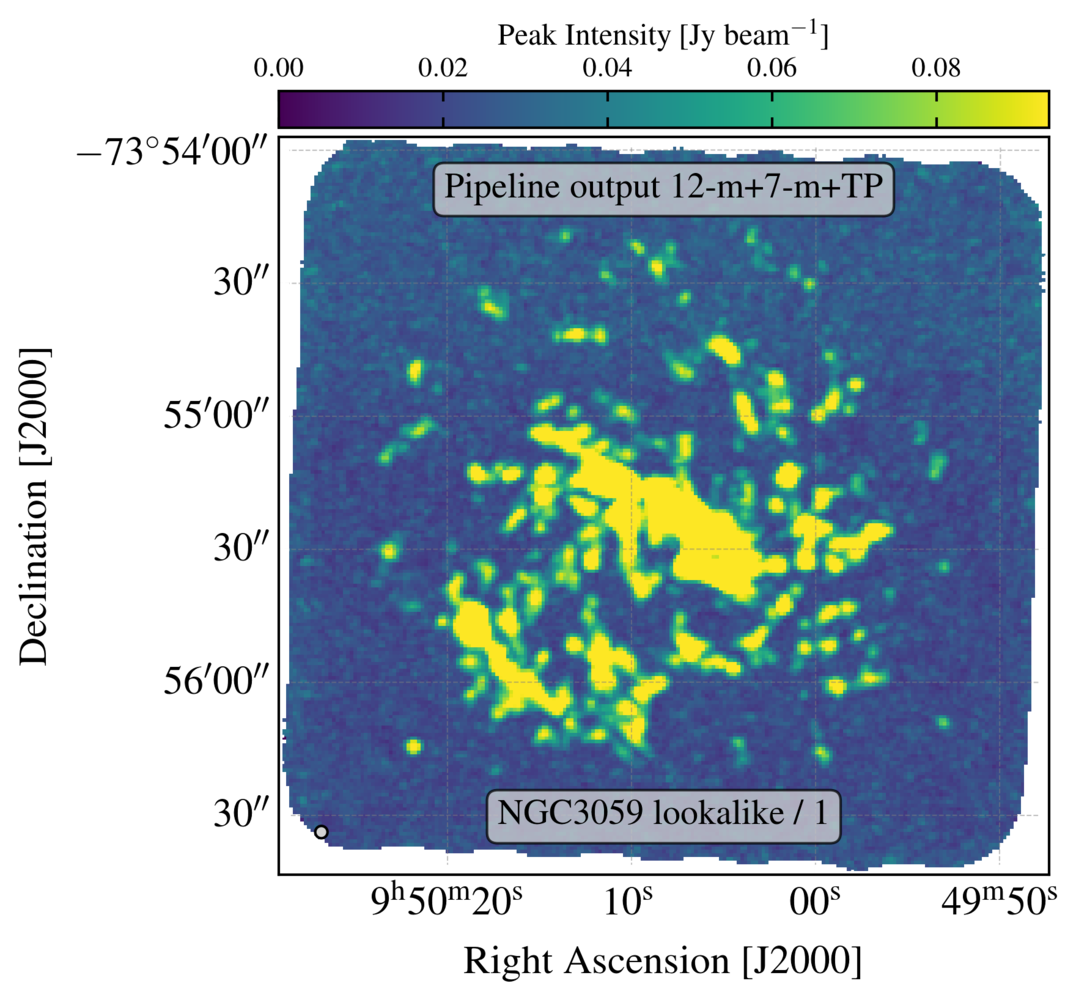}
\end{center}
\vspace{+0.2in}
\caption{
\textbf{Example results from end-to-end tests of the pipeline using simulated data.} The \textit{left} panels show peak intensity images constructed from model input for two of the cases used to test the performance of the pipeline. The \textit{right} panels show the peak intensity maps constructed from the pipeline output image using the same data. To provide a fair visual comparison, the model data have been convolved to match the resolution and astrometric grid of the pipeline output data. All panels show results for the \mbox{12-m}+\mbox{7-m}+TP data, i.e., imaging the combined \mbox{12-m}+\mbox{7-m} data and then feathering with the simulate total power data. The top row shows results for our NGC~1097 lookalike with the original model scaled down a factor of $3$ in intensity. The bottom row shows results for our NGC~3059 lookalike.
\label{fig:endtoendim}
}
\end{figure*}

\begin{deluxetable*}{lccccc}
\tabletypesize{\small}
\tablecaption{
End-to-End Imaging Tests
\label{tab:endtoend}}
\tablewidth{0pt}
\tablehead{
\colhead{Flux\tablenotemark{a}} & 
\multicolumn{5}{c}{Model} \\
\colhead{(Jy~km~s$^{-1}$)} &
\colhead{$1/1$ Model\tablenotemark{b}} &
\colhead{$1/3$ Model} &
\colhead{$1/10$ Model} &
\colhead{$1/30$ Model} &
\colhead{$1/100$ Model} 
}
\startdata
\multicolumn{6}{c}{NGC~1097 Lookalike} \\
\hline
... characteristic \mbox{7-m} model S/N\tablenotemark{c} & $94$ & $34$ & $10$ & $3.5$ & $1.1$ \\
... model & 6,087 (100\%) & 2,029 (100\%) & 609 (100\%) & 203 (100\%) & 61 (100\%) \\
... \mbox{7-m} pipeline clean & 5,931 (\phn97\%) & 1,901 (\phn94\%) & 526 (\phn86\%)& 149 (\phn73\%) & 44 (\phn69\%)\\
... \mbox{12-m}+\mbox{7-m} pipeline clean & 6,202 (102\%) & 2,087 (103\%)& 610 (100\%) & 190 (\phn94\%) & 60 (\phn94\%)\\
... \mbox{7-m}+TP pipeline image & 6,111 (100\%) & 2,045 (101\%) & 624 (102\%) & 217 (107\%) & 77 (120\%)\\
... \mbox{12-m}+\mbox{7-m}+TP pipeline clean & 6,112 (100\%) & 2,050 (101\%) & 627 (103\%)& 220 (108\%) & 80 (125\%)\\
\hline
\multicolumn{6}{c}{NGC~3059 Lookalike} \\
\hline
... characteristic \mbox{7-m} model S/N\tablenotemark{c} & $18$ & $6.1$ & $1.8$ & $0.6$ & $0.2$ \\
... model & 924 (100\%) & 308 (100\%) & 92 (100\%) & 31 (100\%) & 9.2 (100\%) \\
... \mbox{7-m} pipeline clean & 745 (\phn81\%) & 219 (\phn71\%) & 70 (\phn76\%)& 42 (135\%) &  38 (413\%)\\
... \mbox{12-m}+\mbox{7-m} pipeline clean & 941 (102\%) &  343 (111\%)& 153 (165\%) & 118 (381\%) & 111 (1200\%)\\
... \mbox{7-m}+TP pipeline image & 934 (101\%) & 318 (103\%) & 103 (112\%) & 41 (132\%) & 20 (217\%)\\
... \mbox{12-m}+\mbox{7-m}+TP pipeline clean & 937 (101\%) & 321 (104\%) & 106 (115\%)& 45 (145\%) & 23 (250\%)\\
\enddata
\tablenotetext{a}{Integrated flux calculated after continuum subtraction in the model. For the \mbox{12-m}+\mbox{7-m} and \mbox{7-m} data we report the total cleaned flux. For the feathered data we report the flux in the final cube after feathering.}
\tablenotetext{b}{See text. The nominal $1/1$ model is the version created as described in the text based on the strictly masked NGC~1097 or NGC~3059 imaging. The scaled versions reduce the intensity of all voxels by factors of $3$, $10$, $30$, and $100$.}
\tablenotetext{c}{Intensity-weighted intensity value of emission in the input model after convolution to the resolution of the \mbox{7-m}, divided by the $1\sigma$ noise in that \mbox{7-m} cube. This is a characteristic signal-to-noise value for the data and gives an indication of the brightness of emission in the cube, though the detailed brightness distribution is complex and resolution dependent.}
\tablecomments{Summary of inferred flux (model input, clean flux, integrated flux) for our five simulated \cotwo\ measurement sets.}
\end{deluxetable*}

We also tested the performance of the pipeline by applying it to simulated data. For this test, we created a series of simulated \cotwo\ measurement sets using CASA's \texttt{simdata} task. We consider two source intensity distributions, and assume the same observing conditions across all simulated observations, but we vary the overall amplitude of the signal in each input model to create a suite of data sets with differing S/N. We then ran the simulated measurement sets through the staging, imaging, and post-processing parts of the pipeline. Finally, we compared the output from the pipeline to the input model image to assess the performance of the pipeline.

\medskip

\noindent \textbf{Simulation setup:} We simulated observations of \cotwo\ emission from two sources: (1) a modified, more distant version of NGC~1097 and (2) NGC~3059 with no modification to the distance. The two cases span the range of structure that we see in the real data set. The modified NGC~1097 has compact, bright structure in each channel, partially because it displays a very strong velocity gradient and hosts strong features in the form of a compact circumnuclear ring, a strong bar, and well-defined spiral arms. NGC~3059 shows more extended structure in individual channels and a more flocculent overall structure. Reflecting this, the real data for the two targets show different results when comparing the \mbox{12-m}+\mbox{7-m} and \mbox{7-m}-only results in Appendix~\ref{sec:arrays}. In NGC~1097, we find almost no discrepancy between the two cases, while NGC~3059 shows much higher flux in the cleaned \mbox{12-m}+\mbox{7-m} compared to \mbox{7-m}-only data. We chose these targets in part to help us understand this effect (see more in Appendix~\ref{sec:arrays}).

Specifically, we produced the model data cube following these steps:

\begin{enumerate}
\item We began with the ``strictly masked'' \mbox{12-m}+\mbox{7-m}+TP \cotwo\ data cube for each galaxy. For NGC~1097, this cube combined two individual mosaics (``parts''), observed separately. NGC~3059 was observed in a single part.  

\item We rotated each image to align the major axis of the cube with the declination axis. We also resampled each cube to have channel width ${\sim}0.6$~km~s$^{-1}$, using cubic interpolation in CASA's \texttt{imregrid} to interpolate from the cube's $2.54$~km~s$^{-1}$ channels. Then we converted to units of Jy~pix$^{-1}$.

\item Only for NGC~1097, we adjusted the pixel scale of the image. In order to ensure that the fine scale structure in our input model is sharper than our observed beam, we shrunk the pixel scale of the model cube by a factor of two, i.e., a factor of four in area. This effectively places the model image at two times the real distance to NGC~1097, moving it from $13.6$~Mpc \citep{SHAYA17,ANAND21} to $27.2$~Mpc. During this step, we leave the intensity in Jy~pix$^{-1}$ unchanged. Thus the initial NGC~1097 look-alike model can be thought of as a target with twice the distance and four times the luminosity of NGC~1097. 

We do not apply any such rescaling to NGC~3059. This means that in this target we simulate observing structure that has already been convolved with the telescope beam. Because this case is intended to test the imaging performance for more extended sources, we do not view this as a problem.

\item In both targets, we added a continuum with $1/30$ the peak intensity along each line of sight. This is significantly brighter than our typical continuum, but should be removed by our continuum subtraction.

\item We also created additional versions of each model by dividing the intensity in each pixel by $3$, $10$, $30$, and $100$. Because we use a fixed simulated observing time and fixed weather conditions, these different versions correspond to cases with the same structure but different signal-to-noise values. We report characteristic signal-to-noise values for each case in Table~\ref{tab:endtoend}. In practice, the NGC~3059 cube scaled down by a factor of $100$ yields no meaningful results because the galaxy becomes too faint to be detected using our simulated observations.
\end{enumerate}

\medskip 

We simulate interferometric observations of each model image using CASA's \texttt{simobserve} task. In detail, we simulate observing for 6~hours using the ALMA Cycle~5 ACA and $1.5$~hours using the most compact Cycle~5 \mbox{12-m} configuration (i.e., \mbox{C43-1}). All simulations occurred around transit and included simulated thermal noise appropriate for 1~mm of precipitable water vapor. We allowed the simulator to place mosaic fields that would cover the target using the default spacing. For NGC~1097, the simulator placed 33 ACA \mbox{7-m} pointings and 92 \mbox{12-m} pointings. For NGC~3059, the simulator placed 67 ACA \mbox{7-m} pointings and 203 \mbox{12-m} pointings\footnote{This is slightly larger than the ALMA observatory limit of 150 pointings, but this should have no effect on the test.}.

We also created corresponding simulated single dish cubes. To do this, we convolved each model to the resolution of the single dish data using the CASA task \texttt{imsmooth}. For this step, we did not include any continuum, in order to simulate the baseline subtraction during the single dish processing (Section~\ref{sec:totalpower}). Then, we added noise to each simulated single dish cube. The noise that we added was first convolved to the resolution of the single dish data and then scaled so that the rms amplitude of the simulated noise matched the measured $1\sigma$ noise level of the real NGC~1097 single dish cube.

When these steps were finished, we had simulated $u{-}v$ data and single dish cubes for NGC~1097 and NGC~3059 lookalikes with five signal-to-noise levels (see Table~\ref{tab:endtoend}). These data resemble typical observations obtained within the PHANGS-ALMA survey. They allow us to assess the pipeline performance because they correspond to known input images.

\medskip

\noindent \textbf{Pipeline imaging:} We configured the pipeline to process the simulated data in a manner that closely followed the real PHANGS-ALMA imaging. We staged the data, subtracted the continuum, regridded, and rebinned to a data set ready for imaging. Then we imaged and cleaned each data set and applied the postprocessing steps described in Section~\ref{sec:postprocess}. These included feathering with the simulated single dish data.

Next, we convolved each model input image to the resolution of the pipeline-produced output image. Then, we reprojected the model to the astrometric and velocity grid of the pipeline-produced data. Thus, at the end of this process, we had \mbox{12-m}+\mbox{7-m}, \mbox{12-m}+\mbox{7-m}+TP, \mbox{7-m}, and \mbox{7-m}+TP pipeline-imaged, simulated images for both the NGC~1097 and NGC~3059 lookalike, each at five signal-to-noise levels.

\medskip

\begin{figure*}[ht!]
\begin{center}
\includegraphics[width=0.45\textwidth]{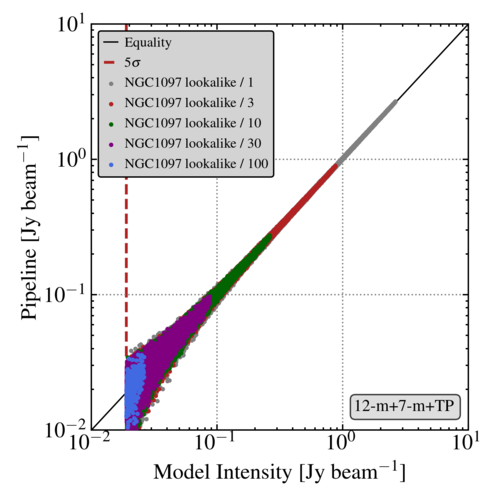}
\includegraphics[width=0.45\textwidth]{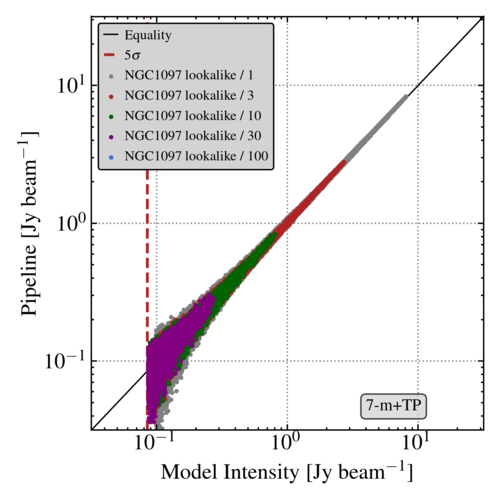}
\includegraphics[width=0.45\textwidth]{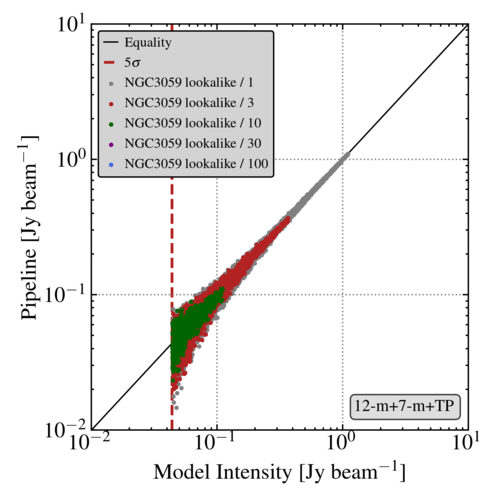}
\includegraphics[width=0.45\textwidth]{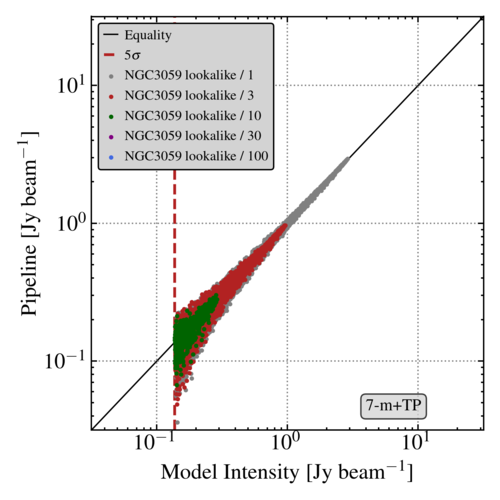}
\end{center}
\vspace{-0.2in}
\caption{
\textbf{Input model and pipeline output results.}
\label{fig:scattersim} Scatter plots showing the pipeline-imaged output ($y$-axis) as a function of the input model intensity ($x$-axis) for (\textit{top row}) the NGC~1097 lookalike models and (\textit{bottom row}) the NGC~3059 lookalike models and (\textit{left column}) \mbox{12-m}+\mbox{7-m}+TP and (\textit{right column}) \mbox{7-m}+TP results. In each case, the input model has been convolved and aligned to the resolution and astrometric grid of the output cube before comparison. The diagonal solid line shows equality, and the vertical dashed line shows a typical $5\sigma$ noise value in the simulated image. We apply this $5\sigma$ value as a threshold in the model. Different colors show individual voxels for each scale version of the input model. Overall, the pipeline results closely match the input model. We investigate the scatter more in Figures~\ref{fig:resids1} and~\ref{fig:resids2}.
}
\end{figure*}

\begin{figure*}[ht!]
\begin{center}
\includegraphics[width=0.4\textwidth]{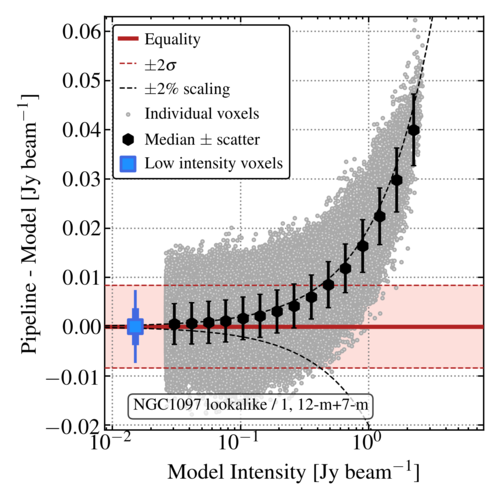}
\includegraphics[width=0.4\textwidth]{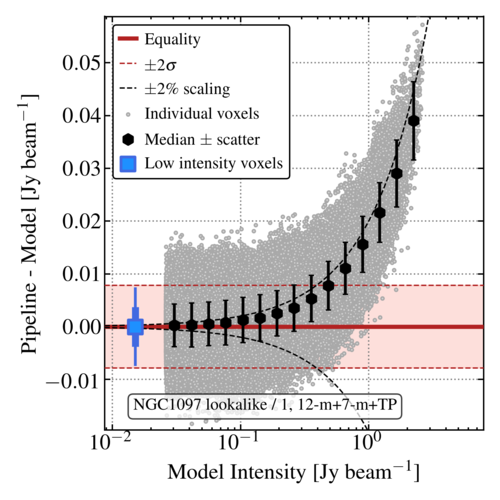}
\vspace{-24pt}
\includegraphics[width=0.4\textwidth]{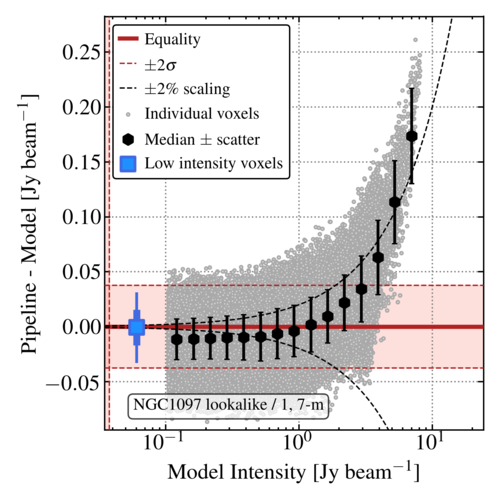}
\includegraphics[width=0.4\textwidth]{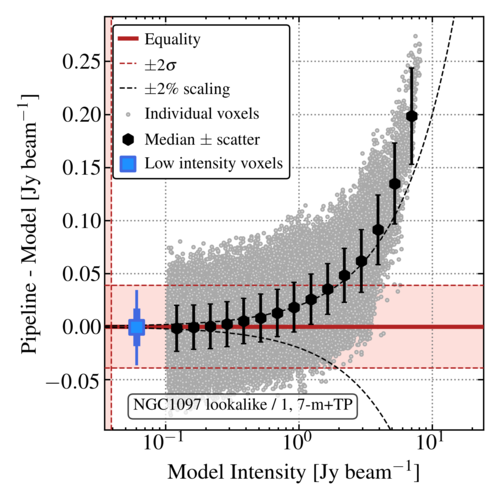}
\end{center}
\vspace{+0.2in}
\caption{
\textbf{Difference between pipeline output and model input for the NGC~1097 lookalike.} Difference between pipeline output and model input values ($y$-axis) as a function of model intensity ($x$-axis) for imaging of our NGC~1097 lookalike model. The panels show results for four cases, imaging with (\textit{top left}) the \mbox{12-m}+\mbox{7-m} arrays together; (\textit{top right}) the full \mbox{12-m}+\mbox{7-m}+TP results, i.e., \mbox{12-m}+\mbox{7-m} imaging with feathering; (\textit{bottom left}) imaging with only the \mbox{7-m} array; and (\textit{bottom right}) the full \mbox{7-m}+TP results, i.e., \mbox{7-m} imaging with feathering. Gray dots show results for those individual voxels that contribute 95\% of the total flux in the model after sorting by intensity. Black points and error bars indicate the median and $1\sigma$ scatter in the residual. The blue point shows the median, $1\sigma$ (thick error bar), and $2\sigma$ (thin error bar) for lower intensity voxels. The red lines show $2\sigma$ thermal noise in the image, and black lines show $\pm 2\%$ fractional scatter. This figure shows results running our fiducial model through end-to-end imaging tests. The fiducial model resembles a brighter version of NGC~1097 at twice the actual distance to that galaxy. Overall, the agreement between pipeline result and model is excellent, but we do see evidence of a ${\sim}2\%$ multiplicative bias such that the pipeline results appear high in all panels.
\label{fig:resids1}
}
\end{figure*}

\begin{figure*}[ht!]
\begin{center}
\includegraphics[width=0.4\textwidth]{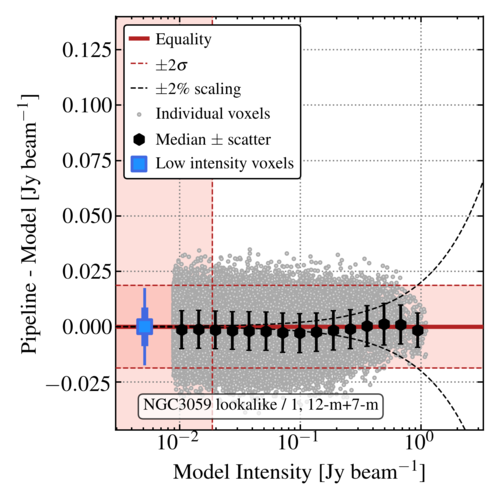}
\includegraphics[width=0.4\textwidth]{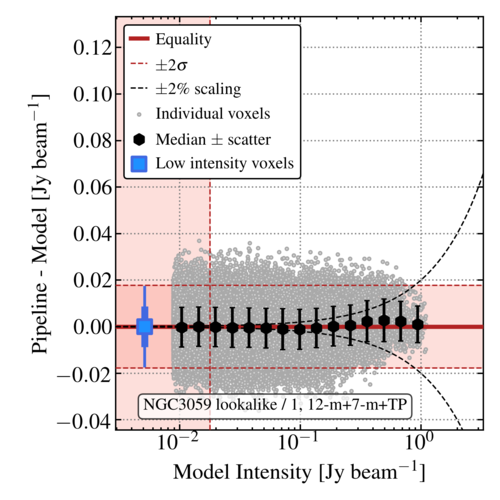}
\vspace{-24pt}
\includegraphics[width=0.4\textwidth]{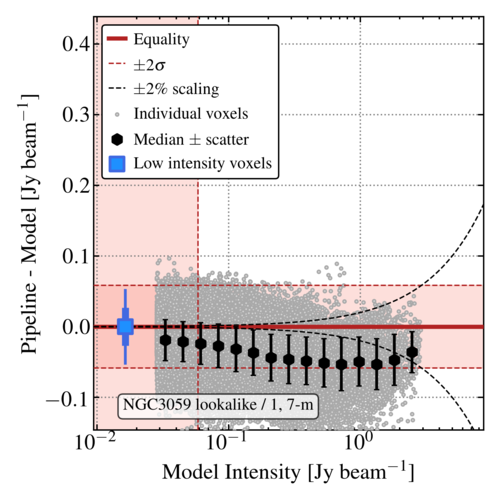}
\includegraphics[width=0.4\textwidth]{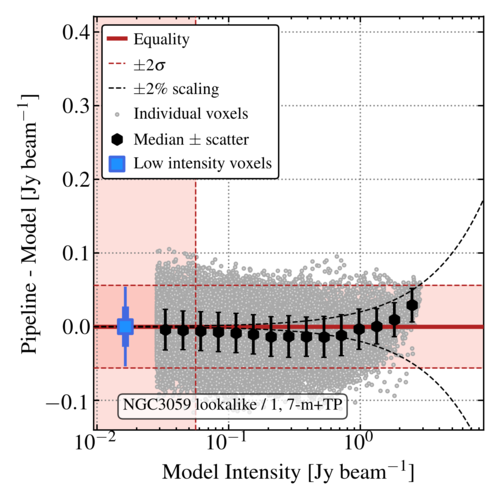}
\end{center}
\vspace{+0.2in}
\caption{
\textbf{Difference between pipeline output and model input for the NGC~3059 lookalike.} As Figure~\ref{fig:resids2}, but for our NGC~3059 lookalike, reflecting a fainter galaxy with more extended source structure. This case does not show any notable bias with the \mbox{12-m}+\mbox{7-m} data, but the \mbox{7-m}-only imaging shows a bias low. For the \mbox{12-m}+\mbox{7-m}+TP imaging, a mild bias plus thermal noise represents a good model for the small offset of the pipeline results from the model. For the \mbox{7-m}-only data, the imaging struggles to recover the model. See Appendix~\ref{sec:arrays} for more discussion.
\label{fig:resids2}
}
\end{figure*}

\noindent \textbf{Results:} Figure~\ref{fig:endtoendim} shows the peak intensity from the beam-matched, aligned, input models and the output from the pipeline. We plot results for the \mbox{12-m}+\mbox{7-m}+TP imaging of our NGC~1097 lookalike scaled down by a factor of three and our brightest NGC~3059 lookalike. In both cases, the imaging shows excellent recovery of the detailed features and large-scale morphology of the input image.

Table~\ref{tab:endtoend} and Figures~\ref{fig:scattersim}, \ref{fig:resids1}, and~\ref{fig:resids2} show these results in more detail. Figure~\ref{fig:scattersim} shows the most basic result, the scaling between input model image ($x$-axis) and PHANGS pipeline output image ($y$-axis) for the \mbox{12-m}+7-{m}+TP and \mbox{7-m}+TP data for both source models and all scalings. Before the comparison we match the resolution and astrometric grid of the model to that of the pipeline output cube. We restrict the comparison to regions that are above $5\times$ the noise in the observed cube in the matched model image (dashed red line). Overall the figure demonstrates excellent performance of the pipeline, with data from all models lying almost exactly along the line of equality. Note that for the two lowest-brightness versions of the NGC~3059 lookalike, the signal-to-noise drops to such low levels that the deconvolution cleans only noise. These imaging cases essentially fail because the simulated observations are not deep enough to see the galaxy.

Of course, the pipeline images do not perfectly match the input model, and Figures~\ref{fig:resids1} and~\ref{fig:resids2} explore the offset between the input and output in more detail. These figures plot the difference between the pipeline and model for individual voxels in the cubes. We show results for the brightest version of the NGC~1097 lookalike (Figure~\ref{fig:resids1}) and the brightest version of the NGC~3059 lookalike (Figure~\ref{fig:resids2}). In each panel, we plot individual (gray) and binned (black) data for the voxels that account for 95\% of the emission in the model image. That is, we construct a CDF of the flux as a function of intensity, and show results for the top 95\%. The exact 95\% value of the threshold is arbitrary, we only need some cut to select regions of interest where the model is positive. The figure also shows the median and scatter for the remaining pixels with a blue point. We indicate the $2\sigma$ level of the statistical, predominantly thermal noise in the data cube using red lines and shading.

Both figures show the same good agreement seen in Figure~\ref{fig:scattersim}. Differences between the model and the pipeline output remain small compared to the intensity value in the cube. We do find low-level systematic deviations, however. As expected, the thermal noise contributes to the scatter, this provides the main explanation for the width of the distributions in Figures~\ref{fig:scattersim}, \ref{fig:resids1}, and~\ref{fig:resids2}. 
We observe a modest positive bias in our pipeline imaging of NGC~1097 but not NGC~3059. The sense of the bias is that the pipeline yields results that are biased high relative to the model by ${\sim}2\%$. To see this, compare the binned results (black points) to the dashed black lines, which illustrate the case where the model has been multiplied (upward curve) or divided (downward curve) by $1.02$. The binned data match the upward-curved line well, indicating that to first order a 2\% positive multiplicative bias and thermal noise provide a good description of the pipeline's output compared to the input model.

The bottom left panels in both figures show the \mbox{7-m}-only imaging results. For both input models, the \mbox{7-m}-only imaging shows a negative offset at a wide range of intensities. For NGC~1097 this offset appears mild, but for NGC~3059 the pipeline image is ${\sim}1{-}2\sigma$ lower than the input model over a wide range of intensities. This reflects the poorer imaging performance of the \mbox{7-m}-only data compared to the \mbox{12-m}+\mbox{7-m} data discussed in Appendix~\ref{sec:arrays}. As mentioned above, the difference between the \mbox{7-m} imaging performance for the two targets is expected. For the real NGC~1097 imaging, the \mbox{12-m}+\mbox{7-m} imaging and \mbox{7-m} imaging agree well and little flux is lost to spatial filtering.  Meanwhile for NGC~3059, the real data show significant differences between the \mbox{12-m}+\mbox{7-m} and \mbox{7-m} images.

We also report the total fluxes in the model, the cleaned images, and the feathered images for each case in Table~\ref{tab:endtoend}. For all of the feathered cases, the pipeline matches the model input within the uncertainty expected from directly summing the total power cube\footnote{Note that for expediency we used the same noise realization for all TP simulations, so they are all biased high relative to the model by roughly the same $+20$~Jy~km~s$^{-1}$.}. This must be the case: the feathering operation will fix the integral of the cube to match the input simulated total power data, which in turn is simply the model galaxy plus noise. 

Table~\ref{tab:endtoend} also gives insight into the performance of the deconvolution. For both the NGC~1097 and NGC~3059 lookalike, the \mbox{12-m}+\mbox{7-m} imaging recovers results within $2{-}3$\% of the model flux, even before feathering. For lower brightness versions of the NGC~1097 lookalike, the \mbox{12-m}+\mbox{7-m} imaging continues to reconstruct almost all of the emission. For the NGC~3059 lookalike, as mentioned above, the imaging fails for the two faintest cases, which have typical \mbox{7-m} signal-to-noise levels of ${\sim}0.6$ and $0.2$. 

The situation with the \mbox{7-m} data is more mixed. In the NGC~1097 lookalike, the deconvolution recovers 94\%, 86\%, 73\%, and 72\% of the model emission for the models scaled down by factors of 3, 10, 30, and 100. For the NGC~3059 lookalike, the situation is even worse, with the \mbox{7-m}-only imaging recovering only $81\%$ and $71\%$ of the total flux in the two brightest cases. In short, for low signal-to-noise ACA-only data sets, the PHANGS pipeline struggles to achieve a full deconvolution of the \mbox{7-m}-only data. In our end-to-end tests, the PHANGS pipeline ACA-only imaging misses $20{-}30\%$ of the flux in the worst cases. We explore this issue with both simulations and real data in Appendix~\ref{sec:arrays}.

\subsection{Comments on quality assurance results}

Both the automated regression and the manual quality assurance tests played an important role in refining the PHANGS pipeline algorithms and catching several important bugs. In the final round of imaging using the latest pipeline, we still found a few cases where the imaging with the default parameters diverged or declared convergence too early (${\sim}4$ out of ${\sim}250$ total cases) in one or more planes. In these cases, we adjusted the pipeline parameters and reran the imaging. Usually, adjusting the convergence criteria or the primary beam cutoff improved the situation.

%% file: summary.tex
\section{Summary} 
\label{sec:summary}

We have presented the PHANGS-ALMA data processing pipeline, explaining the key steps in the processing and our motivation for many of the key decisions. We do not review these here but highlight a few points that may be of general interest to those working on similar problems:

\begin{enumerate}
\item We note that issues related to regridding and interpolation can lead to patterns in the rms noise amplitude along the spectral dimension. The specific issue that we highlight is related to CASA, and will hopefully be addressable in future releases, but the concern that data processing affects the spectral noise pattern and line spread function is general.

\item We present a robust two-stage approach to deconvolving spectral line observations. This approach employs a multiscale deconvolution down to a signal-to-noise threshold of around four. Then it creates a mask based on bright signal and uses a classic \citet{HOGBOM74} deconvolution approach to clean the bright emission ``into the noise.'' We have found this to run robustly and yield good results on a wide variety of nearby galaxy data.

\item We adopt a two-track approach to masking of spectral line data cubes. We create a high confidence, low false-positive, but potentially low completeness ``strict'' mask. This is appropriate for calculations that perform poorly in the presence of noise. We also create a high completeness but noisier ``broad'' mask that will include many false positive but also encompass almost all emission in the cube. Leaving aside the specifics of their creation, we suggest that this two-track approach to masking is a good general approach.

\item We have found that a ``matched-line width filtered peak temperature map'' does an outstanding job of highlighting detailed structure in line data cubes. This is simply a conventional peak temperature (sometimes referred to as ``moment~8'') map constructed from a cube that has been convolved spectrally with a matched-line width filter. 

\item We have compared the imaging results for different arrays and vetted the performance of our pipeline using simulated observations with known input. These tests show that, after the inclusion of total power data, the pipeline does an excellent job of recovering known input. They also show that the \mbox{12-m+7-m} imaging performs significantly better than \mbox{7-m} only imaging in many cases, even after matching the resolution of the output images. The differences, which are explored in detail in the appendix, are a function of signal-to-noise and source structure.

\end{enumerate}

The PHANGS-ALMA pipeline has so far been applied successfully to roughly $1{,}000$ individual measurement sets and is publicly available on \texttt{github}.

\clearpage

%% file: acknowledgments.tex
\acknowledgments
We gratefully acknowledge a prompt and constructive review by the anonymous referee during a difficult time. This work also benefited immensely from helpful discussions with Crystal Brogan, Jeffrey Mangum, and the North American ALMA Science Center and European Southern Observatory staff, including Dirk Petry. The computing and software infrastructure used to process PHANGS--ALMA and develop this pipeline at OSU was built and supported by David Will, who for years was the most supportive and welcoming person in a supportive and welcoming department. He will be sorely missed.

This work was carried out as part of the PHANGS collaboration.
The work of A.K.L., J.S., and D.U. is partially supported by the National Science Foundation (NSF) under Grants No.1615105, 1615109, and 1653300, as well as by the National Aeronautics and Space Administration (NASA) under ADAP grants NNX16AF48G and NNX17AF39G.
ER acknowledges the support of the Natural Sciences and Engineering Research Council of Canada (NSERC), funding reference number RGPIN-2017-03987, and computational support from Compute Canada.
D.L., T.S., E.S., C.M.F., K.S., and T.G.W. acknowledge funding from the European Research Council (ERC) under the European Union’s Horizon 2020 research and innovation programme (grant agreement No. 694343).
CH, AH,and JP acknowledge support by the Programme National “Physique et Chimie du Milieu Interstellaire” (PCMI) of CNRS/INSU with INC/INP co-funded by CEA and CNES. AH acknowledges support by the Programme National Cosmology et Galaxies (PNCG) of CNRS/INSU with INP and IN2P3, co-funded by CEA and CNES. AU and AG-R acknowledge support from the Spanish funding grants AYA2016-79006-P (MINECO/FEDER) and PID2019-108765GB-I00 (MICINN). AU acknowledges support from the Spanish funding grant PGC2018-094671-B-I00 (MCIU/AEI/FEDER).
C.M.F. acknowledges support from the NSF under Award No. 1903946.
M.C. and J.M.D.K. gratefully acknowledge funding from the German Research Foundation (DFG) through an Emmy Noether Research Group (grant number KR4801/1-1). M.C., J.M.D.K., and J.J.K. gratefully acknowledge funding from the DFG Sachbeihilfe (grant number KR4801/2-1). J.M.D.K. gratefully acknowledges funding from the European Research Council (ERC) under the European Union's Horizon 2020 research and innovation programme via the ERC Starting Grant MUSTANG (grant agreement number 714907).
FB, ATB, IB, JdB, JP acknowledge funding from the European Union’s Horizon 2020 research and innovation programme (grant agreement No 726384/EMPIRE).
R.S.K.\ and S.C.O.G.\ acknowledge financial support from the DFG via the collaborative research center (SFB 881, Project-ID 138713538) ``The Milky Way System” (subprojects A1, B1, B2, and B8). They also acknowledge subsidies from the Heidelberg Cluster of Excellence {\em STRUCTURES} in the framework of Germany’s Excellence Strategy (grant EXC-2181/1 - 390900948) and funding from the ERC via the ERC Synergy Grant {\em ECOGAL} (grant 855130).
KK and FS gratefully acknowledge funding from the DFG in the form of an Emmy Noether Research Group (grant number KR4598/2-1). EW acknowledges support from the Deutsche Forschungsgemeinschaft (DFG, German Research Foundation) -- Project-ID 138713538 -- SFB 881 (``The Milky Way System'', subproject P2). CE acknowledges funding from the Deutsche Forschungsgemeinschaft (DFG) Sachbeihilfe, grant number BI1546/3-1. A.S. is supported by an NSF Astronomy and Astrophysics Postdoctoral Fellowship under award AST-1903834.

This paper makes use of the following ALMA data, which have been processed as part of the PHANGS-ALMA \cotwo\ survey: \\

ADS/JAO.ALMA\#2012.1.00650.S, \linebreak % (N628/M74)
ADS/JAO.ALMA\#2013.1.00803.S, \linebreak % (N5128/CenA)
ADS/JAO.ALMA\#2013.1.01161.S, \linebreak % (N1365 + N5236/M83)
ADS/JAO.ALMA\#2015.1.00121.S, \linebreak % (N5236/M83)
ADS/JAO.ALMA\#2015.1.00782.S, \linebreak % (N1313 + N7793)
ADS/JAO.ALMA\#2015.1.00925.S, \linebreak % (pilot low mass)
ADS/JAO.ALMA\#2015.1.00956.S, \linebreak % (pilot high mass)
ADS/JAO.ALMA\#2016.1.00386.S, \linebreak % (N5236/M83)
ADS/JAO.ALMA\#2017.1.00392.S, \linebreak % (low mass follow-up)
ADS/JAO.ALMA\#2017.1.00766.S, \linebreak % (early-type)
ADS/JAO.ALMA\#2017.1.00886.L, \linebreak % (large program)
ADS/JAO.ALMA\#2018.1.01321.S, \linebreak % (N253, N300, Circinus)
ADS/JAO.ALMA\#2018.1.01651.S. \linebreak % (main sample follow-up)
ADS/JAO.ALMA\#2018.A.00062.S. \linebreak % (ACA-only nearby)
ALMA is a partnership of ESO (representing its member states), NSF (USA), and NINS (Japan), together with NRC (Canada), NSC and ASIAA (Taiwan), and KASI (Republic of Korea), in cooperation with the Republic of Chile. The Joint ALMA Observatory is operated by ESO, AUI/NRAO, and NAOJ. The National Radio Astronomy Observatory is a facility of the National Science Foundation operated under cooperative agreement by Associated Universities, Inc.

\software{
\texttt{ALMA Calibration Pipeline} (L. Davis et al. in preparation),
\texttt{CASA} \citep{MCMULLIN07},
\texttt{numpy} \citep{NUMPY2006}, 
\texttt{scipy} \citep{SCIPY2020}, 
\texttt{astropy} \citep{ASTROPY1,ASTROPY2} 
IDL Astronomy User's Library \citep{IDLASTRO}, 
\texttt{cprops} \citep{ROSOLOWSKY06},
GILDAS \citep{PETY2005},
R \citep{RMANUAL},
PHANGS-ALMA Total Power Pipeline \citep{HERRERA20},
\texttt{spectral-cube},
\texttt{radio-beam}
}

%% file: contributions.tex
\section{Contributions}
\label{sec:contrib}

The processing of the PHANGS-ALMA data and creation of the pipeline was a team effort, with major contributions from many people and input from the entire team. This paper also reflects major direct and indirect contributions from many people. We summarize some of the key contributions here.

\smallskip

\noindent \textbf{The PHANGS ALMA Data Reduction (ADR) Group:} The group has been led by J.~Pety since the beginning of the PHANGS collaboration and met weekly for most of the time since 2016. Key contributors to tests, discussions, and development over the course of the project include: M.~Chevance, C.~Faesi, C.~Herrera, A.~Hughes, A.~Hygate, D.~Liu, A.~Leroy, T.~Saito, E.~Rosolowsky, E.~Schinnerer, K.~Sliwa, A.~Schruba, A.~Usero.

\smallskip

\noindent \textbf{Interferometric and Post-processing Pipeline:} The code was mostly developed by A.~Leroy, D.~Liu, E.~Rosolowsky, and T.~Saito with code review at several stages by A.~Schruba and key input from J.~Pety, E.~Schinnerer, A.~Schruba, and A.~Usero. Additional tests related to many aspects of data processing were carried out by D.~Liu and T.~Saito. E.~Koch and C.~Wilson offered important input on algorithms. Tests of short spacing correction algorithms were led by K.~Sliwa during the pilot programs and then T.~Saito during the Large Program and beyond. T.~Saito led the research described in Appendix \ref{sec:ssc} of this paper. A.~Leroy and T.~Saito led the work described in Appendix \ref{sec:arrays} with major input from A. Hughes, J. Pety, E. Rosolowsky, and E. Schinnerer. The pipeline was deployed for PHANGS-ALMA by A.~Leroy and T.~Saito.

\smallskip

\noindent \textbf{Total Power Pipeline:} C.~Herrera developed most of the total power pipeline, with major input from J.~Pety, and A.~Usero and code review by E.~Rosolowsky. K.~Sliwa played a key role in prototyping approaches to the total power processing using the pilot data. A.~Usero developed and deployed the telluric ozone correction algorithm described in Section \ref{sec:totalpower} and led investigation and communication of this issue in close collaboration with C.~Faesi, C.~Herrera, and J.~Pety. A.~Usero also led investigation of the calibration stability and gain in the total power data, described in Sections \ref{sec:uvquality} and \ref{sec:tp-cal}. A.~Weiss, J.~Pardo, and C.~de Breuck also provided important input on this topic. A.~Usero led investigation of the flux stability of the total power observations. The total power pipeline was deployed on PHANGS-ALMA by C.~Faesi, C.~Herrera, and A.~Usero.

\smallskip

\noindent \textbf{Quality Assurance (Cubes):} A.~Hughes developed the IDL version of the quality assurance software for cubes described in Section \ref{sec:quality}. D.~Liu wrote the  python version. A. Hughes led regression testing and coordinated cube quality assurance efforts at several stages. A. Leroy and T. Saito developed and deployed the end-to-end tests described in Section \ref{sec:endtoend}. Many members of the team contributed careful review of data products, including: I.~Beslic, M.~Chevance, J. den Brok, C.~Eibensteiner, C.~Faesi, A. Garc\'{i}a-Rodr\'{i}guez, C.~Herrera, A.~Hygate, M.~Jimenez~Donaire, J.~Kim, A.~Leroy, D.~Liu, J.~Pety, J.~Puschnig, M.~Querejeta, E.~Rosolowsky, T.~Saito, E.~Schinnerer, A.~Schruba, A.~Sardone, J.~Sun, A.~Usero, D.~Utomo, T.~Williams.

\smallskip

\noindent \textbf{Quality Assurance (Visibility Data):} T.~Saito developed the $u{-}v$ quality assurance procedures and software described in Section \ref{sec:uvquality}. T.~Saito and C.~Herrera carried out most of the inspection using these tools. D. Liu carried out the analysis of flux calibration scale described in Section \ref{sec:uvquality}.

\smallskip

\noindent \textbf{Infrastructure:} E.~Rosolowsky created and maintained the PHANGS server and shared archive, which was crucial to distributing the work and results. D.~Will created and maintained the computing and software environments used for building and executing the pipeline at OSU.

\smallskip

\noindent \textbf{Observatory and Community Support:} The Joint ALMA Observatory and North American ALMA Science Center offered extensive support. They were responsive and flexible regarding re-observations of total power data affected by the telluric ozone feature (Section \ref{sec:totalpower}) and responsive and helpful when issues related to imaging in \texttt{CASA} arose early in the project. We specifically acknowledge helpful communication with Anthony Remijan regarding imaging, Crystal Brogan regarding the noise behavior in Band 6, and Jeffrey Mangum regarding several aspects of data processing. More broadly, this work builds on the hard work of the \texttt{CASA} team, the ALMA pipeline team, and the ALMA observatory effort to provide excellent quality assurance. We acknowledge the hard work of the developers, the scientists who support and guide the effort, and the data analysts. Similarly, we build on the large work by the \texttt{astropy} and broader scientific \texttt{python} community and also acknowledge the astronomical IDL community, which laid the foundation for much of this work.

%% file: tpinternalcal.tex
\begin{figure}
\centering
\includegraphics[width=0.6\textwidth]{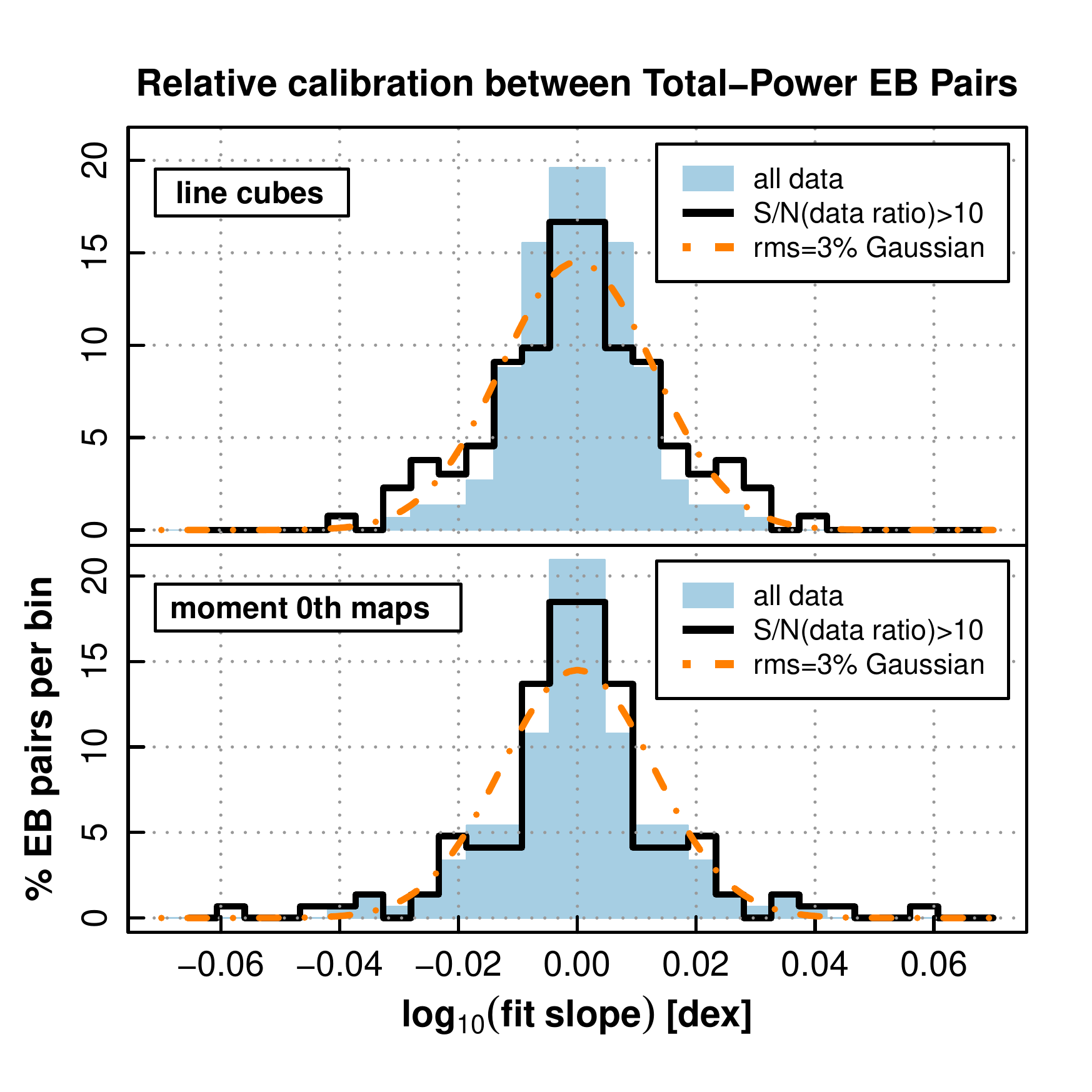}
\vspace{-18pt}
\caption{\textbf{Assessment of the internal calibration of ALMA's total power observations.} Each histogram illustrates one of the four regression tests described in the text that we run on $74\times2$ EB pairs. The top and bottom panels show the two tests on line cubes and moment~0 maps, respectively. The blue filled histograms correspond to tests where all valid data are taken. The black histogram correspond to tests including only data where the nominal S/N in the ratio between the two measurements is $>10$. The orange dot-dashed curve is a reference Gaussian distribution with $3\%$ rms ($0.013$~dex).
\label{fig:tp-cal}}
\end{figure}

\section{Internal stability of the calibration for the PHANGS-ALMA total power data}
\label{sec:tp-cal}

The total power observations set the overall flux in our final data cubes. Our processing assumes that both the total power and interferometric data are correctly calibrated (Section~\ref{sec:feather} and Appendix~\ref{sec:ssc}). The observatory calibration scheme anchors the amplitude calibration of the total power data to interferometric observations with the \mbox{7-m} array, and so to the ALMA calibrator database. In principle this should yield stable, high quality calibration. 

To check this, we assessed the internal consistency of the PHANGS-ALMA total power observations. Over the course of our ALMA Large Program, we observed the same targets repeatedly, and the individual observations are already deep enough that they detect most galaxies at high significance. This allows us to compare how the line brightness towards the same target varies when measured on different days. The magnitude of these variations gives us an upper limit to the stability of the ALMA total power flux calibration.

For this test, we selected a subset of six galaxies. Using the procedures described in \citet{HERRERA20} and Section~\ref{sec:totalpower}, we generated a \cotwo\ line cube for every ``execution block'' (EB), each of which corresponds to an individual ${\sim}1$~hour long observation. This resulted in 31 independent data cubes, with $N=2{-}7$ cubes per galaxy. Each target galaxy was observed on $2{-}3$ different dates, with the spread between observations spanning from 4~days to 11~months. Typically two consecutive EBs were observed on any given day.  

For each galaxy, we consider every possible pair of EBs. For each pair, we fitted a linear function with no intercept to the scatter plot of intensities measured at the same voxel position in the two data cubes. For the fit, we used a total-least-squares linear regression scheme, taking into account the noise level in both maps or cubes. Because this exercise requires comparing detected emission between the two EBs, we only consider velocities within the galaxy's velocity range. 

As a crosscheck, we repeated the exercise using the integrated intensity (moment~$0$) maps obtained by integrating over the galaxy's velocity range. We also carried out fits both using all data and restricting to data with ${\rm S/N}>10$ in the ratio $Y/X$, where $Y$ and $X$ refer to the intensities or integrated intensities in the two data sets. We also repeated each comparison considering $X|Y$ and $Y|X$, i.e., swapping which data set we treat as the reference variable, in order to prevent any biases from the way EBs are processed. Thus we ended up with four different slope measurements: maps and cubes, each with and without a S/N cut. In total, we compared $148$ EB pairs and also considered each with $Y|X$ and $X|Y$.

Figure~\ref{fig:tp-cal} shows the distribution of fit slopes from each of our four tests. We expect a slope of unity ($\log_{10} \mathrm{fit~slope} = 0$) if the calibration is identical between the two EBs. Departures from this capture the variations in the relative calibration between the two EBs. We report results in log scale because it is natural to expect calibration uncertainties to be multiplicative. 

The rms of the distribution of fit slopes shown in Figure~\ref{fig:tp-cal} is $0.010{-}0.017$~dex ($\approx 2{-}4\%$), regardless of how we calculate it: directly or by fitting the histograms with a Gaussian function. Although the histograms hint at some low-level wings, almost all the slope measurements deviate by less than ${\sim}7\%$ ($0.03$~dex) from unity. We adopt a $3\%$ rms scatter on the EB-to-EB scaling as a reasonable description of our results. We did not find any convincing indication that the slopes depend on the difference between the observing dates of the EB pairs. This suggests that the calibration uncertainties affecting different EBs are mutually uncorrelated.

We interpret this scatter as indicative of the stability of the ALMA total power calibration. It will not reflect underlying uncertainties in the ALMA calibrator database, but most other uncertainties should be captured by this test. Assuming that calibration uncertainties affecting different EBs are uncorrelated, the uncertainty in individual EBs is about $\sqrt{2}$ times lower than the measured $0.03~\mathrm{dex} \approx 3\%$ difference between two independent EBs. 

The final total power cube for each galaxy typically results from averaging several EBs. If the calibration uncertainty is uncorrelated, this would reduce the uncertainty from $3\%$ by an additional $\sqrt{N}$ factor, where $N$ is the number of EBs. Thus we conclude the internal calibration of the PHANGS-ALMA data is robust at the ${\sim}1 \%$ level. Given the link to the interferometric calibration scheme, this experiment also bolsters our confidence in adopting the observatory-provided calibration without rescaling when combining the total power and interferometric data.

\medskip

\noindent \textbf{A note on absolute calibration for PHANGS-ALMA ``version~4'':} As part of our quality assurance, we examined the stability of the total power antenna gain. This is the observatory-provided number used to translate from the ``chopper wheel''-based Kelvin scale to an absolute flux scale. It is expressed in units of Jansky-per-Kelvin, or Jy-per-K. We found that the observatory-provided Jy-per-K of individual observations varied systematically by ${\sim}7\%$ based on delivery date. Consultation with the ALMA observatory revealed that this reflects delivery of some incorrect gain values during the time period 2017--2018. Surface improvements to the total power telescopes improved the gain, but there was some lag in reflecting these improvements within the delivered products. The correction for this effect is straightforward, but requires reprocessing the data from the original execution block stage, and so will not be reflected in PHANGS-ALMA's initial public delivery, ``version~4.'' Taking into account the averaging of multiple blocks, we estimate that this effect implies a $2{-}5$\% bias high for the overall flux scale of the data set for data delivered during 2018. The issue was not severe enough to disrupt the internal stability tests described here, and we expect it to be addressed in future releases. 

%% file: arrays.tex
\section{Relative Performance of \mbox{7-m} and \mbox{12-m}+\mbox{7-m} Imaging}
\label{sec:arrays}

\begin{figure*}[ht!]
\begin{center}
\includegraphics[width=0.475\textwidth]{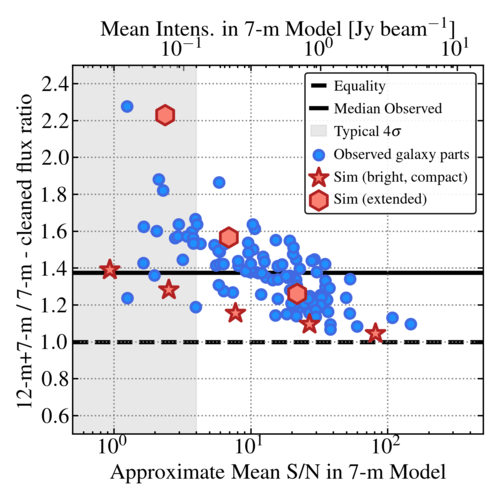}
\includegraphics[width=0.475\textwidth]{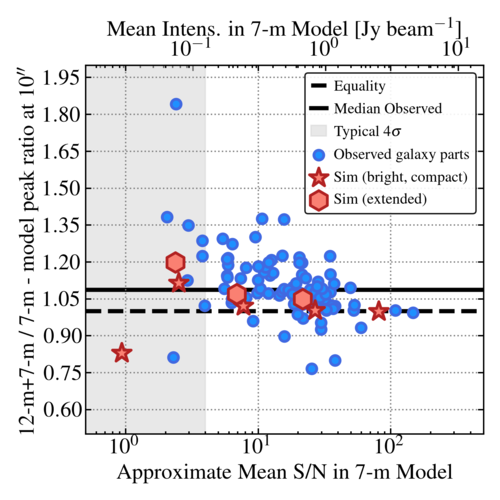}
\end{center}
\vspace{-0.1in}
\caption{
\textbf{Deconvolved flux in different arrays.} (\textit{left}) Ratio of cleaned flux between the \mbox{12-m}+\mbox{7-m} and \mbox{7-m}-only array images of the same galaxy ($y$-axis) as a function of the typical signal-to-noise ratio of deconvolved emission in the 7-m model image ($x$-axis; see text). (\textit{right}) Ratio of intensity at the peak of the 12-m+\mbox{7-m} model to intensity at the same location in the \mbox{7-m} model, after convolving both to a matched $10\arcsec$ resolution. Again, we plot this as a function of the approximate typical signal-to-noise in the \mbox{7-m} model image. Blue points in both panels show PHANGS-ALMA targets. Red points show results for the two sets of simulations described in Section~\ref{sec:endtoend}, with the bright, compact NGC~1097-based galaxy appearing as a star and the more extended NGC~3059-based galaxy as a hexagon. The \mbox{12-m}+\mbox{7-m} imaging recovers more flux than the \mbox{7-m}-only imaging, and the difference depends on the typical signal-to-noise of the emission. Both the integrated flux and the peak flux show this effect, but the effect is much stronger with regards to the integrated flux.}
\label{fig:arraydeconvolve}
\end{figure*}

We image most of our targets using both the \mbox{7-m}-only and the combined \mbox{12-m}+\mbox{7-m} data. In Section~\ref{sec:endtoend} we do the same with simulated galaxies in order to test the performance of the pipeline. These tests consistently show that the \mbox{7-m}-only imaging tends to deconvolve less flux than the \mbox{12-m}+\mbox{7-m} imaging. The effect appears strongest for extended sources and at modest signal-to-noise.

\begin{figure*}[ht!]
\begin{center}
\includegraphics[width=0.475\textwidth]{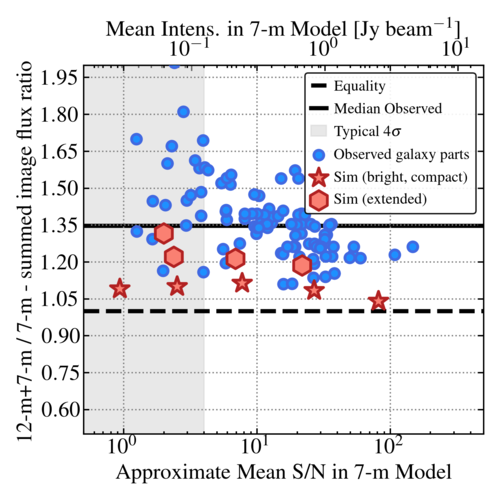}
\includegraphics[width=0.475\textwidth]{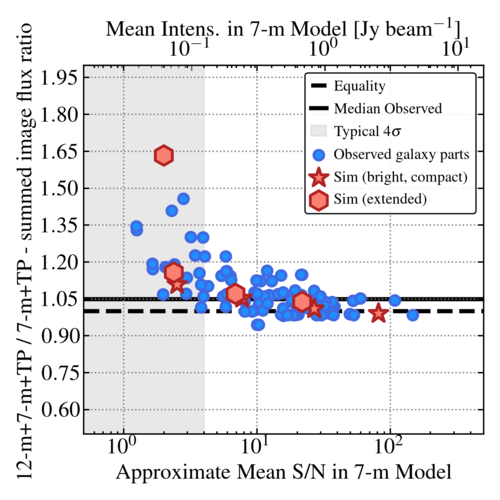}
\includegraphics[width=0.475\textwidth]{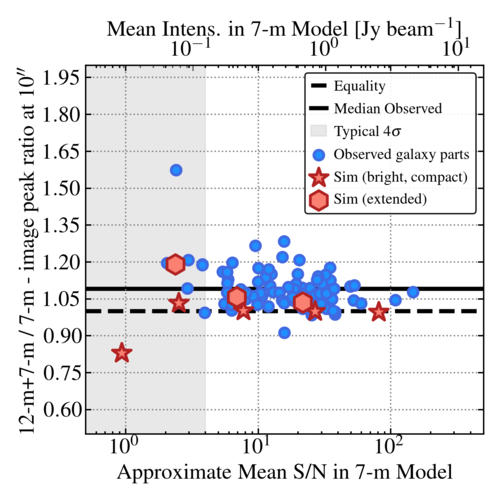}
\includegraphics[width=0.475\textwidth]{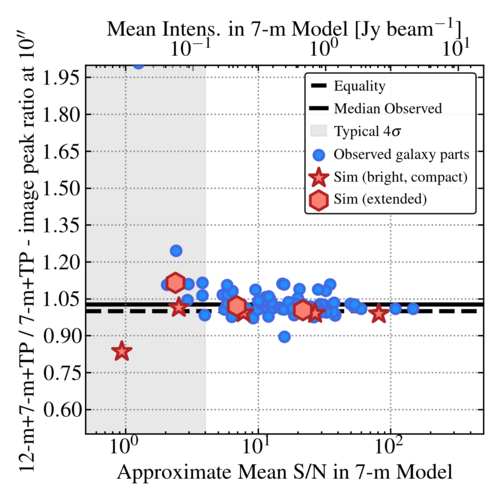}
\end{center}
\vspace{-0.1in}
\caption{
\textbf{Sum and peak flux in different arrays with and without feathering.} Similar to Figure~\ref{fig:arraydeconvolve} but now showing (\textit{top row}) the ratio of the total summed flux in the \mbox{12-m}+\mbox{7-m} image to that in the \mbox{7-m}-only image (\textit{top left}) without feathering and (\textit{top right}) with feathering. The \textit{bottom row} shows the ratio of \mbox{12-m}+\mbox{7-m} to \mbox{7-m} intensities at the peak of the \mbox{12-m}+\mbox{7-m} image after matching the resolution of the two images at $10\arcsec$. The \textit{bottom left} image shows the case before feathering and the \textit{bottom right} image shows the case after feathering. Again blue points show results for PHANGS-ALMA targets and red points show simulation results (Section~\ref{sec:endtoend}). The \mbox{12-m}+\mbox{7-m} imaging recovers more flux than the \mbox{7-m}-only imaging, with the two becoming more consistent as the signal-to-noise increases. After feathering, the two images are much more consistent for targets with reasonably high signal-to-noise, but still show some divergence in very low signal-to-noise cases. The effect appears much stronger in the integrated emission but is present at a lower level in the peak intensity of the image. Both simulations and real data show the effect, though it appears stronger in the real data due to the idealized nature of the simulations.
\label{fig:arraycompare}
}
\end{figure*}

We demonstrate this effect for both the real PHANGS-ALMA data and the simulation in Figures~\ref{fig:arraydeconvolve} and~\ref{fig:arraycompare}. The left panel in Figure~\ref{fig:arraydeconvolve} shows the ratio of cleaned flux in the \mbox{12-m}+\mbox{7-m} imaging to that in the \mbox{7-m}-only imaging. That is, the $y$-axis shows the ratio of the summed fluxes of the model image. The $x$-axis shows a quantity that traces the typical signal-to-noise of reconstructed emission in the image. Specifically, we calculate the intensity-weighted mean intensity in the \mbox{7-m} model via:

\begin{equation}
\label{eq:iwi}
\langle I_\nu \rangle = \frac{\sum_{\rm model} I_\nu^2}{\sum_{\rm model} I_\nu}~,
\end{equation}

\noindent where $I_\nu$ is the intensity of a voxel in the model. Then $\langle I_\nu \rangle$ is just the weighted average intensity in the model, and so captures the typical brightness in the image. In Figures~\ref{fig:arraydeconvolve} and~\ref{fig:arraycompare}, we plot $\langle I_\nu \rangle$ on the upper $x$-axis. The lower $x$-axis shows $\langle I_\nu \rangle / \sigma$, where $\sigma = 0.031$~Jy~beam$^{-1}$ is a typical noise level across our full \mbox{7-m} imaging data set. In both figures, we plot each imaged galaxy part with both \mbox{12-m}+\mbox{7-m} and \mbox{7-m}-only as a blue point. The simulations described in Section~\ref{sec:endtoend} appear as red points, with different shapes reflecting our two model distributions at different signal-to-noise levels. 

\medskip

\noindent \textbf{Imaging the \mbox{7-m}-only data deconvolves less flux than imaging the \mbox{12-m}+\mbox{7-m} data:} The left panel of Figure~\ref{fig:arraydeconvolve} shows that in both the simulations and the real data, the \mbox{12-m}+\mbox{7-m} imaging consistently deconvolves more flux than the \mbox{7-m}-only imaging. The effect shows a clear anti-correlation with the typical signal-to-noise in the data, such that bright targets show a much better match between the two deconvolved images than fainter targets. For low brightness sources, the \mbox{7-m} model can contain as little as $\sim 60\%$ of the \mbox{12-m}+\mbox{7-m} flux. Across the whole sample, the \mbox{7-m}-only imaging deconvolves a median $73\%$ of the flux deconvolved by the \mbox{12-m}+\mbox{7-m} imaging.

The red points in Figure~\ref{fig:arraydeconvolve} show that this effect occurs in the simulated data, too. The deconvolved fluxes in Table~\ref{tab:endtoend} also highlight this result. The simulations also show significant spread at fixed signal-to-noise, demonstrating that source structure plays a large role in deconvolution. The NGC~3059 lookalike galaxy has extended structure within individual channels and shows more severe discrepancies at a given signal-to-noise than the compact NGC~1097 lookalike. Similarly, the large spread in the real data at any given signal-to-noise level likely reflects differences in the source structure within individual channels. Consistent with this interpretation, the model for our compact simulation, NGC~1097, shows some of the best agreement between the \mbox{12-m}+\mbox{7-m} and \mbox{7-m}-only data across the whole sample. 

The right panel in Figure~\ref{fig:arraydeconvolve} shows that the effect is even present, though with smaller magnitude, at the peak of the map. Here we plot the ratio of peak intensities in the two models after convolving both to $10\arcsec$ resolution. On average, the \mbox{7-m}-only imaging achieves a peak intensity ${\sim}93\%$ of that found using the \mbox{12-m}+\mbox{7-m} data at matched scales. Again, we observe an anti-correlation of the effect with signal-to-noise, but with significant source-to-source scatter.

The \mbox{7-m}-only imaging shows these shortcomings despite the fact that we clean the \mbox{7-m} data to the $1\sigma$-level during the single scale clean (Section~\ref{sec:imaging}), and none of our by-hand attempts to improve the cleaning yielded systematically better results for the \mbox{7-m}-only data. Although we do require nearby $4\sigma$ emission to conduct the single scale clean, our by-eye assessment of the residuals does not reveal any systematic isolated $<4\sigma$ emission that could explain this behaviour. Similarly, the \mbox{7-m} images produced by our pipeline appear to compare favorably to the observatory-delivered products. In short, we have no reason to think that a simple algorithmic fix can address the issue despite exploring several options. On the other hand, the \mbox{12-m}+\mbox{7-m} data show excellent match to the input model in simulations and do not exceed the overall flux constraints set by the total power data. These \mbox{12-m}+\mbox{7-m} images do, in fact, appear to represent our best images and the combined arrays do a better job at flux recovery than the \mbox{7-m} data alone.

Our best explanation for the issue is that the \mbox{12-m}+\mbox{7-m} data have significantly better sensitivity on the relevant spatial scales to reconstruct emission from galaxies. Even at short $u{-}v$ separation distances, the \mbox{12-m}-only baselines add significant sensitivity and lead to a synthesized beam with fewer strong sidelobes. The \mbox{7-m} data on the other hand, have less sensitivity on scales matched to the emission and poorer $u{-}v$ coverage than the \mbox{12-m}+\mbox{7-m} data. As a result, our deconvolution recovers less flux in the \mbox{7-m}-only image than the \mbox{12-m}+\mbox{7-m} image.

While this qualitative explanation seems reasonable, we were surprised by the magnitude of the effect in the real data. Our current understanding is that for realistic structures in nearby galaxies, the non-linear nature of the deconvolution procedure interacts in a destructive way with the limited $u{-}v$ coverage and sensitivity of the \mbox{7-m} array. This agrees qualitatively with previous investigations on similar topics, which show a strong non-linearity in interferometric image reconstruction of complex sources when dealing with limited coverage and sensitivity \citep[in particular, see][]{HELFER02}. The fact that the idealized simulations show qualitatively similar results to the data underscore that this is not an effect driven solely by calibration issues, limited knowledge of the ALMA primary beams, or some similar issue that would only affect the observations. These issues may still play important secondary roles, however. In particular, we expect that the combined impact of phase and amplitude noise might also lead to non-linearities in the sensitivity of the interferometer to any extended strutures at low signal-to-noise \citep[e.g., see][for a discussion of the nature of amplitude noise]{LAY94}. More careful analysis of low signal-to-noise simulations with phase noise might help this situation.

\medskip

\noindent \textbf{In the real data, the images show the same effect:} The top left panel of Figure~\ref{fig:arraycompare} shows the ratio of total flux in the \mbox{12-m}+\mbox{7-m} image to that in the \mbox{7-m}-only image. The lower left panel shows the ratio of peak intensities between these two images at matched $10\arcsec$. In other words, the left column of Figure~\ref{fig:arraycompare} shows the same results as Figure~\ref{fig:arraydeconvolve} but for the real data, with the residuals added back in to the deconvolved model.

Including the residual emission improves the situation by a small amount for the real data. However, a significant overall offset between the total flux in the \mbox{12-m}+\mbox{7-m} and \mbox{7-m} images remains visible in the top left panel of Figure~\ref{fig:arraycompare}. The simulations show a much larger improvement when the residuals are included. We understand this to reflect that the residuals in the simulations are idealized relative to those in the real data. We expect this because our simulations neglect phase noise. The simulations also achieve better rotation synthesis, and so better $u{-}v$ coverage, than the real data because the simulations observe a long continuous \mbox{7-m} block around transit while the real observations observe short blocks at random times. Still, the simulations continue to show a significant effect in the final images. The lower panel shows that effect also appears for the peaks of the image in the real data. 

\medskip

\noindent \textbf{Feathering largely corrects the issue, but not at the lowest signal-to-noise levels:} In PHANGS-ALMA, we always attempt to include total power observations. These serve to anchor the total flux in the images and also to provide short spacing information in the final image. The right column of Figure~\ref{fig:arraycompare} shows the ratios of flux (top) and peak intensity (bottom) between the \mbox{12-m}+\mbox{7-m} and \mbox{7-m} data after feathering.

The figures show that in cases with mean intensity levels $\gtrsim 4\sigma$ the flux discrepancy almost vanishes after feathering. In the lower significance cases, we expect that the overall signal-to-noise in the cube is so low that statistical uncertainties in, e.g., masking to calculate the total flux may drive some of the visible scatter. The bottom right panel shows that the agreement in the peak intensity also becomes much better with feathering, implying that the short spacing correction helps locally, not only for the global flux. Overall, feathering reduces discrepancies in the integrated flux to $\lesssim 5\%$.

\medskip

\noindent \textbf{Synthesis:} These results clearly demonstrate that the deconvolution of \mbox{7-m}-only data suffer from significant shortcomings and that these persist into the final images for the real \mbox{7-m}-only data. Because the discrepancies are largely resolved by feathering, we expect our final data products, even those involving only the \mbox{7-m} array, to be largely correct. However, \mbox{7-m}-only images of nearby galaxies should be viewed with caution. Even the feathered data seem likely to harbor second-order fidelity issues, though a detailed investigation beyond what we present in Section~\ref{sec:endtoend} will have to wait for future work.

This analysis highlights a somewhat unexpected point. It would have been be easy to attribute the flux missed from the \mbox{7-m} deconvolution to ``spatial filtering'' that can only be addressed by including short spacing data. This does not appear to be the case. Instead, in galaxies with clumpy structure and strong velocity gradients, including sensitive \mbox{12-m} data significantly improves our flux recovery. 

%% file: ssctest.tex
\section{Testing Short Spacing Correction Methods}
\label{sec:ssc}

Correcting interferometric data for missing short and zero spacings is a key part of reconstructing the true intensity distribution on the sky. Several SSC methods have been proposed, and there is not yet a clear consensus on the best approach. In this appendix, we test the suitability of three of these methods for PHANGS--ALMA. Our goal here is not a thorough assessment of each method, which would require a large amount of research~\citep[e.g., see ALMA memos 398 and 488:][]{PETY01c,TSUTSUMI04}. Instead, we test how three popular methods work in a test case constructed to suffer from extreme spatial filtering. 

To conduct this test, we use the \texttt{CASA} task \texttt{simobserve} to simulate interferometric and total power observations of artificial intensity distributions. Then we image and combine these observations using different SSC techniques. We compare the recovered image to the known input image to evaluate the performance of each SSC technique.

\begin{figure}[t!]
\gridline{\fig{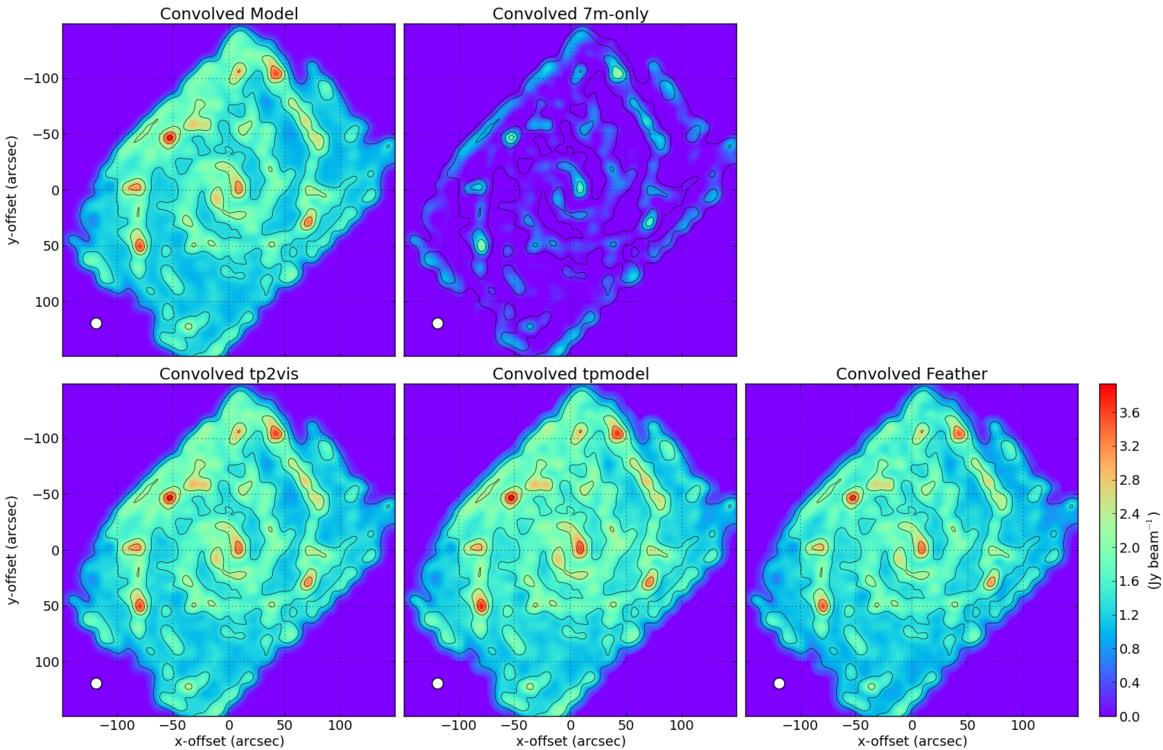}{1.0\textwidth}{}
}
\vspace{-24pt} % reduce excessive gaps between figure rows
\caption{\textbf{Results of short spacing correction tests for NGC~0628.} The \textit{top left} panel shows our model image convolved to the comparison resolution of $10''$. We construct the model image from a clipped version of the peak intensity map of the real combined \mbox{12-m}+\mbox{7-m}+TP PHANGS--ALMA \cotwo\ observations for this target. Because we drop all velocity information for this exercise and focus on analysis of relatively low resolution 7{-}m observations, the model shows positive emission essentially everywhere with very smooth structure. The \textit{top middle} panel shows the result of simulated observations using this input model and only the ACA \mbox{7-m} array with no total power information.  Significant missing flux can be seen in the image, indicating strong spatial filtering by the interferometer. The \textit{bottom row} shows short spacing corrected simulated observations. Each panel shows a different SSC method, from left to right: joint deconvolution using \texttt{tp2vis}, CLEANing using the simulated total power data as a model (\texttt{tpmodel}), and Fourier plane combination after deconvolution (\texttt{feather}, our adopted method in the PHANGS--ALMA pipeline). All images have the same $10''$ beam and the color scale is fixed across images to allow direct comparison. The white circle in the bottom left of each panel shows the beam.}
\label{fig:ssc_0628}
\end{figure}

\begin{figure}[t!]
\gridline{\fig{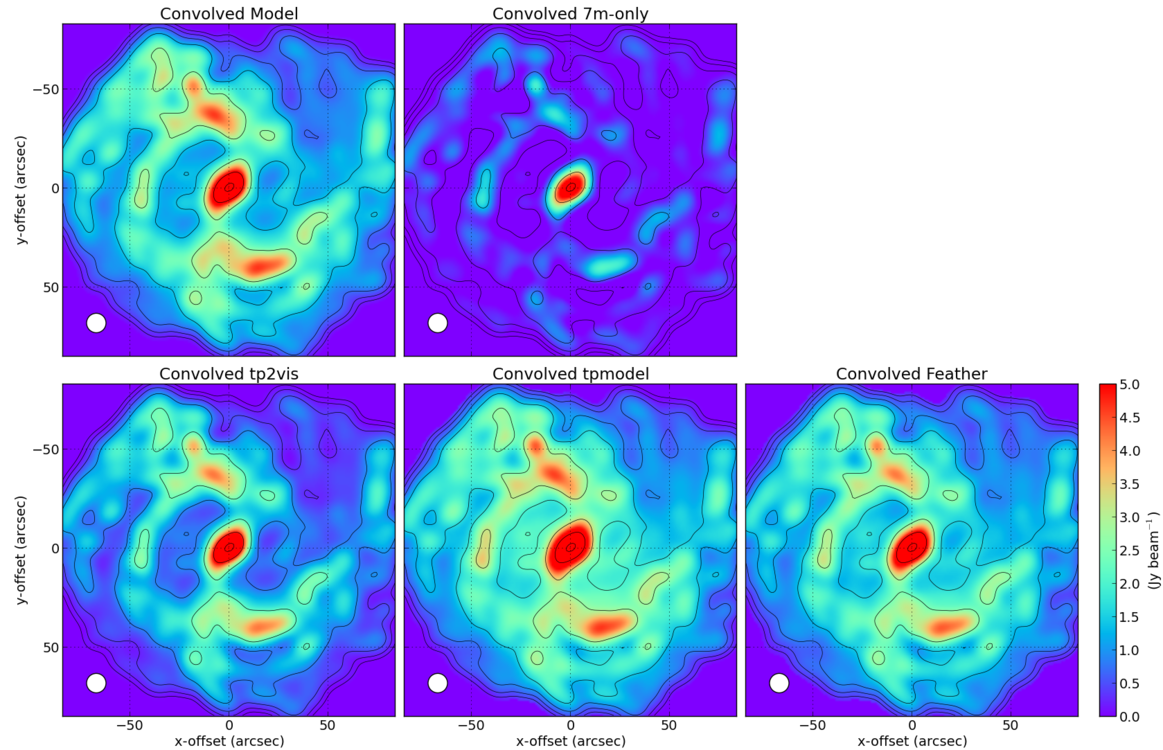}{1.0\textwidth}{}
}
\vspace{-24pt} % reduce excessive gaps between figure rows
\caption{\textbf{Results of short spacing correction tests for NGC~4303.} Similar to Figure~\ref{fig:ssc_0628} but for NGC~4303.}
\label{fig:ssc_4303}
\end{figure}

We conduct these tests using a kind of ``worst case'' scenario for spatial filtering in a data set like PHANGS--ALMA. For our model input images, we adopt peak intensity (see Section~\ref{sec:products}) maps derived from the real PHANGS--ALMA data. We apply some clipping to isolate significant emission, but remove all velocity structure from the map. This leads to a model with widespread positive emission across each entire mosaic, e.g., as seen in the top left panel in Figures~\ref{fig:ssc_0628} and~\ref{fig:ssc_4303}). This does not represent a truly realistic simulation of a galaxy. Our real targets show clumpy, sharp structure in individual velocity channels, while these models often show extended, relatively smooth structure on the scales accessed by the 7{-}m array used to carry out these tests. But these models should present a case that can serve as a useful test of short spacing correction algorithms. In that sense, this calculation complements the more realistic simulations in Section~\ref{sec:endtoend}, in which the interferometric imaging recovers a larger fraction of the flux seen by the total power data.

\medskip

\noindent \textbf{Simulations:} The simulations were set up as follows:

\begin{enumerate}
\item \textbf{Sample}: We run simulations considering 64 PHANGS--ALMA targets that had full \mbox{12-m}+\mbox{7-m}+TP \cotwo\ imaging available and clear CO detections as of Fall 2019.

\item \textbf{Sky Model}: For each galaxy, we use the \mbox{12-m}+\mbox{7-m}+TP peak \cotwo\ intensity map (Section~\ref{sec:products}) as the input sky model for \texttt{simobserve}. Before inputting them to the simulation, we clipped these images at a threshold corresponding to $3$~times the rms noise in the map. By eye, this did a reasonable job of mostly including real structure emission associated with the galaxy. Pixels with values below this threshold had their values set to zero. After clipping, we convert the units of the maps to Jy~beam$^{-1}$, appropriate for use with \texttt{simobserve}. We keep the sky coordinates the same as the true galaxy. For convenience, we set the source velocity to 0~km~s$^{-1}$. This choice should have no impact on the imaging.

\item \textbf{Interferometric Simulations}: We use \texttt{CASA} $5.4.0$ for the simulations, employing the task \texttt{simobserve} to construct the simulated measurement set and \texttt{tclean} for the subsequent imaging.  We define our own hexagonally-spaced mosaic grid to cover each input image. The spacing between neighboring pointings is set to $24\arcsec$, i.e., half-beam sampling for the \mbox{7-m} array. For this exercise, galaxies observe with multiple independent mosaics in the real data set, e.g., NGC~2903 (Figure~\ref{fig:linmos}), were treated as a single image and not simulated in separate parts.

We simulate observations with the Cycle~5 ACA \mbox{7-m}-array configuration (aca.cycle5.cfg\footnote{\url{https://almascience.eso.org/tools/casa-simulator}}) reflecting that most PHANGS--ALMA {7-m} observations were obtained during Cycle~5 in the context of our Large Program. 

The integration time of each $u{-}v$ data point is set to 10~seconds. The total observing time is set to 4~hours, which represents a reasonable match to the typical PHANGS--ALMA \mbox{7-m} observing time per target. The source transits at the midpoint of the simulated observations. The observations thus happen at the highest possible elevation for each source.

We did not add thermal or phase noise. Instead we concentrate on evaluation of SSC methods in the case of ideal observations. Aside from simply creating scatter in the measurements, we expect that that the main effect of adding thermal and phase noise would be to add uncertainty to the deconvolution procedure. Thus we prefer to focus on only the short-spacing correction here.
\item \textbf{Interferometric imaging:} 
We image the simulated visibility data using \texttt{tclean}. As in the PHANGS--ALMA pipeline (see Section~\ref{sec:imaging}), we start with a multiscale devoncolution using \texttt{tclean}, setting the \texttt{scale} parameter set to \mbox{[0,\,2,\,4]} pixels, with the pixel scale set to $1''$. We clean until the peak residual is $\leq 4$ times the noise determined from the input image before clipping. We then continue with a single-scale (``H\"{o}gbom'') \texttt{tclean}. This proceeds until the peak of the residuals  $\leq 1$ times the rms noise from the original image.

The imaging adopts Briggs weighting with $\texttt{robust} = 0.5$. We use $\texttt{cyclefactor} = 4$, $\texttt{cell} = 1.0\arcsec$, $\texttt{cycleniter} = 100$, and $\texttt{gain} = 0.2$ (resp.\ $0.1$) for multiscale clean (resp.\ single-scale clean). We do use a clean mask, which we create by convolving the input image to the synthesized beam size and detecting the significant pixels in this smoothed image. 

\item \textbf{Total power data for simulations}: We simulate ideal single dish observations by convolving the input sky image with the beam of the total power telescopes, $28.6\arcsec$ at $230.5$~GHz. 
\end{enumerate}

\medskip

\noindent \textbf{Short Spacing Correction:} There are currently at least three popular methods for short spacing correction for ALMA. \texttt{CASA}'s \texttt{feather} routine represents the path recommended by many observatory guides and documentation. Alternatively, the \texttt{tp2vis} method offers the most advanced current implementation of joint deconvolution. A third path uses the input clean model to incorporate information on extended emission. Variations exist on each of these techniques, but broadly they span the range of current approaches in wide use. Before describing our results, we briefly describe each:

\begin{enumerate}
\item \textbf{Joint Deconvolution (\texttt{tp2vis})}:
\citet{KODA19} present and give a full description of the \texttt{tp2vis}. This method converts a total power map into visibilities via a ``simple'' deconvolution of the total beam from the data in the Fourier plane. Then the data are multiplied by the primary beam of the interferometric dish in the image plane, in this case the \mbox{7-m} antenna. Finally, the Fourier plane is fully sampled to produce a visibility table that reflects the total power data. The weight density of the total power visibilities is adjusted to match that of the interferometric visibilities. The total power visibilities are then merged with the interferometric ones using the \texttt{CASA} task \texttt{concat}. 

We use the same imaging scheme described above to image the combined interferometric and total power data set produced by \texttt{tp2vis}.
\item \textbf{Model-assisted CLEAN (\texttt{tpmodel})}:
This method utilizes a user-defined image as an initial CLEAN model. This has been used in various forms for several years (see, e.g., \citealt{DIRIENZO15}). 

In our implementation, we pass the simulated total power image to \texttt{tclean} via the \texttt{startmodel} parameter. This initializes the CLEAN model, i.e., the deconvolved image, to the total power image. This input model is Fourier transformed and subtracted from the interferometric visibilities before the imaging proceeds. The imaging proceeds as normal, modifying the initial model until it achieves a good match to the interferometric data. Thus, this procedure essentially gives priority to the interferometer data and uses the total power as an additional guess to fill in missing information. 

In practice, we first convert the units of the input total power image from Jy~beam$^{-1}$ to Jy~pixel$^{-1}$, because \texttt{CASA} tracks the CLEAN model in these units.
\item \textbf{Feathering (\texttt{feather})}:
In feathering, the total power and interferometric images are combined in the Fourier domain after imaging and deconvolution \citep[e.g.,][]{BAJAJA79,COTTON17}. The main difference between \texttt{feather} and \texttt{tp2vis} is that with \texttt{feather} data combination happens after deconvolution, so that feather does not represent ``joint'' deconvolution. The advantage of this approach is its simplicity and robustness. This combination approach is adopted in much of the \texttt{CASA} documentation and has been commonly used in \texttt{MIRIAD} as well, where a version is implemented as \texttt{immerge}.

In practice, we run \texttt{feather} following the default \texttt{CASA} approach, matching the \texttt{CASA} guides. We set $\texttt{sdfactor} = 1.0$, $\texttt{effdishdiam} = -1.0$, and $\texttt{lowpassfiltersd} = {\rm False}$. As in the main PHANGS--ALMA pipeline processing, we apply the primary beam correction before feathering. We found that this leads to better recovery of the source structure near the edge of the mosaicked field of view.
\end{enumerate}

For each target, we use all three methods to create a short spacing corrected image. The \texttt{tp2vis} method produces images with a slightly larger synthesized beam size than the other two methods. Therefore we convolve all of the short spacing corrected images to $10\arcsec$ to allow a direct comparison of the methods.

\medskip

\noindent \textbf{Evaluation of Results:} 
Figures~\ref{fig:ssc_0628} and~\ref{fig:ssc_4303} show the results from our experiment for two galaxies: NGC~0628 and NGC~4303. The figures show the convolved input images, the output of imaging using only the \mbox{7-m} data with no SSC, and the results for each of the three SSC approaches. All panels have the same beam and intensity scale to allow direct comparison.

 The top row of Figures~\ref{fig:ssc_0628} and~\ref{fig:ssc_4303} show examples of the input model and output imaging, both convolved to $10''$ resolution, or slightly coarser than the typical resolution of the 7{-}m array at 230~GHz. In both targets, the \mbox{7-m}-only image shows extended negative sidelobes, or ``negative bowls,'' in place of extended positive emission in the input image. This illustrates how our choice to collapse the emission to a single plane yields an image that appears positive essentially everywhere. In both galaxies, significant, positive emission pervades the map at the resolution of the 7{-}m array. As a result, the interferometric images visible in the top middle panel show much stronger spatial filtering than we observe in our actual PHANGS--ALMA data. This highlights that the total power information is crucial, though we again caution that we have created a scenario that made these effects severe.

To the eye, all the short spacing corrected images appear fairly similar. This suggests that all three SSC methods generally do a good job of recovering the input sky model with minor differences. However, when examined in detail, there are differences between the three results. In the rest of this appendix, we quantify the differences among the different SSC results using three metrics. 

We quantify the results for each SSC method using three measurements: the ``image fidelity,'' the fraction of the total model flux recovered, and the difference in peak intensity between the output and the model.

We calculate the image fidelity at each pixel defined, e.g., following ALMA memo 386~\citep{PETY01a}, as:
\begin{equation}
\label{eq:fidelity}
{\rm fidelity} = \frac{\rm |input\:model|}{\rm |input\:model - output\:image|}.
\end{equation}
Thus fidelity of 10 means that the difference between the image and the model is $9\%$, fidelity of 100 corresponds to a difference of $\sim 1\%$, and fidelity of 1 corresponds to a  50\% difference. More generally, the higher the fidelity, the better the output image matches the model.  In practice, we compute the median fidelity over the entire image to quantify the overall quality of each SSC method.

\begin{figure}[t!]
\gridline{\fig{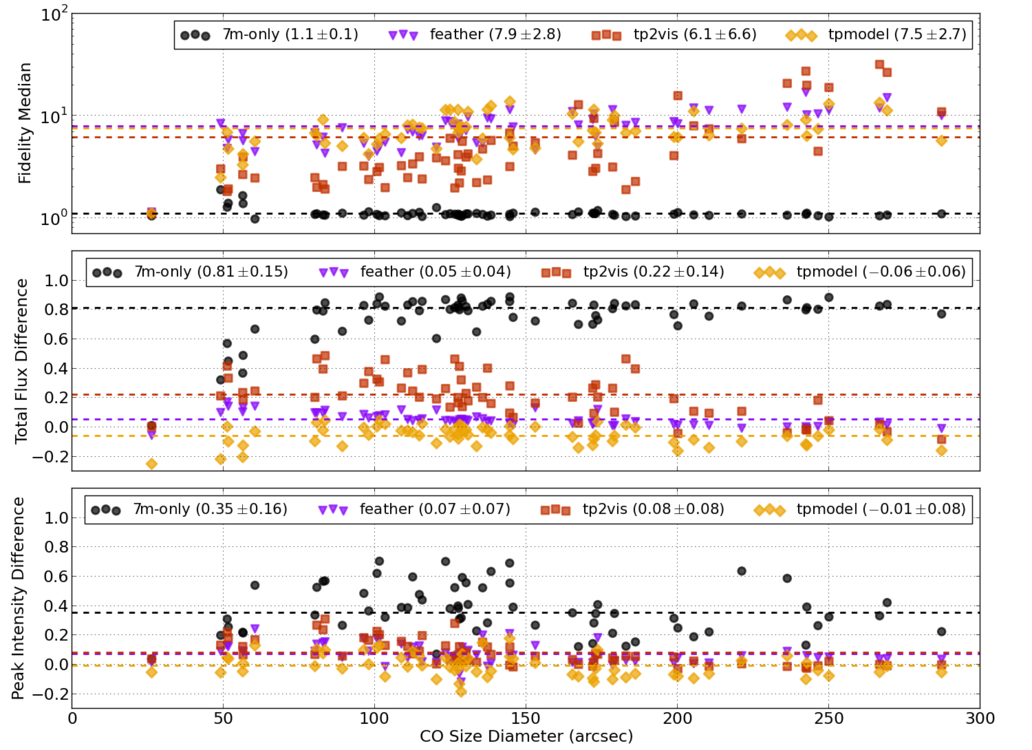}{1.0\textwidth}{}
}
\vspace{-24pt} % reduce excessive gaps between figure rows
\caption{\textbf{Measurements of the effectiveness of short spacing corrections.}
Results for simulated imaging of PHANGS--ALMA data using three short spacing correction schemes (\texttt{tp2vis}, \texttt{tpmodel}, and \texttt{feather}) and imaging without any short spacing correction (``\mbox{7-m}-only''). Each point shows a result for one reconstruction algorithm applied to simulated images based on one PHANGS--ALMA galaxy. The color of the point indicates the algorithm used for short spacing correction. The points are sorted by the diameter of CO emission from that galaxy. The \textit{top} row shows median image fidelity (Equation~\ref{eq:fidelity}). The \textit{middle} row show the fraction of flux missed in the output image compared to the input model; e.g., $0.0$ means no missing flux while $0.25$ means $25$\% of the flux is missed in the output. The \textit{bottom} row shows the fractional difference in peak intensity between the output image and the model; again $0.25$ indicates a $25$\% lower peak intensity in the output compared to the model. Dashed lines indicate the mean performance of each method for each metric and the legend reports the mean and standard deviation for each method and metric. }
\label{fig:ssc_fidelity}
\end{figure}

The top panel of Figure~\ref{fig:ssc_fidelity} shows the median image fidelity as a function of the size of the CO disk for the 64 PHANGS--ALMA targets in our sample. We pick CO disk size as our independent variable because we expect targets with more extended emission, i.e., a larger CO disk size, to show more spatial filtering and thus be more dependent on the SSC. We computed the size as the diameter that contains all the non-zero pixels in the model image. 

As expected, the \mbox{7-m}-only images show extremely low  median fidelity, $1.1$ with rms scatter $0.1$ across the sample. This corresponds to the output image differing by ${\sim}50$\% from the input image, on average. This is consistent with the visual appearance of strong spatial filtering seen comparing the top left and top middle panels in Figures~\ref{fig:ssc_0628} and~\ref{fig:ssc_4303}. The interferometer misses the large-scale structure that contributes much of the flux in the model.

All of the three SSC methods show much higher median fidelity than the interferometer data alone, with average values in the range $6.1{-}7.9$). This implies that the reproduced images are consistent with the model within a ${\sim}10{-}15$\% accuracy. They show high scatter from target-to-target, however. In general, small targets show lower fidelity, even after SSC correction, compared to large targets.

The middle panel of Figure~\ref{fig:ssc_fidelity} shows the difference in total flux between the output image and the model as a function of the CO disk size. We define:
\begin{equation}
\label{eq:fluxdiff}
{\rm total~flux~difference} = \frac{{\rm sum~of ~input~\:model - sum~of\:~output~image}}{\rm sum~of~input\:model}~.
\end{equation}
In this case $0.0$ corresponds to a perfect match between input and output images.

On average, the 7-{m}-only images miss $\sim 80\%$ of the flux. The smallest sources show better flux recovery in these images, but for virtually every source with CO disk diameter $\gtrsim 75''$ roughly $80\%$ of the flux is missed. Recall that our test cases have discarded velocity information and so represent a worst case. In the actual PHANGS--ALMA data, the 7{-}m array recovers 50{-}80\% of the flux observed by the single dish (Figure~\ref{fig:recovery}) and some of the missed emission appears to be due to deconvolution effects rather than spatial filtering (Appendix~\ref{sec:arrays} and Section~\ref{sec:endtoend}).

The bottom panel in Figure~\ref{fig:ssc_fidelity} shows the difference in peak intensity between the output image and model. Again we define:
\begin{equation}
\label{eq:peakdiff}
{\rm peak~intensity~difference} = \frac{{\rm peak~in~input~model - peak~in~output~image}}{\rm peak~in~input~model}~.
\end{equation}
This statistic captures the ability of the output image to capture the brightest emission in the image. On average the 7{-}m-only imaging recovers a peak flux $35\%$ lower than in the model. Almost all 7{-}m images show a depressed peak flux. This shows that including total power data can be crucial even for studying compact sources. This will be especially true when, e.g., as is the case for galactic nuclei, these sources are surrounded by extended diffuse structures.

The peak fluxes of the images reconstructed by the three SSC methods agree with the model peak flux within ${\sim}10$\%. The recovered peak flux does not depend on CO disk size or other galaxy parameters (e.g., total flux, peak flux, and average flux).

Combining all three metrics, we can evaluate each of the SSC methods. Before noting a few shortcomings, we emphasize that all three methods represent a marked improvement over the imaging using only the 7{-}m data with no SSC.

\texttt{tpmodel}: Images reconstructed using the \texttt{tpmodel} method show comparable median fidelity to those reconstructed via \texttt{feather}. They show the best overall match to the data in peak intensity. However, the middle panel of Figure~\ref{fig:ssc_fidelity} shows that the \texttt{tpmodel} method does tend to recover slightly too much flux compared to the model. The total flux is overestimated by ${\sim}6$\% on average and by as much as ${\sim}25$\% in the most extreme cases. This can be attributed to the fact that the total power data input as a model have a much larger beam size than the CLEAN product, which creates an extended artifact surrounding strong peaks\footnote{\href{https://github.com/teuben/dc2019/blob/master/talks/Kauffmann.pdf}{https://github.com/teuben/dc2019/blob/master/talks/Kauffmann.pdf}}. Some groups have adopted iterative strategies to address this, e.g., using feathered data from a previous iteration of clean as a model \citep[e.g.,][]{BOLATTO13B,LEROY15A}. We experimented with these methods and did not find a stable, general approach, but this might represent an interesting future direction.

\texttt{tp2vis}: Images reconstructed using \texttt{tp2vis} show a high median fidelity, but with a larger dispersion than the other two SSC methods. The middle panel also shows that with \texttt{tp2vis} one tends to underestimate the total flux in our simulations by ${\sim}22$\% on average and by up to ${\sim}50$\% in the most extreme cases. \texttt{tp2vis} does a good job of recovering the peak intensity, comparable to \texttt{feather} and slightly worse than the \texttt{tpmodel} approach. 

We interpret the large scatter to mean that \texttt{tp2vis} may require some fine-tuning of parameters for each source. Given the computing requirements for one round of imaging and the fact that \texttt{tp2vis} did not offer a clear improvement over the other algorithms, it was not realistic to apply this fine-tuning to the current round of PHANGS--ALMA imaging. This might represent a useful future direction.

\texttt{feather}: Images reconstructed using \texttt{feather} yield a high median fidelity, ${\sim}7.9$, comparable to \texttt{tpmodel} and with a similar scatter. On average, \texttt{feather} recovers the total flux with similar accuracy to \texttt{tpmodel} and somewhat better than \texttt{tp2vis}. It shows lower scatter in total flux recovery than either of the other two methods. Feather tends to recover peak intensities ${\sim}7\%$ too low, similar to \texttt{tp2vis} and slightly worse than \texttt{tpmodel}.

\medskip

\noindent \textbf{Summary:} 
In summary, all three SSC methods represent a marked improvement over using only the \mbox{7-m} data for these cases. They yield results consistent with one another at the ${\sim}10$\% level. For PHANGS--ALMA, we ultimately utilized \texttt{feather} because it is stable and simple with consistent performance across the sample. For ALMA, which has good overall flux calibration and a consistent flux calibration scheme for both the interferometer and total power, \texttt{feather} has the additional advantage of not requiring additional human supervision or intervention. 

On average, a feathered image in our experiment has ${\sim}5\%$ bias in recovering the total flux and ${\sim}7\%$ underestimate of the peak flux. The median deviation from the input image is ${\sim}12\%$ based on the image fidelity calculation. As emphasized above, these calculations significantly overstate the uncertainty because we discard velocity information (Section~\ref{sec:postprocess}). Still, they suggest a $5{-}10$\% uncertainty associated with image reconstruction. This will be comparable to the uncertainty due to calibration uncertainty and thermal noise in many cases, and highlights the need for continued work on this topic.